\documentclass[a4paper,twocolumn,accepted=2025-09-10]{quantumarticle}
\pdfoutput=1
\usepackage{silence}
\usepackage{amsmath, amsfonts, amssymb}
\usepackage{soul}
\usepackage{dsfont}
\usepackage{enumitem}
\usepackage[colorlinks={true}, citecolor={blue}, filecolor={blue}, linkcolor={blue}, urlcolor={blue}]{hyperref}
\usepackage{braket}
\usepackage{color}
\usepackage{amsmath,amsfonts}
\usepackage{amssymb,amsthm,times}
\usepackage{bbold}
\usepackage{listings}
\usepackage{float}
\usepackage{braket}
\usepackage[utf8]{inputenc}
\usepackage[english]{babel}
\usepackage[T1]{fontenc}
\usepackage{graphicx}
\usepackage{dcolumn}
\usepackage{bm}
\usepackage{microtype}
\usepackage{xcolor}
\usepackage[font=footnotesize,caption=false, justification=centerlast]{subfig}
\usepackage[normalem]{ulem}

\mathchardef\mhyphen="2D

\DeclareMathOperator{\Tr}{Tr}

\newcommand{\abs}[1]{\left|#1\right|}
\newcommand{\XsqRMT}{X^{2}_{\rm RMT}}

\newcommand{\norm}[1]{\left\lVert#1\right\rVert}

\renewcommand{\Re}{\mathop{\mathrm{Re}}}
\renewcommand{\Im}{\mathop{\mathrm{Im}}}

\begin{document}

\title{Quantum Chaos and Universal Trotterisation Behaviours in Digital Quantum Simulations}

\author{Cahit Kargi}
\email{cahit.kargi@student.uts.edu.au}
\affiliation{Centre for Quantum Software and Information \& School of Mathematical and Physical Sciences, Faculty of Science, University of Technology Sydney, New South Wales 2007, Australia}
\affiliation{Sydney Quantum Academy, Sydney, New South Wales, Australia}

\author{Angsar Manatuly}
\email{angsar.manatuly@student.uts.edu.au}
\affiliation{Centre for Quantum Software and Information \& School of Mathematical and Physical Sciences, Faculty of Science, University of Technology Sydney, New South Wales 2007, Australia}
\affiliation{Sydney Quantum Academy, Sydney, New South Wales, Australia}

\author{Lukas M. Sieberer}
\affiliation{Institute for Theoretical Physics, University of Innsbruck, 6020 Innsbruck, Austria}

\author{Juan Pablo Dehollain}
\affiliation{Centre for Quantum Software and Information \& School of Mathematical and Physical Sciences, Faculty of Science, University of Technology Sydney, New South Wales 2007, Australia}

\author{Fabio Henriques}
\affiliation{Centre for Quantum Software and Information \& School of Mathematical and Physical Sciences, Faculty of Science, University of Technology Sydney, New South Wales 2007, Australia}
\affiliation{Sydney Quantum Academy, Sydney, New South Wales, Australia}

\author{Tobias Olsacher}
\affiliation{Center for Quantum Physics, University of Innsbruck, 6020 Innsbruck, Austria}
\affiliation{Institute for Quantum Optics and Quantum Information of the Austrian Academy of Sciences, 6020 Innsbruck, Austria}

\author{Philipp Hauke}
\affiliation{INO-CNR BEC Center and Department of Physics, University of Trento, Via Sommarive 14, I-38123 Trento, Italy}

\author{Markus Heyl}
\affiliation{Max Planck Institute for the Physics of Complex Systems, N\"{o}thnitzer Str. 38, 01187 Dresden, Germany}
\affiliation{Theoretical Physics III, Center for Electronic Correlations and Magnetism, Institute of Physics, University of Augsburg, D-86135 Augsburg, Germany}

\author{Peter Zoller}
\affiliation{Center for Quantum Physics, University of Innsbruck, 6020 Innsbruck, Austria}
\affiliation{Institute for Quantum Optics and Quantum Information of the Austrian Academy of Sciences, 6020 Innsbruck, Austria}

\author{Nathan K. Langford}
\email{nathan.langford@uts.edu.au}
\affiliation{Centre for Quantum Software and Information \& School of Mathematical and Physical Sciences, Faculty of Science, University of Technology Sydney, New South Wales 2007, Australia}


\begin{abstract}

Digital quantum simulation (DQS) is one of the most promising paths for achieving first useful real-world applications for industry-scale quantum processors.
Yet even assuming continued rapid progress in device engineering and successful development of fault-tolerant quantum processors, extensive algorithmic resource optimisation will long remain crucial to exploit their full computational power.
Currently, among leading DQS algorithms, Trotterisation provides state-of-the-art resource scaling.
And recent theoretical observations of a distinct breakdown threshold in empirical performance for Trotterised Ising models suggest that even better performance than expected may be possible prior to the threshold.

Here, to start exploring this possibility, we study 
multiple paradigmatic DQS models with experimentally realisable Trotterisations, and 
provide strong evidence for universality of a range of Trotterisation performance behaviours, including not only the threshold, but also new features in the pre-threshold regime that is most important for practical applications.
In each model, we observe a distinct Trotterisation threshold shared across widely varying performance signatures;
we further show that an onset of quantum chaotic dynamics causes the performance breakdown and is directly induced by digitisation errors.
In the important pre-threshold regime, we are able to identify new distinct regimes displaying qualitatively different quasiperiodic performance behaviours, and show analytic behaviour for properly defined operational Trotter errors.
Our results rely crucially on diverse new analytical tools,
and provide a previously missing unified picture of Trotterisation behaviour across local observables, the global quantum state, and the full Trotterised unitary.
This work provides new insights and tools for addressing important questions about the algorithm performance and 
underlying theoretical principles of sufficiently complex Trotterisation-based DQS, that will 
help in extracting maximum simulation power from future quantum processors.

\end{abstract}

\maketitle
 


\section{Introduction}\label{Sec:Intro}


Quantum simulation offers exciting opportunities for quantum processors to tackle the ``in-silico'' modelling of complex quantum systems that is often intractable for even the most advanced classical computing technology~\cite{Georgescu2014QS}.
In analogue quantum simulation, a controllable, specialised surrogate is engineered to directly emulate the dynamical properties of the full target system, with rapid scaling of noisy intermediate-scale quantum (NISQ) processors being achieved for models in condensed-matter physics~\cite{GrossC2017AQSUC, BernienH2017MB51, ZhangJ2017AQS53, SepehrE2021QP256, PascalS2021QSRA} and lattice gauge theories~\cite{YangB2020ogi, MilA2020srl, BanulsMC2020slg}.
In digital quantum simulation (DQS)~\cite{LloydS1996UQS}, complex target dynamics is instead constructed in discrete steps from a sequence of simpler interaction primitives (gates) on a more generic processor.
This versatile approach offers universal programmability that is not limited to models which can be mapped \emph{in toto} onto systems realisable naturally in a laboratory environment, and has been used to simulate highly varied target models across quantum chemistry~\cite{AspuruGuzik2005SQME, OMalley2016QSME, Kandala2017HEQE, GoogleAI2020HF}, ultrastrong coupling~\cite{MezzacapoA2014DQRS, LangfordNK2017DQRS, LamataL2017DQSD}, condensed-matter~\cite{Lanyon2011UDQS, Heras2014DQS, Salathe2015DQS, BarendsR2015dqs, NeillC2016EDQS} and high-energy~\cite{Martinez2016LGTS} physics.

\begin{figure}[t]
	\centering
 	\includegraphics[scale=0.95]{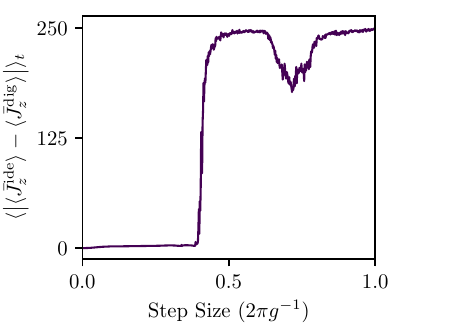}
 	\caption{Trotterisation performance for the A2A-Ising model (with $j=256$; see Sec.~\ref{Sec:Models} for definition), quantified in this figure with the average point-to-point absolute error in expectation values over $t = 200~(2\pi g^{-1})$ (see Sec.~\ref{Sec:ErrorAnalyses} for definitions and more detailed error analyses; data from Fig.~\ref{Fig:7_TrotterErrors}~(g)), breaks down rapidly after reaching a certain Trotter step size, observed as the large increase in the errors.}\label{Fig:1_ThresholdConcept}
\end{figure}

A digital gate-based approach offers compatibility with future universal quantum computers~\cite{LloydS1996UQS} and the capacity for error correction to overcome the noise and decoherence limiting the power of NISQ devices~\cite{Preskill2018NISQ}.
Yet it also introduces digitisation errors intrinsic to the algorithm which are reduced through finer and more sophisticated discretisation~\cite{LloydS1996UQS, TrotterHF1959PSGO, SuzukiM1976GTF, Suzuki1985DFQM, SUZUKI1990DMC, Suzuki1991GFMB, Childs2012LCU, Berry2015TTS, Low2017QSP, Low2019hamiltonian, Childs2018QSQS, Childs2020TTE}.
Typically, this increases the required experimental resources (in terms of gates, qubits or runtime) and the exposure to noise, thus creating a trade-off between sufficient discretisation (algorithmic errors) and acceptable hardware noise (gates, decoherence, etc).
In medium-term NISQ applications, DQS resource requirements can be somewhat reduced by combining the discretisation that increases versatility and accessible complexity, with analogue toolbox primitives to maximally exploit architecture-specific efficiencies~\cite{Lanyon2011UDQS, Heras2014DQS, Salathe2015DQS, NeillC2016EDQS, MezzacapoA2014DQRS, LangfordNK2017DQRS, LamataL2017DQSD}.
Ultimately, however, for both NISQ devices and even fully error-corrected quantum processors, simulation power will for a long time be limited by such resource constraints.
It is therefore an essential open problem to improve our fundamental understanding of digitisation error and how much can be tolerated for a given application, and if possible, how to compensate or correct it.

A key ongoing goal for DQS research is to better understand the computational performance of DQS algorithms with a view to bringing tasks of interest within reach of practical quantum processors and generally maximising their achievable computational power.
In deriving worst-case error bounds from the full Trotterised unitary, many early works~\cite{LloydS1996UQS, Berry2007EQA, Hastings2014QSQC, Wecker2014GCQS} predicted very poor scaling of Trotter errors with simulation time and system size, giving rise to very pessimistic assessments about the practical feasibility of DQS.
Over time, it has been shown that these bounds can be improved by means of more sophisticated implementations of Trotterised gate sequences~\cite{Raeisi2012ESMB, Hastings2014QSQC, Campbell2017SGMU, Campbell2019RCHS, Childs2019fasterquantum}, even to the extent that Trotterisation can match or outperform other state-of-the-art digitisation algorithms~\cite{Childs2018QSQS, Childs2020TTE}.
Yet exact numerical calculations for specific DQS models have shown that even the best rigorous bounds are often orders of magnitude "loose" compared with empirical performance~\cite{Raeisi2012ESMB, Poulin2014TSS, Babbush2015CBTE, Childs2018QSQS, Childs2019fasterquantum, Childs2020TTE}.
Indeed, recent studies of Trotterised Ising models~\cite{HeylM2019QLTE, SiebererLM2019DQC} have even observed a rather sharp empirical performance threshold (Fig.~\ref{Fig:1_ThresholdConcept}) that was not previously predicted by rigorous, worst-case bounds, which they link to a transition between regions of localised dynamics and quantum chaotic dynamics.
In addition, their observation of controllable, perturbative errors for some asymptotically time-averaged observables in the small-step regime, suggests that Trotterised simulations using large (but pre-threshold) step sizes may provide access to near-ideal (error-free) performance results through extrapolation.
These works highlight the need for a much better understanding of operationally relevant Trotterisation performance (not just worst case) in the context of concrete DQS models.

In this work, we carry out a broad and systematic study of empirical Trotterisation performance across diverse, paradigmatic models and a wide range of Trotter step sizes.
Synthesising the results from numerous different and complementary performance signatures, our work provides new insights into Trotterisation performance, including in the ``good simulation'' regime that is of key importance for DQS applications, as well as demonstrating marked universality for a range of performance behaviours, including both the threshold itself and new features in the pre- and post-threshold regimes.
Crucial to these new insights are both our systematic approach and our introduction of diverse new analytical tools, including new approaches for exploring detailed system dynamics, well-defined error metrics for observables, and quantitatively rigorous global signatures of quantum chaos based on random matrix theory (RMT).
These tools enable us to study the same behaviours in situations that cannot be straightforwardly generalised to a well-defined classical limit, which is critical for DQS, since interesting use cases most commonly correspond to models which cannot be simulated classically~\cite{BoixoS2018QSNT, Arute2019QS, WuY2021QASC, ZhuQ2021QCA60}.
In addition to providing a deeper understanding of Trotterisation performance, the advances in this work motivate a range of new and interesting directions.

The paper is organized as follows.
In Sec.~\ref{Sec:resultsOverview}, we provide, up front, an overview of our key results and contributions.
In Sec.~\ref{Sec:Background}, we provide key background information about Trotterisation, its Floquet interpretation, and quantum chaos (with more detailed background on quantum chaos in Appendix~\ref{App:RMTandQuantumChaos}).
Sec.~\ref{Sec:Models} describes the physical models and their experimentally accessible Trotterisations.
After presenting our detailed findings from individual analytical approaches in Sec.~\ref{Sec:Results}, we discuss in Sec.~\ref{Sec:Conclusion} the broader, overarching conclusions we can establish by combining our results, along with new and open questions raised by our work, and the outlook for addressing them.
Due to the significant extent of analysis and results reported in Sec.~\ref{Sec:Results} and the relevant appendices, Sec.~\ref{Sec:Conclusion} is written to be rather self-contained, so that the reader may skip to this section and understand how our results support the broad conclusions of our work, and delve into the detailed results as required.

\subsection{Overview of Results}\label{Sec:resultsOverview}

\subsubsection{Universality}
In this paper, we analyse varied, detailed aspects of Trotterisation performance for a wide range of concrete models to further explore the potential for improvements offered over generic bounds~\cite{Raeisi2012ESMB, Poulin2014TSS, Babbush2015CBTE, Childs2018QSQS, Childs2019fasterquantum, Childs2020TTE, HeylM2019QLTE, SiebererLM2019DQC}.
By studying experimentally realisable Trotterisations of the Heisenberg~\cite{Heras2014DQS, Salathe2015DQS}, Rabi-Dicke~\cite{MezzacapoA2014DQRS, LangfordNK2017DQRS, LamataL2017DQSD}, and all-to-all Ising (A2A-Ising)~\cite{SiebererLM2019DQC} models, we draw conclusions that apply to a wide range of contexts, such as influence of different symmetries, simulation with non-qubit components, systems with infinite-dimensional Hilbert spaces, and systems without clear classical limits.
We identify and provide strong evidence for the universality of a range of performance behaviours across system contexts, especially in the important pre-threshold regime of Trotterised DQS.
For example, we demonstrate that quantum chaotic dynamics emerges even in a highly structured and symmetric Heisenberg spin model that has no well-defined classical limit.
We show these behaviours are reflected consistently across diverse physical signatures and system properties, and use new tools to provide a consistent understanding across all signatures.
Importantly, our systematic and detailed analysis also allows us to identify new pre- and post-threshold behaviours in Trotterisation performance.

\subsubsection{Analysis techniques}
The systematic analysis underpinning our results was crucially enabled by the introduction of a range of new analytical tools and techniques covering both evolution dynamics and static properties of the Trotterised step unitary.
As well as filling in important gaps in results, these new approaches have opened up important new questions about DQS performance, while also providing valuable new tools for addressing them.
Similarly to techniques introduced to characterise quantum processors in terms of targeted signatures such as entanglement witnesses, gate and channel fidelities, much of our focus in this work is on more easily operationally measurable signatures (such as local observables).
These offer better experimental accessibility and scaleability over inefficient full-system characterisation, such as state or process tomography.
Such focussed approaches are particularly relevant and well suited to DQS, where the usual aim is to analyse and estimate the values or evolution of physical quantities of interest, such as ground-state energies~\cite{OMalley2016QSME, Kandala2017HEQE, GoogleAI2020HF} or a phase transition order parameter~\cite{BernienH2017MB51, SepehrE2021QP256, PascalS2021QSRA}.

\paragraph{Full system dynamics}
Studying the full system dynamics is arguably the most natural way to study DQS in experiments~\cite{Lanyon2011UDQS, Salathe2015DQS, BarendsR2015dqs, LangfordNK2017DQRS}, but can require some ingenuity to extract useful understanding from a large volume of data.
Our analysis shows that detailed dynamics can already directly reveal valuable insights and information both qualitative and quantitative in nature.
For example, we show that a shared threshold can be observed across all signatures studied, including the simulation fidelity, which was previously thought not to exhibit a persistent threshold based on analysis from time averages only~\cite{SiebererLM2019DQC}.
Studying the full dynamics, together with the corresponding Fourier spectra (only accessible from the full dynamics), also shows very clearly how quasiperiodic dynamics (with strongly clustered spectral frequencies) before the threshold, is destroyed after the threshold (with spectra becoming more equally spaced in frequency and flat in amplitude).
Arguably even more important are the insights we are able to obtain about Trotterisation performance in the pre-threshold regime, discussed further below.

\paragraph{Operational error metrics}
Finding the best ways to extract useful quantitative information about simulation errors in a condensed, digestible form from detailed system dynamics nevertheless remains an important challenge, especially for experiments where errors from unitaries~\cite{LloydS1996UQS, Berry2007EQA, Childs2020TTE} or the outcomes of phase estimation~\cite{AspuruGuzik2005SQME, OMalley2016QSME, Wecker2014GCQS} are often either inaccessible or resource intensive.
Previous work on the Trotterisation threshold therefore focussed primarily on time-averaged and asymptotic-time errors and local observables~\cite{HeylM2019QLTE, SiebererLM2019DQC}.
Yet practical DQS requires understanding performance over finite times, and we show here that the time-averaged quantities do not distinguish the targeted information from discrete sampling errors, which arise under time averaging even for ideal system dynamics.
We therefore introduce new point-to-point time-averaging metrics which properly disentangle Trotter errors from sampling effects.
These new metrics allow us to show that both local observables and simulation fidelities exhibit analytic, perturbative behaviours in the all-important pre-threshold regime, and demonstrate they depend in a sensible way on system size and simulation time, contradicting conclusions suggested by long-time error metrics~\cite{HeylM2019QLTE}.

\paragraph{Conclusive agreement with RMT signatures}
Given the suggested link~\cite{HeylM2019QLTE, SiebererLM2019DQC} between Trotterisation performance breakdown and the emergence of quantum chaos, a key aim of our research was to identify a system- and model-agnostic way to objectively diagnose agreement with quantum chaotic dynamics directly from the Trotterised unitary.
In particular, we were interested in obtaining definitive answers for cases at the modest system sizes relevant for currently accessible experiments, and cases without meaningful classical limit (of particular relevance for quantum computing and DQS, where quantum advantage demands systems cannot be represented classically~\cite{BoixoS2018QSNT, LeoneL2021QCQ}).
Unfortunately, while common dynamical signatures of quantum chaos (like fidelity decay and out-of-time-order correlators) can be intuitive, they provide only indicative and inconclusive evidence for quantum chaos~\cite{EmersonJ2002FDEI, HaakeF2018QSC, LucaD2016ETH}.
The results can also be highly initial-state dependent, making it unwieldy to use dynamical signatures to extract information about the global dynamics.
References~\cite{HeylM2019QLTE, SiebererLM2019DQC} also study signatures based on the underpinning RMT description of quantum chaotic dynamics in terms of random unitary evolution, using characteristic derived RMT quantities such as average level (eigenphase) spacings.
In this work, we introduce a new, quantitatively rigorous diagnostic signature for quantum chaos in unitary quantum evolution, based on RMT predictions of the eigenvector statistics for a random unitary operator~\cite{IZRAILEVFM1987CSEF, KusM1988UEV} (see also Appendix~\ref{App:EigenVecTheory}).
Our approach defines a reduced chi-squared goodness-of-fit test statistic, $\XsqRMT$, to provide a statistically objective, ``black-box'' assessment of how well a given unitary operator agrees with the predictions for a particular RMT universality class.
This has two important benefits: it diagnoses agreement at the full statistical distribution level, and therefore avoids ambiguities that arise from studying individual statistical moments such as mean values, and benefits from the existence of closed-form analytical predictions for RMT eigenvector statistics that do not exist for RMT eigenvalue statistics.
$\XsqRMT$ allows robust statistical RMT comparisons for individual unitary operators, almost arbitrary system dimensions, all system and model types, and irrespective of whether a meaningful classical limit exists.
Our work lays the groundwork for broader use of this technique in diverse areas such as quantum complexity in quantum computing~\cite{BoixoS2018QSNT, Arute2019QS, WuY2021QASC, ZhuQ2021QCA60, DowlingN2022qcv}, Floquet physics and many-body physics~\cite{Abanin2015, Mori2016, Kuwuhara2016, Abanin2017}.

\subsubsection{Threshold to quantum chaos}
The main results from the two previous works on Ising model DQS by some of the authors can be broadly categorised as follows:
Reference~\cite{HeylM2019QLTE} demonstrated the existence of a previously unobserved performance threshold marking a transition between localised and delocalised dynamical regimes, along with indirect dynamical signatures of quantum chaotic behaviour in the delocalised regime.
Then, by considering the special case of the A2A-Ising model, Ref.~\cite{SiebererLM2019DQC} showed that the Trotterised model and its threshold could be understood in terms of quantum chaos in the kicked top based on its well-studied classical limit, but this connection does not generalise to other models.
In this work, using the results from our new tools across varied model types, we demonstrate a deeper and more conclusive generic connection between DQS performance thresholds and the onset of quantum chaos than previous works~\cite{HeylM2019QLTE, SiebererLM2019DQC}, including for modest system sizes and for systems with no clear classical limit.
Specifically, we see both quasiperiodic (non-quantum-chaotic) behaviour (from full dynamics and Fourier spectra) and clear analytic errors (from properly defined operational metrics) before the threshold, and introduce a new analysis method showing quantitative and objective agreement with quantum chaos in Trotterised unitaries after the threshold.
Thus, we see that the onset of quantum chaos at the threshold marks a fundamental transition from a regime of ``meaningful'' approximate simulation of some target unitary, to one categorised by uncorrectable and uncontrollable errors, in other words a total breakdown in simulation performance.
And by deliberately studying regular (non-quantum-chaotic) target models, we demonstrate that this onset is caused directly by Trotterisation errors.

\subsubsection{Unified picture of Trotterisation behaviours}
Combining the results from our systematic analysis, across diverse system and model types and signatures, has also allowed us to develop a new, unified understanding of Trotterisation performance and behaviour, across local and global signatures, and state-specific versus process level measures.
Previous analysis suggested that errors in local observables were surprisingly robust against both system size and simulation time before the threshold, in stark contrast to both the full unitary dynamics and the simulation fidelity of the global quantum state, with a fragile simulation fidelity threshold being swamped by pre-threshold errors and no longer even observable at larger system sizes~\cite{HeylM2019QLTE, SiebererLM2019DQC}.
In this work, however, we not only demonstrate clear signatures of the shared threshold for local observables, the global quantum state, and the full Trotterised unitary, but also show that observables and state fidelities both exhibit clear analytic behaviours and mutually consistent size- and time-dependence.
We also show that seemingly inconsistent simulation fidelity behaviours can be explained in terms of clear new pre-threshold performance regimes that we can identify from the full dynamics.
Also, the coherent periodic behaviour of the simulation fidelity before the threshold provides suggestive opportunities for studying possible perturbative mitigation of Trotterisation errors at the state level (not only for the observables).
These results provide significant new insight about markedly universal Trotterisation performance behaviours especially in the pre-threshold regime that is most important for DQS.

\paragraph{Stable regions}
The full system dynamics also reveals a series of stable (non-quantum-chaotic) regions beyond the threshold, where the quantum chaotic (non-quasiperiodic) dynamics are to some degree interrupted.
In the Heisenberg DQS model, we account for these regions analytically as a form of digital aliasing.
In the A2A-Ising DQS model, we connect the appearance of these islands phenomenologically with the re-emergence of stable islands in the collective spin phase space, by exploiting the equivalence of the Trotterised model~\cite{SiebererLM2019DQC} to a quantum kicked top.
Based on these observations, we term them ``stable regions''.
In general, these revivals in stability occur for step sizes close to some periodicity in the Trotterised dynamics, usually in one of the Trotter step's constituent gates, and ``freezes out'' that component of the contributing dynamics.
They can be more or less pronounced depending on the choice of initial state and the prominence of underlying symmetries in the system Hilbert space and Trotterisation recipe.
The stable regions are typically also much less prominent in the simulation fidelities, because the digitisation-induced regular dynamics does not necessarily in any way match the underlying, non-digitised target dynamics.
We leave a more detailed analysis of these digitisation-induced stable regions as a goal for future research.

\section{Background}\label{Sec:Background}


This section summarises the relevant concepts from Trotterisation and quantum chaos, introducing the notations and definitions that are used in the rest of the manuscript.
In all the equations here, we take $\hbar = 1$.

\subsection{Trotter Errors}\label{Sec:TrotterBackground}

In Trotterisation, the target model Hamiltonian $H_{M}=\sum_{l=1}^{L}H_{l}$ is decomposed into a sum of experimentally accessible Hamiltonians $\{H_{l}\}$.
Then, the desired time evolution $U_{M}(t) = e^{-iH_{M} t}$ is approximated by a sequence of discrete time evolution operators $U_{l}(\tau) = e^{-iH_{l} \tau}$ for corresponding Hamiltonians $H_{l}$, according to~\cite{LloydS1996UQS}
\begin{equation}\label{Eq:TrotterisationDef}
	U_{M}(t) \approx \underbrace{\left[\prod_{l} U_{l}(\tau)\right]^{r}}_{:=U_{\tau}^{r}},
\end{equation}
where we define the (Trotter) step size $\tau = t/r$ as the ratio of total simulated time $t$ and number of (Trotter) steps $r$.
Defining the unitary operators $U_{\tau}$ and $U_{\tau}^{r}$ for a single and $r$ Trotter steps, respectively, the approximation errors $\epsilon$ can be defined using a Taylor series expansion~\cite{LloydS1996UQS}
\begin{equation}\label{Eq:GenericErrors}
	U_{M}(t) - U^{r}_{\tau} = \frac{t^{2}}{r}\sum_{l>m}\left[H_{l}, H_{m}\right] + \mathcal{O}(\frac{t^{3}}{r^{2}}),
\end{equation}
where $\mathcal{O}(t^{3}/r^{2})$ subsumes the higher-order corrections.
Eq.~\eqref{Eq:TrotterisationDef} is a generic example of first-order Trotterisation, but in general the unitary operations in a Trotter step $U_{\tau}$ are determined also by a Trotterisation order.
The approximation errors $\norm{U_{M}(t) - U^{r}_{\tau}} \rightarrow 0$ for $r \rightarrow \infty$ (i.e. $\tau \rightarrow 0$) for fixed $t$, and the rate of convergence to continuum depends on the order of the Trotterisation~\cite{SuzukiM1976GTF, Suzuki1985DFQM, SUZUKI1990DMC, Suzuki1991GFMB}.
Trotter-Suzuki approximations (at any order) become exact when all the summands $\{H_{l}\}$ commute, but this is commonly not satisfied.
Additionally, in practice, $\tau$ is always finite, so we target a maximum error $\epsilon$ (for fixed simulated time $t$) and try to determine the Trotter step size (for a particular Trotterisation order) required to achieve the desired accuracy.

The usual assumption in Eq.~\eqref{Eq:GenericErrors} is that the first-order error term dominates (and consequently bounds) the errors, but this intuition does not always hold~\cite{Childs2020TTE}.
A more concrete bound on the error can be obtained by using a tail bound of the Taylor expansion~\cite{Berry2007EQA}.
This worst-case bound scales, at best, linearly with simulation time $t$~\cite{Berry2007EQA}, and can be expressed in terms of the required number of steps $r_{2k}$ that ensures an error $\norm{U_{M}(t) - U^{r}_{\tau}} \leq \epsilon$ for a $2k$(th)-order Trotterisation~\cite{Berry2007EQA, Childs2019fasterquantum}:
\begin{equation}\label{Eq:GateCountEstimate}
	r_{2k} = \mathcal{O}(\frac{(\max_{l}\norm{H_{l}} Lt)^{1+\frac{1}{2k}}}{\epsilon^{\frac{1}{2k}}}),
\end{equation}
where the $\norm{.}$ is the spectral-norm, $L$ is the number of summands, and $k$ is any positive integer.

Though Trotterisations provide convenient implementations of direct Hamiltonian decompositions, the scaling of the gate estimate in Eq.~\eqref{Eq:GateCountEstimate} with time, system size, and other parameters draws a pessimistic picture for the product formulas.
Consequently, alternative, so-called post-Trotter DQS algorithms~\cite{Childs2012LCU, Berry2015TTS, Low2017QSP, Low2019hamiltonian}, with different overheads such as requirements for auxiliary qubits, have been developed to overcome the apparent inefficiency of Trotterisation.
But while the generic estimates of gate complexities, such as Eq.~\eqref{Eq:GateCountEstimate}, provide sufficient conditions for the desired accuracy~\cite{Berry2007EQA}, empirical analyses of Trotterisation algorithms show that such rigorous estimates are often loose, and the desired accuracy is typically already achieved for orders of magnitude fewer steps than the estimates~\cite{Raeisi2012ESMB, Hastings2014QSQC, Wecker2014GCQS, Poulin2014TSS, Babbush2015CBTE, Childs2019fasterquantum, Childs2020TTE}.
Additionally, recent rigorous theory results~\cite{Childs2020TTE}, providing improved bounds on Trotter errors, showed that Trotterisation can match or even outperform the state-of-the-art post-Trotter algorithms, suggesting that Trotterisation may even be the optimal algorithm.
Also note that the performance of Trotterisation algorithms might be improved even further by exploiting commutations between summands~\cite{Raeisi2012ESMB, Childs2018QSQS} or by randomised ordering of unitary evolutions in each Trotter step~\cite{Campbell2017SGMU, Campbell2019RCHS, Childs2019fasterquantum}.

Since we focus on Trotterisation algorithms in this paper, we may, unless otherwise specified, use DQS to refer specifically to Trotterisation.

\subsection{Trotterisation Sequences as Floquet Systems}

In addition to gate complexities, it is equally important to know the nature of the errors and the minimum $r$ (maximum $\tau$) at which the approximation is still valid.
For example, before the complete breakdown point of the approximation due to coarse digitisation~\cite{HeylM2019QLTE, SiebererLM2019DQC}, it may still be possible to correct the errors.
Alternatively, it may also be possible to predict the $r \rightarrow \infty$ (i.e. $\tau \rightarrow 0$) value for a particular choice of observable from a smaller number of steps.

An alternative approach to Trotter error analyses~\cite{HeylM2019QLTE, SiebererLM2019DQC} is to interpret the Trotterised evolution as the stroboscopic dynamics under a Floquet Hamiltonian $H_{F}$,
\begin{equation}\label{Eq:FloquetDef}
	U_{\tau} = \prod_{l} U_{l}(\tau) = e^{-iH_{F} \tau}.
\end{equation}
The digitisation error in the effective Floquet Hamiltonian is then quantified by the Floquet-Magnus (FM) expansion, 
\begin{equation}\label{Eq:HamiltonianErrors}
	H_{F} = H_{M} + i\frac{\tau}{2}\sum_{l>m}\left[H_{l}, H_{m}\right] + \mathcal{O}(\tau^2),
\end{equation}
where $\mathcal{O}(\tau^2)$ subsumes any higher-order corrections.
The Floquet-Magnus expansion has a finite radius of convergence $\tau^{*}$, beyond which the approximation breaks down.
Theoretically, this represents the largest step size for which Trotterisation will realise a meaningful approximate simulation (that is, where the Trotterisation's effective Hamiltonian can still be understood in terms of the target Hamiltonian plus some controllable error perturbation that increases with Trotter step size).
Taking the Floquet-Magnus expansion as a special case of a more general Magnus expansion for a periodically time-dependent Hamiltonian, a rigorous sufficient condition of convergence can be derived that suggests that the radius of convergence should scale in inverse proportion to system size~\cite{Casas2007MEC, Moan2008CMS, BLANES2009MEA}.
If this lower bound on $\tau^*$ were tight, its scaling with inverse system size would suggest that realising large system sizes should be very challenging.

As discussed in the introduction, two recent numerical works studying DQS of the Ising model by some of the authors~\cite{HeylM2019QLTE, SiebererLM2019DQC} identified a performance threshold in Trotter step size (a threshold $\tau$) separating a region of controllable errors from a quantum chaotic regime.
A work investigating where the Floquet-Magnus expansion diverges saw a similar threshold in Floquet heating~\cite{TakashiI2018hit}.
Moreover, by interpreting the Trotterised evolution as a periodically time-dependent quantum system, Refs~\cite{HeylM2019QLTE, SiebererLM2019DQC} showed that pre-threshold Trotter errors for the Ising DQS in dynamical quantities such as expectation values, rather than errors in the full time evolution operator ($\norm{U_{M}(t) - U^{r}_{\tau}}$), agree with the perturbative corrections predicted by the Floquet-Magnus expansion~\cite{HeylM2019QLTE}.
Some localised instability regions were observed more recently in~\cite{ChinniK2022ted} only for specific mean-field, kicked multispin models, but no such instabilities have emerged in either Refs~\cite{HeylM2019QLTE, SiebererLM2019DQC} or in the much wider range of models considered in this paper.
The findings of Refs~\cite{HeylM2019QLTE, SiebererLM2019DQC} therefore open up interesting potential avenues for achieving better practical results with accessible performance, even for NISQ-era DQS.
For example, perturbative treatments could enable extrapolation of ideal dynamics (i.e. $\tau \rightarrow 0$) from the simulations with finite Trotter step sizes $\tau$.
This would effectively circumvent the impact of Trotter errors by allowing accurate simulation results to be calculated from simulations run over a range of comparatively large Trotter step sizes, and would simultaneously also help minimise the impact of errors extrinsic to the algorithm, such as decoherence and imperfect gates.
In addition to our broader results about new Trotterisation performance behaviours, our work also extends the scope of these findings and helps elevate the observations (that were mainly limited to Ising models) from Refs~\cite{HeylM2019QLTE, SiebererLM2019DQC} to the status of a universal DQS principle.

\subsection{Quantum Chaos}

In contrast to classical chaos, where infinitesimal differences in initial states diverge exponentially in time, the unitarity of dynamics in quantum mechanics ensures that the overlap between two arbitrary initial states stays the same for all times $t$,
\begin{eqnarray}
	\mathcal{F}(\psi_{1}(0), \psi_{2}(0)) &=& 
	|\braket{\psi_{1}(0)|\overbrace{U^{\dagger}(t)U(t)}^{=\mathbb{1}}|\psi_{2}(0)}|^{2} \\
	&=& \mathcal{F}(\psi_{1}(t), \psi_{2}(t)), \nonumber
\end{eqnarray}
where $\mathcal{F}(\psi_{1},\psi_{2}) := \abs{\braket{\psi_{1}|\psi_{2}}}^{2}$ is the fidelity between the states $\ket{\psi_{1}}$ and $\ket{\psi_{2}}$.
But while the notion of quantum chaos does not na\"ively follow popular concepts of classical chaos, similar intuitive signatures can still be found in quantum dynamics.
For example, the perturbation fidelity between two states evolved from the same initial state $\ket{\psi}$ under two slightly different unitaries $U(t)$ and a perturbed $U_{p}(t)$, $\mathcal{F}(\psi_{p}(t), \psi(t)) := \abs{\braket{\psi|U_{p}^{\dagger}(t)U(t)|\psi}}^{2}$,
decays exponentially for quantum chaotic systems~\cite{PeresA1984SQM}.
Unfortunately, although this interpretation is intuitive, this fidelity decay signature is reliable only for appropriate perturbations~\cite{EmersonJ2002FDEI}.

There are many other dynamical signatures of quantum chaos~\cite{HaakeF2018QSC}, and all of them (including the reliability of a perturbation for fidelity decay~\cite{EmersonJ2002FDEI}) derive from certain features governing the eigenvectors of either a unitary evolution operator, or equivalently its corresponding Hamiltonian.
For quantum chaotic systems, the eigenvectors show certain random matrix properties, namely that the distribution of their components in another basis follow specific distributions from random matrix theory (RMT)~\cite{IZRAILEVFM1987CSEF, KusM1988UEV}.
This type of analysis, known as eigenvector statistics, underlies the study of quantum chaos that will be described in this work.
A more detailed overview of quantum chaos, its relation with RMT, and technical subtleties in calculating the signatures of quantum chaos are provided in Appendix~\ref{App:RMTandQuantumChaos}.

Throughout this paper, we use quasiperiodic, stable, or regular interchangeably to mean \emph{non-quantum-chaotic}.

\section{Physical Models and Their Trotterisations}\label{Sec:Models}


\begin{figure}[ht]
	\centering
 	\subfloat[DQS of all-to-all (A2A) Ising system of $N$ qubits maps to the quantum kicked top of a single spin with spin number $j = N/2$.
	Here, ZZ represents the Ising interaction.]{\includegraphics[scale=1]{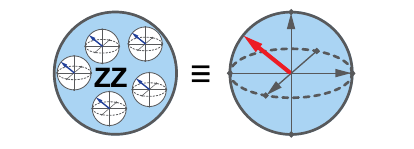}} 
 	\newline
 	\subfloat[The nearest-neighbour Heisenberg model is a chain of $N$ qubits with XYZ coupling between each neighbouring qubits.]{\includegraphics[scale=1]{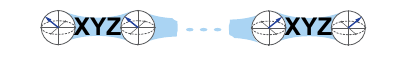}}
	\newline
 	\subfloat[The collective behaviour of the qubit ensemble inside a cavity is represented as a large spin for the Dicke model.]{\includegraphics[scale=1]{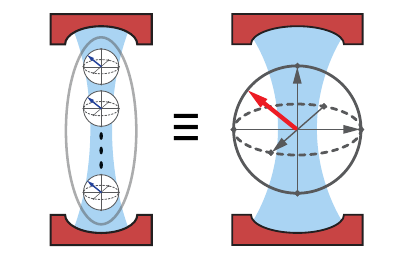}}
 	\caption{Pictorial schematics of the physical models analysed in this paper.}\label{Fig:2_Models}
\end{figure}

To obtain a more comprehensive understanding of the effects of Trotterisation on DQS performance (including not only the threshold observed in Refs~\cite{HeylM2019QLTE, SiebererLM2019DQC}, but also, e.g., pre-threshold behaviours), we consider a set of paradigmatic model systems, which share the feature that their Trotterised dynamics could be realisable experimentally in the near future (Fig.~\ref{Fig:2_Models}): the (i) Ising (in all-to-all interactions limit), (ii) Heisenberg, and (iii) Rabi-Dicke models.
In this section, we provide descriptions of the model Hamiltonians $H_{M}$ and their experimentally realisable Trotter-Suzuki decompositions.
Since our goal is to investigate Trotterisation performance behaviours, we do not provide either an extensive or a complete summary (or references) of these models.
They are extensively studied fundamental quantum models describing magnetism (Ising and Heisenberg~\cite{Nolting2009QTM}) and light-matter interactions (Rabi-Dicke~\cite{Rabi1936PSQ, Dicke1954CSRP, JaynesCummings, TavisCummings, DiazF2019USC}), and they and their various generalisations and modifications offer rich physical phenomena and practical applications, such as quantum phase transitions (in Ising~\cite{Zurek2005QPT, Heyl2013PTI, Heyl2019QPT}, Heisenberg~\cite{Dyson2976PTHM}, and Dicke and Rabi models~\cite{Dicke1954CSRP, Hwang2015QPTR}), and even mappings to NP problems~\cite{Lucas2014INP}.

\subsection{Spin Models}

The general spin-spin interaction Hamiltonian~\cite{Nolting2009QTM} 
\begin{equation}
	H = \sum_{k,l}g_{k,l} \left[ \alpha\left(\sigma^{k}_{x}\sigma^{l}_{x} + \sigma^{k}_{y}\sigma^{l}_{y} \right) +
	\beta\sigma^{k}_{z}\sigma^{l}_{z} \right]
\end{equation}
defines the Heisenberg interaction for $\alpha = \beta = 1$, Ising interaction for $\alpha = 0$ \& $\beta = 1$, and the exchange (or XY-) interaction for $\alpha = 1$ \& $\beta = 0$, where $\sigma_{\mu}$ are the Pauli spin operators with $\mu\in\{x,y,z\}$.
The structure of the coefficients $g_{k,l}$ defines further variations in each case~\cite{Nolting2009QTM}.
Here, we consider all-to-all ($g_{k,l} \neq 0$ $\forall k,l$) and one-dimensional nearest-neighbour ($g_{k,l}\neq 0 \text{ iff } \abs{k-l} = 1$) interactions for the Ising and Heisenberg models, respectively.

\subsubsection{The Ising Model}

Trotterisation thresholds in Ising-model DQS have been considered for various regimes of power-law interactions in Refs~\cite{HeylM2019QLTE, SiebererLM2019DQC}.
Here, we revisit the all-to-all (A2A) case, and reveal new insight into Trotterisation performance and the threshold behaviours.
In this limit, the DQS model maps to the quantum kicked top~\cite{SiebererLM2019DQC}, and reduces to a collective spin system with total spin $j$, as depicted in Fig.~\ref{Fig:2_Models}~(a), facilitating the study of Trotterisation performance behaviours up to relatively large system sizes.
Here, we describe the corresponding quantum kicked-top model and the effective target Hamiltonian, and refer to Ref.~\cite{SiebererLM2019DQC} for further details of the model.
Specifically, the corresponding Trotter-Suzuki decomposition is given by the kicked-top Floquet operator
\begin{equation}\label{Eq:A2AIsingDecomp}
	U_{I}(t) = e^{-iH_{I}t} \approx \left( e^{-iH_{z}t/n}e^{-iH_{x}t/n} \right)^{r} = U_{\tau}^{r},
\end{equation}
where $H_{I} = H_{z} + H_{x}$ is the effective target-model Hamiltonian.
The individual summand Hamiltonians are
\begin{equation}\label{Eq:IsingHamiltonian}
	H_{\mu} = \omega_{\mu}J_{\mu}  + g_{\mu}J^{2}_{\mu}/(2j + 1),
\end{equation}
for $\mu\in\{x, z\}$, with standard angular momentum operators $J_{\mu}$ of a spin $j=N/2$ system, corresponding to the (maximal) collective spin of $N = 2j$ spin-1/2 systems in an A2A-Ising system~\cite{SiebererLM2019DQC}.
The quantum kicked-top models have already been implemented in various platforms, including in a single Caesium atom with $j = 3$ \cite{ChaudhuryS2009QSC}, the collective spin of an ensemble of three superconducting qubits~\cite{NeillC2016EDQS} corresponding to $j = 3/2$, and others (including novel proposals~\cite{MourikV2018QCNS}) discussed in Ref.~\cite{SiebererLM2019DQC}.

For ease of comparison and consistency, we show our results for the same simulation parameters as in Ref.~\cite{SiebererLM2019DQC}, that is $g_{z} := g$ (with time measured in units of $2\pi/g$ for all models), $\omega_{x} = 0.1g_{z}$, $g_{x} = 0.7g_{z}$, and $\omega_{z} = 0.3g_{z}$, and the initial state is $\ket{\theta=0, \phi=0} = \ket{j, j}$, where $\ket{\theta, \phi} = \exp\left(i\theta(J_{x} \sin(\phi) - J_{y} \cos(\phi))\right)\ket{j, m=j}$ is the spin coherent state.
The qualitative behaviour of our results does not depend on the precise values of these parameters~\cite{SiebererLM2019DQC}.

\subsubsection{The Heisenberg Model}

The Heisenberg model is an extensively studied model in the quantum theory of magnetism~\cite{Nolting2009QTM}, and, in addition to its fundamental importance, the random-field version of this model is proposed to allow demonstrations of quantum speed-up in quantum simulations~\cite{Childs2018QSQS}. Here, we consider the uniform-field case with parameters and system sizes readily available in current experiments~\cite{Salathe2015DQS}.

Specifically, we consider a nearest-neighbour interacting spin chain version of the Heisenberg model with open boundary conditions, isotropic and homogeneous couplings, and a longitudinal external field (as depicted in Fig.~\ref{Fig:2_Models}~(b)):
\begin{equation}\label{Eq:HeisenbergModel}
	H_{H} = \underbrace{\frac{\omega}{2}\sum_{k=1}^{N} \sigma_{z}^{k}}_{H_{z}} 
	+ g\sum_{k=1}^{N-1} \left( \sigma_{x}^{k}\sigma_{x}^{k+1} + 
	\sigma_{y}^{k}\sigma_{y}^{k+1} + \sigma_{z}^{k}\sigma_{z}^{k+1} \right),
\end{equation}
where $g$ is the nearest-neighbour coupling strength, and $\omega$ is the Zeeman splitting due to the homogeneous magnetic field applied in the $z$-direction.
All our results, including the existence of stable regions, qualitatively hold also for the cases with a transverse (i.e. $\sigma_{y}$ or $\sigma_{x}$ term) or no external field (data available, but not shown).
With isotropic couplings, the Heisenberg interaction is rotationally symmetric, and the distinction between longitudinal and transverse fields is not strictly relevant for it.
However, this rotational symmetry is broken by the chosen Trotterisation, so the distinction is relevant for analysing the Trotterised unitary and is reflected in the structure of its eigenvector statistics (see Appendix~\ref{App:HeisebergEigVec}).

The Heisenberg model Hamiltonian in Eq.~\eqref{Eq:HeisenbergModel} can be decomposed into $H_{H} = H_{z} + H_{xy} + H_{xz} + H_{yz}$~\cite{Heras2014DQS,Salathe2015DQS}, where the exchange interaction
\begin{equation} 
	H_{xy} = g^{\prime}\sum_{k=1}^{N-1}\left( \sigma_{x}^{k}\sigma_{x}^{k+1} + \sigma_{y}^{k}\sigma_{y}^{k+1} \right),
\end{equation}
with interaction strength $g^{\prime} = g/2$, is the workhorse interaction for many experimental platforms~\cite{Salathe2015DQS, MonroeC2021QSSS, DehollainJP2014EISC, VeldhorstM2014EIQD}.
$H_{z}$ is just the free-evolution, and the other two terms of the decomposition are obtained from the exchange interaction by applying local transformations:
\begin{equation}
	H_{xz} = \mathcal{R}_{x,\tfrac{\pi}{2}}^{\forall k}H_{xy}\mathcal{R}_{x,\tfrac{\pi}{2}}^{\forall k \dagger}
	= g^{\prime}\sum_{k=1}^{N-1}\left( \sigma_{x}^{k}\sigma_{x}^{k+1} + \sigma_{z}^{k}\sigma_{z}^{k+1} \right),
\end{equation}
and
\begin{equation}
	H_{yz} = \mathcal{R}_{y,\tfrac{\pi}{2}}^{\forall k}H_{xy}\mathcal{R}_{y,\tfrac{\pi}{2}}^{\forall k \dagger}
	= g^{\prime}\sum_{k=1}^{N-1}\left( \sigma_{y}^{k}\sigma_{y}^{k+1} + \sigma_{z}^{k}\sigma_{z}^{k+1} \right),
\end{equation}
where $\mathcal{R}_{\mu,\theta}^{\forall k} = \prod_{k=1}^{N}e^{-i\frac{\theta}{2} \sigma^{k}_{\mu}}$ (for $\mu\in\{x, y, z\}$) are locally applied, single-qubit rotations by an angle $\theta$ around the $\mu$ axis.

This decomposition, without the Zeeman term, was proposed~\cite{Heras2014DQS} and experimentally realised in circuit-QED~\cite{Salathe2015DQS}.
The generalisation we introduce here can be implemented in any platform~\cite{Salathe2015DQS, MonroeC2021QSSS} with the exchange interaction and single-qubit rotations, via the following Trotterisation
\begin{equation}\label{Eq:HeisenbergDecomposition}
	U_{\tau} = U_{z}(\tau)\prod_{k=1}^{N-1}U_{xy}^{k,k+1}(\tau)\prod_{k=1}^{N-1}U_{xz}^{k,k+1}(\tau)
	\prod_{k=1}^{N-1}U_{yz}^{k,k+1}(\tau),
\end{equation}
where the $U_{ab}^{k, k+1}(t) = \exp(-iH_{ab}^{k, k+1}t)$ are unitary operators $\forall ab \in \{xy, xz, yz\}$, $U_{z} = \exp(-iH_{z}t)$, and the qubit rotations are implicit.

In our numerical simulations, we choose the highly symmetric parameters of $\omega = g$, and the initial state for the results presented in the main text of the paper is $\ket{1}_{i=1} \otimes \ket{0}_{i>1}$, with all qubits in the ground state except for a single excitation in the first qubit.
We explored the Heisenberg DQS for a range of different parameters, and our results do not depend on the precise choice of these parameters, but the quasiperiodic dynamics before the threshold may show variations depending on the initial state.
We discuss initial-state effects further in our results (Sec.~\ref{Sec:Results}), and support these discussions in Appendix~\ref{App:HeisenbergAppendix_initialState} by analysing the results for spin coherent, product, and random initial states.

\subsection{The Rabi-Dicke Model}

The quantum Rabi (QR) model~\cite{Rabi1936PSQ} describes the fundamental interaction between a two-level system (i.e., a qubit) and a single mode of the quantised electromagnetic field
\begin{equation}
	H_{R} = \omega_{c} a^{\dagger}a + \frac{\omega_{q}}{2}\sigma_{z} + g(a^{\dagger} + a)\sigma_{x},
\end{equation}
where $a^{\dagger}$ and $a$ are the creation and annihilation operators for the field mode, and $\omega_{c}$, $\omega_{q}$, and $g$ are the cavity-field, qubit, and coupling frequencies.
Both natural and typical experimental systems typically satisfy $g \ll \omega_{c}, \omega_{q}$, where a rotating-wave approximation (RWA) applies, reducing the QR model to the Jaynes-Cummings (JC) model~\cite{JaynesCummings} 
\begin{equation}
	H_{JC} = \omega_{c} a^{\dagger}a + \frac{\omega_{q}}{2}\sigma_{z} + g(a^{\dagger}\sigma_{-} + a\sigma_{+}),
\end{equation}
where $\sigma_{\pm} = (\sigma_{x} \pm i\sigma_{y})/2$ are raising/lowering operators for a two-level system.  
However, DQS provides a flexible platform to explore the full QR model without the need for challenging, direct experimental implementations. This was recently demonstrated in circuit QED~\cite{LangfordNK2017DQRS}, and can easily be generalised.

One important generalisation is the $N$-qubit case described by the Dicke model~\cite{Dicke1954CSRP}:
\begin{equation}
	H_{D} = \omega_{c} a^{\dagger} a + \frac{\omega_{q}}{2}\sum_{k=1}^{N} \sigma_{z}^{k} + 
	\frac{g}{\sqrt{N}}\sum_{k=1}^{N}(a^{\dagger} + a)(\sigma_{+}^{k} + \sigma_{-}^{k}).
\end{equation}
By exploiting the collective spin behaviour of the atomic ensemble (as depicted in Fig.~\ref{Fig:2_Models}~(c)), it can be written in terms of angular momentum operators~\cite{DiazF2019USC}
\begin{equation}
	H_{D} = \omega_{c} a^{\dagger} a + \omega_{j}J_{z} + 
	\frac{g}{\sqrt{2j}}(a^{\dagger} + a)(J_{+} + J_{-}),
\end{equation}
where $j = N/2$ is the total collective spin, the frequency $\omega_{j} = \omega_{q}$, and $J_{\pm} = (J_{x} \pm iJ_{y})/2$ are raising/lowering operators of an angular momentum with total spin $j$.
Under the RWA, the Dicke model reduces to the Tavis-Cummings (TC) model~\cite{TavisCummings},
\begin{equation}\label{Eq:TCHamiltonian}
	H_{TC} = \omega_{c} a^{\dagger} a + \omega_{j}J_{z} + 
	\frac{g}{\sqrt{2j}}(a^{\dagger} J_{-} + aJ_{+}).
\end{equation}

In systems with access to TC coupling and single-qubit rotations, digital quantum simulation of the Dicke Hamiltonian (including the Rabi Hamiltonian) can be achieved via the decomposition
\begin{equation}
	H_{D} = H_{TC}(g, \Delta_{c}, \Delta_{j}^{TC}) + H_{ATC}(g, \Delta_{c}, \Delta_{j}^{ATC}),
\end{equation}
where the anti-Tavis-Cummings (anti-TC) Hamiltonian
\begin{eqnarray}\label{Eq:ATCHamiltonian}
	H_{ATC} &=& \mathcal{R}_{x,\pi}^{\forall k}H_{TC}\mathcal{R}_{x,\pi}^{\forall k \dagger} \\
	&=& \Delta_{c} a^{\dagger} a - \Delta_{j}^{ATC}J_{z} \\ && + \frac{g}{\sqrt{2j}}(a^{\dagger} J_{+} + aJ_{-}) \nonumber
\end{eqnarray} 
contains only the counter-rotating terms and is obtained from the TC interaction by locally applied qubit rotations of an angle $\pi$ (with $\mathcal{R}_{x(y),\theta}^{\forall k} = \prod_{k=1}^{N}e^{-i\theta J_{x(y)}}$ within the collective spin subspace).
The effective simulation frequencies are $\omega_{c}^{\text{eff}} = 2\Delta_{c}$ and $\omega_{j}^{\text{eff}} = \Delta_{j}^{TC}-\Delta_{j}^{ATC}$, where $\Delta_{\mu} = \omega_{\mu} - \omega_{\text{RF}}$ are defined relative to a nearby rotating frame (RF), giving precise control of simulated frequencies (see Ref.~\cite{LangfordNK2017DQRS} for details on the RF and its experimental implementation in circuit QED).
The first-order Trotterisation
\begin{equation}\label{eq:DickeDecomposition}
	U_{\tau} = U_{TC}(\tau)U_{ATC}(\tau),
\end{equation}
where $U_{A/TC}(t) = \exp\left(-i H_{A/TC}t\right)$, is proposed in Ref.~\cite{MezzacapoA2014DQRS}.
Both this and the 2nd-order Trotterisations are experimentally realised for $j = 0.5$, i.e., the Rabi model, in circuit QED~\cite{LangfordNK2017DQRS}.

The Dicke model is known to be quantum chaotic for $g$ values around the critical coupling strength $g_{c} = \sqrt{\omega_{c}\omega_{j}}/2$ of the superradiant phase transition~\cite{CliveE2003QCD, Clive2003CDM}.
In order to study digitisation-induced quantum chaos, we specifically choose cavity and spin frequencies far enough from its quantum chaotic regime, in its normal phase, but still staying in the ultra-strong coupling regime (so that RWA does not apply).
Our results (on qualitative Trotterisation performance behaviours and the existence of the threshold) hold for both the quantum chaotic and regular parameter regimes of the Dicke model, and, in the Appendix~\ref{App:DickeChaos}, we discuss our choice of parameters and also provide example results for the Dicke DQS in its quantum chaotic regime.
In our numerical simulations, the system parameters are $\omega_{j} = \omega_{c} = 3.5g$ (giving $g=g_{c}/1.75$, which is sufficiently smaller than $g_{c}$ for the model to be non-quantum-chaotic), and the initial state is $\ket{0}\otimes\ket{j,j}$: that is, with zero photons in the cavity and the qubit ensemble in the spin coherent state $\ket{j,j} = \ket{\theta=0, \phi=0}$, where $\ket{\theta, \phi} = \exp\left(i\theta(J_{x} \sin(\phi) - J_{y} \cos(\phi))\right)\ket{j, m=j}$.
In our descriptions, we will sometimes refer to the Rabi-Dicke model to emphasise our results encompass both cases.


\begin{figure*}
	\centering
	\includegraphics[scale=0.94]{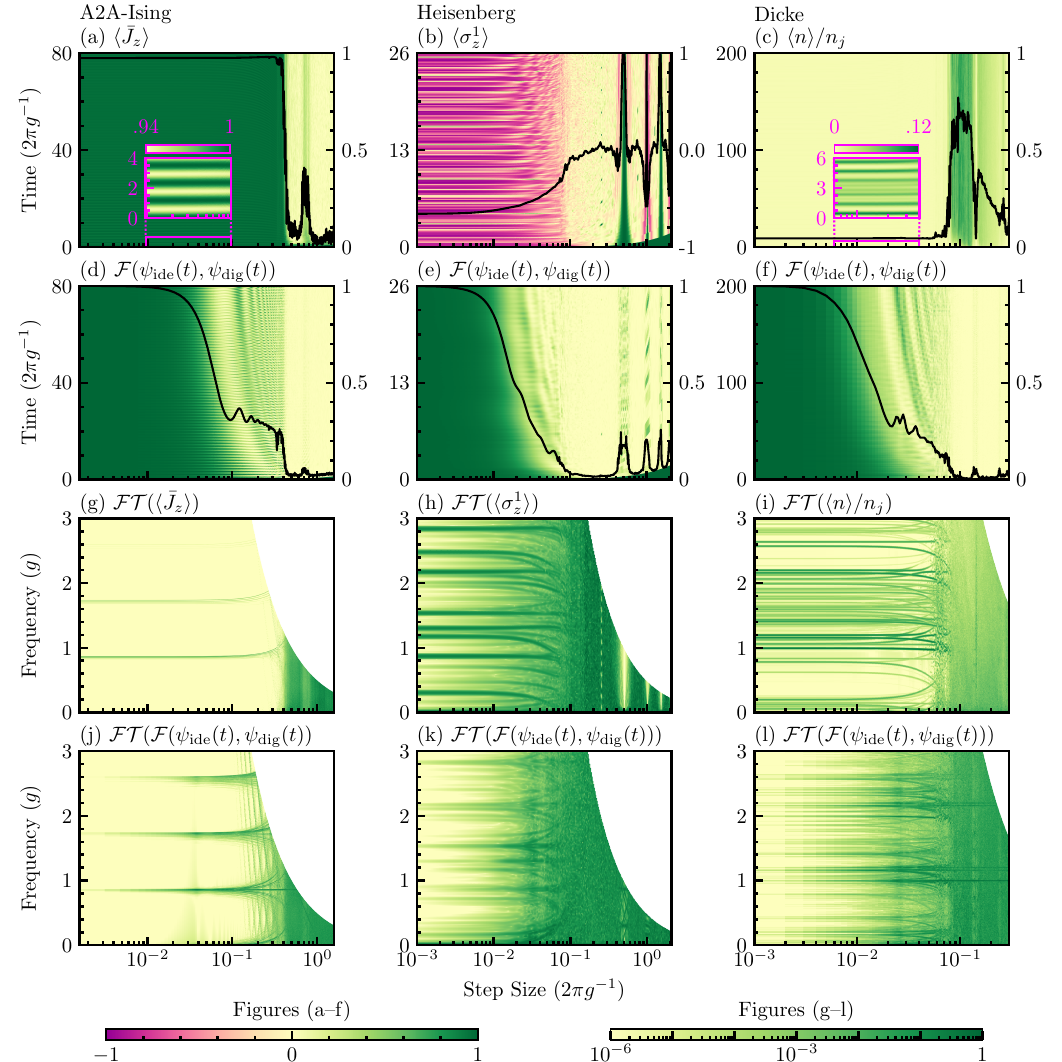} 
	\caption{Full dynamics (with the corresponding Fourier transforms (FTs)) for each model system, showing (a--c) local observables (FTs in (g--i)) and (d--f) simulation fidelity (FTs in (j--l)).
	Left to right, columns show results for A2A-Ising (with $j = 64$), Heisenberg (chain of $N = 8$ qubits), and Dicke (with $j=6$) models, respectively, with x-axes common to each column showing step sizes.
	The y-axes of the colour plots (left axes) are the simulation times in (a--f) and the FT frequencies in (g--l).
	The black lines (right y-axes) in (a--f) are the time averages of the colour plots.
	The colour plots (a--f) are shown using a linear colour scale ranging from -1 to 1, and the FTs (normalised to have a unit maximum amplitude for each $\tau$) in (g--l) are shown using a logarithmically scaled colour map ranging from $10^{-6}$ to 1.
 	In each model, dynamical quantities share a common Trotterisation threshold at a critical step size separating quasiperiodic from quantum chaotic regimes and two pre-threshold regimes of different step size dependencies.
	(a--c) Quasiperiodicity in expectation values of example local observables is clearly observed prior to the threshold, and destroyed after it: shown for (a) magnetisation $\braket{\bar{J}_{z}} = \frac{1}{j}\braket{J_{z}}$, (b) polarization of qubit one $\braket{\sigma_{z}^{1}}$, and (c) normalised photon number $\langle n\rangle/n_{j}$ (with $n_{j} = 7\times\dim_j$).
	(d--f) Simulation fidelity $\mathcal{F}(\psi_{\text{dig}}(t), \psi_{\text{ide}}(t))$ decays rapidly with time beyond the same threshold, with clear quasiperiodic oscillations in the near-threshold region showing why the time-averaged simulation fidelity starts decreasing strongly already before the threshold.
	These markedly universal behaviours across the different models are better demonstrated with the FT in (g--l), which show: (i) distinct stable frequencies at small step sizes, (ii) ``branching'' of frequencies at larger step sizes, and (iii) \emph{effectively flat spectra} beyond the threshold.
	In each model, the dynamical signatures of quantum chaos (a--c) disappear in certain regimes beyond the threshold, referred to as stable regions, showing revival of regular dynamics, without a corresponding return of accurate target system simulation (d--f).}
	\label{Fig:3_FullDynamicsAndFT}
\end{figure*}



\section{Universal Behaviours in Trotterisation Performance}\label{Sec:Results}


In this section, we present the main results of this work,
involving a broad and comprehensive analysis of Trotterisation performance across a range of experimentally achievable physical models and system sizes, using new and expanded characterisation techniques.
For each model, we analyse various dynamical, time-averaged, and static quantities together with error measures.
Our analysis provides new results about highly generic Trotterisation performance behaviours both in the pre-threshold regime that is most important for DQS, and at the Trotterisation threshold previously observed in ~\cite{HeylM2019QLTE, SiebererLM2019DQC}.
In this section, we discuss the detailed individual results obtained for each of the analytical approaches we use, before looking more broadly at the conclusions we can draw by combining these results in the Discussion section following.

We start by analysing the detailed time evolution of different physical signatures at one (representative) system size, providing new insights into general Trotterisation performance and threshold behaviours.
We then compare the time averages of these dynamical quantities for system sizes across multiple orders of magnitude in Hilbert-space dimension, starting from the smallest and scaling up.
In every model, the Trotterised dynamics show signatures of quantum chaos beyond the threshold with additional model-dependent characteristics, such as the structure of stable regions observed inside the quantum chaotic regions.
Following this, we then use RMT to introduce a powerful new approach to identify quantum chaos, based on eigenvector statistics with a chi-squared goodness-of-fit test, and we demonstrate conclusive agreement with the RMT predictions of quantum chaotic dynamics beyond this threshold, observe minimal system-size dependence, and identify the existence of stable regions beyond the threshold.
These results both support and significantly extend the analysis based on dynamical signatures.
Finally, using various error measures, we analyse the relationship of the errors with the system size, total simulation time, and step size.

\subsection{Trotterisation Performance Behaviours in Dynamical Quantities}\label{Sec:DynamicalLand}

Here, we show that studying time evolutions, rather than time averages, provides clear signatures both for the regular and quantum chaotic dynamics revealing improved insight into the Trotterisation performance and threshold behaviours.
We study the time evolutions of expectation values and simulation fidelities (along with other quantum chaos signatures in Appendix~\ref{App:otherSignatures}) as a function of step size.
In particular, signatures of regular (quantum chaotic) dynamics before (after) the threshold, such as observation (destruction) of quasiperiodicity in expectation values, enable identification of Trotterisation thresholds, even when they are not always clear from the time averages of dynamical quantities, especially for the simulation fidelities.
By analysing the full dynamics, we are also able to calculate the corresponding Fourier spectra for each of these quantities to better show both the pre-threshold quasiperiodicity and the different regimes of step-size dependence in the underlying frequencies.
Finally, we note that the dynamical quantities exhibit islands of non-chaotic behaviour (stable regions) inside the quantum chaotic regime, which we discuss in more detail in Sec.~\ref{Sec:stableRegions}.

Colour plots in Fig.~\ref{Fig:3_FullDynamicsAndFT}~(a--f), with the time on (the left) y-axes and the step size on x-axes, show our results for the dynamical quantities given in the titles of each sub-figure.
The black lines on the colour plots are the time average of the same quantities (averaged over the full time-axis), with their y-axes given on the right side of the figures (matching the corresponding colour bar axes).
The colour plots in Fig.~\ref{Fig:3_FullDynamicsAndFT}~(g--l), with the frequency on y-axes and the step size on x-axes, show the Fourier transforms of the quantities plotted in Fig.~\ref{Fig:3_FullDynamicsAndFT}~(a--f), as given in the titles of each sub-figure.
The white regions at large step sizes in Fig.~\ref{Fig:3_FullDynamicsAndFT}~(g--l) mark where the low sampling rates (Trotter step sizes) restrict the maximum frequency that can be resolved.

\subsubsection{Destruction of Quasiperiodicity in the Dynamics of Physical Observables}\label{Sec:quasiperiod}

In quantum simulations, we are mostly interested in studying the expectation values of some physical observables, which are typically periodic or quasiperiodic for regular quantum dynamics, and the destruction of (quasi)periodicity is a characteristic signature of quantum chaotic dynamics.
Here, we show the onset of this destruction in Trotterised dynamics as the step size is increased beyond a clear threshold.
Figures~\ref{Fig:3_FullDynamicsAndFT}~(a--c) illustrate this effect for A2A-Ising (with $j=64$), Heisenberg (chain of $N = 8$ qubits), and Dicke (with $j=6$) models, showing the time evolutions of the observables: (a) magnetisation---the normalised expectation value $\braket{\bar{J}_{z}} = \frac{1}{j}\braket{J_{z}}$ of the $J_{z}$ angular momentum operator, (b) first qubit polarization---the expectation value of the qubit-1 Pauli z-operator $\braket{\sigma_{z}^{1}}$, and (c) normalised photon number $\langle n\rangle/n_{j}$ (with $n_{j} = 7\times\dim_j := 7\times(2\times j + 1)$ based on the numerical observations discussed in the Sec.~\ref{Sec:SystemSize}), respectively.
Because the total simulation time of each model in Fig.~\ref{Fig:3_FullDynamicsAndFT} is chosen to reveal long time details of the dynamics, especially for simulation fidelities discussed in Sec.~\ref{Sec:SimFidDynamics}, some of the fast lower contrast dynamics in quasiperiodic regimes are only revealed when zoomed in.
The insets in Fig.~\ref{Fig:3_FullDynamicsAndFT} show the quasiperiodic behaviour present in each model by magnifying both the temporal features and the colour contrasts.

The transition from quasiperiodic to quantum chaotic dynamics at the threshold is arguably even more pronounced in the corresponding Fourier spectra.
Figures~\ref{Fig:3_FullDynamicsAndFT}~(g--i) show the Fourier transform of the expectation value time traces for the same observables defined above.
These clearly illustrate strongly clustered frequency components with highly varied amplitudes in the dynamics before the threshold and essentially flat and equally spaced Fourier spectra after the threshold.
Figures~\ref{Fig:3_FullDynamicsAndFT}~(g--i) also provide the first indications that there are two distinct pre-threshold regimes for the quasiperiodic dynamics (which will be discussed further in relation to simulation fidelities below), defined by ``branching'' of the spectral frequencies once the step size is sufficiently large.
Note that the slanted ``lines'' appearing near the threshold in Fig.~\ref{Fig:3_FullDynamicsAndFT}~(g) are due to digital aliasing of high frequencies that are not properly resolved due to the low sampling rate ($1/\tau$).

The threshold position is clear from both the destruction of quasiperiodicity and the time averages (shown in black lines) in Fig.~\ref{Fig:3_FullDynamicsAndFT} for (a) Ising and (c) Dicke models.
For the Heisenberg model, however, while the full dynamics clearly exhibit a threshold, the time averages do not.
The reason for this is immediately apparent from the full dynamics and corresponding spectra.
Specifically, the bending of the quasiperiodic frequencies prior to the threshold washes out the time averaging and obscures the threshold, even while the full dynamics still shows a clear transition.
Here, the pre-threshold behaviour is connected with initial-state-dependent Trotterisation behaviours.
In Appendix~\ref{App:HeisenbergAppendix_initialState}, we analyse the Heisenberg DQS for various classes of initial states and show that, while all the states ultimately converge to the same threshold, they show significantly varied Trotter errors prior to the threshold.
Note, our results do not depend on the specific choice of operator for the expectation values or the sub-system.
The first qubit of the Heisenberg chain is only a numerically convenient illustrative example (see Appendix~\ref{App:HeisenbergAppendix_otherQubits} for the other qubits in the Heisenberg chain).

\subsubsection{Simulation Fidelity}\label{Sec:SimFidDynamics}

In this section, we study the overall simulation performance as characterised by the simulation fidelity to the target dynamics.
In previous work~\cite{SiebererLM2019DQC}, where simulation performance was characterised through the use of time-averaged quantities, the simulation fidelity seemed to exhibit no persistent threshold behaviour.
Here, however, by studying full time evolutions, we demonstrate that a clear breakdown threshold also appears at the same location in the simulation fidelity, and can explain the apparently inconsistent behaviour observed in Ref.~\cite{SiebererLM2019DQC}.

The simulation fidelity, defined as the overlap between the states $\psi_{\text{dig}}(t)$ and $\psi_{\text{ide}}(t)$ obtained, respectively, from Trotterised and ideal dynamics,
\begin{equation}
	\mathcal{F}(\psi_{\text{dig}}(t), \psi_{\text{ide}}(t)) := \abs{\braket{\psi_{\text{dig}}(t)|\psi_{\text{ide}}(t)}}^{2},
\end{equation}
directly characterises simulation performance via the error in the simulated output state.
Figures~\ref{Fig:3_FullDynamicsAndFT}~(d--f) respectively for A2A-Ising, Heisenberg, and Dicke models, show that the simulation fidelity is oscillatory in time before the threshold and quickly decays beyond it.
This sudden change in the dynamical characteristics of simulation fidelity identifies the threshold.

Studying the simulation fidelity dynamics in more detail, however, we can identify three qualitatively distinctive regions of dynamical behaviour.
Again, these observations are even more clear in the Fourier spectra in Figures~\ref{Fig:3_FullDynamicsAndFT}~(j--l).
In the first region of each model, at short enough times and small enough step sizes, the simulation fidelity shows a clear fast oscillation with a period that is approximately step-size independent, but an amplitude that gets deeper with increasing step size.
At larger step sizes and times, the slower and larger excursions in simulation fidelity with step-size-dependent quasiperiodicity (characterised by splitting and bending frequencies in the spectra) start to dominate. 
Finally, at the same step-size threshold as observed in other signatures, we see a sudden (simulation-time-independent) transition from quasiperiodic fidelity dynamics to simulation fidelities that rapidly collapse with simulation time (illustrated by nearly flat, evenly spaced spectra), an effect which becomes increasingly stark with increasing system size.

These dynamical results provide clear explanations for the trends observed in the time-averaged values in Ref.~\cite{SiebererLM2019DQC}.
Namely, while the same sudden threshold can be observed in simulation fidelity as in other signatures, significant pre-threshold reductions in time-averaged simulation fidelity result from the increasing amplitude and complexity of quasiperiodic oscillations with both step and system sizes, to the extent that the breakdown threshold in time-averaged fidelity can even be completely obscured at larger system sizes (see also Sec.~\ref{Sec:SystemSize}).
They furthermore show that the fidelity does not irrevocably break down until after the threshold, hinting that pre-threshold errors could be significantly correctable with perturbative treatments, such as the Floquet-Magnus expansion used in Refs~\cite{HeylM2019QLTE, SiebererLM2019DQC} for observable expectations.
This implies that, before the threshold, the states might differ only up to perturbative corrections in the Floquet-Magnus expansion for the specific Trotterisation.
While outside the scope of this paper, our results highlight that a detailed study of full simulation fidelity dynamics and possible connections to perturbative error correction terms may reveal valuable new insights into the nature of Trotter errors.

Our three models cover different systems, system parameters and symmetries.
Yet in terms of both the nature of the Trotterisation threshold and the nature of error perturbations in the quasiperiodic regimes, the observation and behaviour of simulation fidelity in the three qualitatively different regions shown here, remain robust across specific model and parameter choices (including across rather wide-ranging exploratory numerical analyses not presented here).

\subsubsection{Stable Regions}\label{Sec:stableRegions}

In each model, beyond the stability threshold, we also observe clear quasiperiodic regions, as seen in Fig.~\ref{Fig:3_FullDynamicsAndFT} (and in Appendix Figs.~\ref{AppFig:10_OtherDynamicalSignatures} and~\ref{AppFig:13_IsingPhaseSpacePortraits}), where the quantum chaotic dynamics are partially or strongly interrupted.
These stable regions arise as a form of Trotterised digital aliasing.
Specifically, they are localised at step sizes where the Trotterisation (sampling) frequency is sufficiently close to some underlying frequency in the Trotterised dynamics, which then effectively freezes its contribution to the Trotterised evolution.
For example, in our Trotterised Heisenberg model, we observe revivals of stability at step sizes which synchronise with the dynamics of the two-body exchange couplings: that is, for our model parameters, at $\tau_{\text{stable}} = 0.5\,n$~$(2\pi g^{-1})$ for integer $n$, where $U_{xy}(\tau_{\text{stable}}) = \rm{diag}\left(\lambda_1, \ldots, \lambda_\mathcal{D} \right)$, with $|\lambda_j|^2=1$.
The revivals therefore also depend explicitly on the Trotterisation recipe and product formula.
The more components frozen by this sampling resonance, the stronger this effect.
Because of its high parameter symmetry and simple constituent gates, our Trotterised Heisenberg model exhibits very strong revivals in comparison with those of the A2A-Ising and Dicke models.
The revivals are observed much less prominently in simulation fidelity, however, because the digitisation-induced stable behaviour does not generally match the underlying, non-digitised target dynamics.

The Trotterised A2A-Ising model provides a complementary lens through which to understand the stable regions.
By exploiting its equivalence to a quantum kicked top and studying its well-understood classical limit~\cite{SiebererLM2019DQC, HaakeF2018QSC}, we are able to empirically connect the appearance of the stable regions with the re-emergence of stable islands in the phase space of the collective spin (see Appendix Sec.~\ref{App:IsingStableRegions}).
Again, we observe that revivals in stability do not produce an equivalent revival in simulation fidelity, with observed phase-space dynamics being very different from the original target, even within the stable region.

There are many interesting questions still to understand here, and we emphasise that these may have important implications for employing DQS in future state-of-the-art quantum computers.
We plan to explore these effects in more detail in future work.


\begin{figure*}[ht]
	\centering
	\includegraphics[scale=0.94]{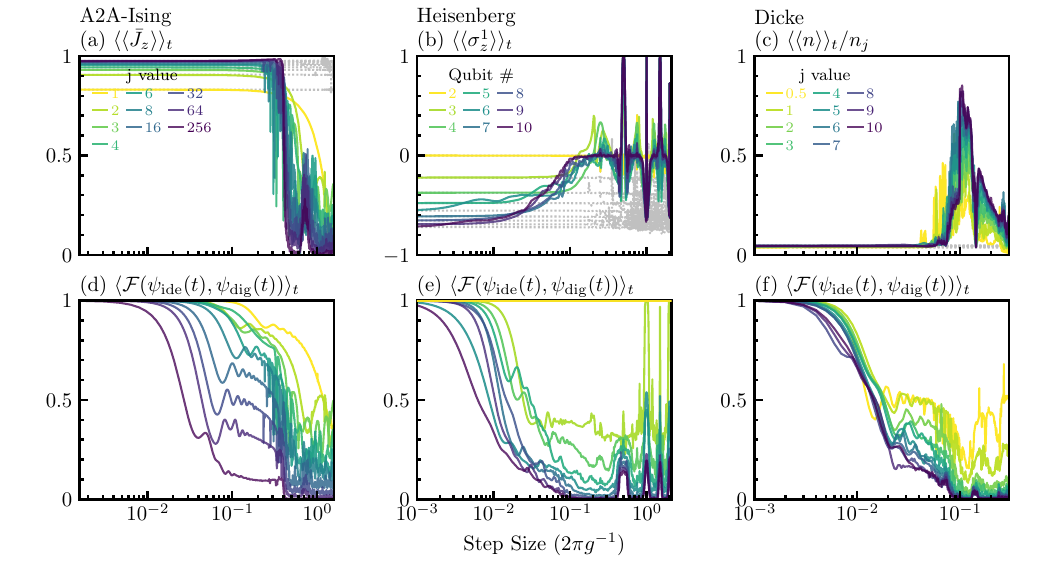}
	\caption{Time-averages of dynamical signatures for each model compared across system size: (a--c) local observables and (d--f) simulation fidelity.
	Left to right, columns show results for A2A-Ising, Heisenberg, and Rabi-Dicke models, respectively, with line colours corresponding to system sizes given in legends in (a--c), and x-axes common to each column showing step sizes.
	$\langle . \rangle_{t}$~represents time averaging over periods $t = 200~(2\pi g^{-1})$ (A2A-Ising), $t = 50~(2\pi g^{-1})$ (Heisenberg), and $t = 200~(2\pi g^{-1})$ (Rabi-Dicke).
	(a--c) Expectation values of example local observables: (a) magnetisation $\langle \braket{\bar{J}_{z}}\rangle_{t}$, (b) polarization of qubit one $\langle \braket{\sigma_{z}^{1}}\rangle_{t}$, and (c) normalised photon number $\langle\langle n\rangle\rangle_{t}/n_{j}$. 
	Dotted silver lines are time averages of sampled ideal dynamics, sampled at intervals of $\tau$, to illustrate the emergence of sampling errors at large step sizes, even in the absence of digitisation errors.
	(d--f) While time-averaged simulation fidelities $\langle\mathcal{F}(\psi_{\text{dig}}(t), \psi_{\text{ide}}(t))\rangle_{t}$ do not show the same sudden threshold because of pre-threshold quasiperiodic oscillations, they show a secondary drop at the threshold.
	Where they occur (e.g., for large enough $j$ values in the A2A-Ising and Rabi-Dicke models), stable regions are readily observed in time-averaged signatures, and are generally observed at the same position independent of size.}
	\label{Fig:4_DynamicalAverages}
\end{figure*}


\subsection{System-Size Dependence}\label{Sec:SystemSize}

In this section, we explore how Trotterisation behaviours change as a function of system size by analysing the time averages of the dynamical quantities, and also study what Trotterisation behaviours can be observed already at modest system sizes.
While we have shown that time-averaged quantities do not always show the threshold as well as their full dynamics, using the full dynamics is generally impractical for studying the dependence of Trotterisation behaviours on some other parameter such as system size.
Here, with the help of the detailed understanding we have developed by studying full dynamics in the previous section, we analyse the changing behaviour of dynamical quantities with system size via their time averages, in each case from the smallest examples to dimensions orders of magnitude larger.
The results~(Fig.~\ref{Fig:4_DynamicalAverages}) show that time averages already exhibit observable threshold behaviour for system sizes within reach of existing experimental platforms.
As discussed above, this is clearest in the Ising and Dicke models, with non-quantum chaotic variations and state dependence in the Heisenberg model pre-threshold region washing out the threshold for time averages.
And while the detailed shape of the threshold changes, its location shows no strong dependence on system size, even over a rather large range of Hilbert-space dimension (limited mainly by numerics), appearing to converge for increasing system size.

The target systems considered here are integrable (and therefore non-quantum-chaotic).
In non-integrable target systems, the Trotterisation threshold can still exist and be observable even when the target dynamics are themselves quantum chaotic irrespective of step size.
(In Appendix~\ref{App:DickeChaos2Chaos}, we show an example in the $j{=}6$ Dicke model where the Trotterised dynamics show stronger signatures of quantum chaos after the threshold than the ideal dynamics, even when the ideal dynamics are already in the quantum chaotic regime of the model.)
As discussed in Ref.~\cite{HeylM2019QLTE}, for non-integrable target systems, the threshold position may move to smaller step sizes for very large system sizes and run times, due to generic Floquet heating effects (see Discussion for more details).
Because we consider integrable target systems, however, these observations of minimal system-size dependence for position are likely to hold even at very large sizes, since the assumptions underpinning generic Floquet heating results may not apply.

It is important to note that subtle issues can arise when calculating time averages from discretely sampled time series.
The light grey lines in Fig.~\ref{Fig:4_DynamicalAverages} show the temporal averages for the ideal evolution sampled at equal time intervals $\tau$, illustrating how undersampling (large step sizes) can cause errors in time-averaged values even for ideal dynamics.
This highlights that there is a key distinction between discrete sampling errors and the Trotterisation errors, which we discuss further in Sec.~\ref{Sec:ErrorAnalyses}.

We now discuss the temporal averages of Trotterised dynamics for each model:

\paragraph{A2A-Ising model}
The first column of Fig.~\ref{Fig:4_DynamicalAverages} shows the time averages of (a) $\braket{\bar{J}_{z}}$ values and (d) simulation fidelities for a selection of system sizes between $j=1$ and $j=256$ (labelled in Fig.~\ref{Fig:4_DynamicalAverages}~(a)), averaged over $t = 200~(2\pi g^{-1})$.
Thresholds in the expectation values appear around the same position for any system size considered, and at the same location as a secondary drop in the simulation fidelities which marks its transition from oscillatory to decaying behaviour.
At the smallest sizes ($j \in \{1,2,3\}$), however, we observe neither the secondary drop in simulation fidelity, nor thresholds in the expectation values, although substantial smooth deviations from the ideal dynamics do start to appear at around the same position.
In Appendix~\ref{App:IsingSmallSystems}, we plot the dynamics of $\braket{\bar{J}_{z}}$ expectation values for $j \in \{1,2,3,4,6,8\}$, showing that destruction of quasiperiodicity in the full dynamics only becomes clearly observable once $j \gtrsim 4$, where the sharper threshold also emerges in the time averages. 
The threshold then sharpens and converges in position as system size increases.

\paragraph{Heisenberg model}
The second column of Fig.~\ref{Fig:4_DynamicalAverages} shows the time averages of (b) $\braket{\sigma_{z}^{1}}$ values and (e) simulation fidelities for the system sizes given in the Fig.~\ref{Fig:4_DynamicalAverages}~(b), and for $t = 50~(2\pi g^{-1})$.
The threshold position is again largely independent of system size, but identification of the threshold from these time-averaged quantities is more subtle, due to initial-state effects discussed in Sec.~\ref{Sec:quasiperiod}.
In Appendix~\ref{App:HeisenbergAppendix_initialState}, we provide time averaging and error analyses for various types of initial states, confirming that different initial state choices with varying amounts of pre-threshold Trotter errors still show similar changes in behaviour at a shared threshold.
In addition, for the same system sizes considered here, our analyses of eigenvector statistics in Sec.~\ref{Sec:QChaosOnset} show that \emph{global} signatures of the full Trotter step unitaries (that is, encompassing all initial states) also show the same distinctive threshold.
Note that the two-qubit case has unit simulation fidelity (yellow line in Fig.~\ref{Fig:4_DynamicalAverages}~(e)), because it has no Trotter errors, even though time-averaged quantities still exhibit sampling errors at large steps.

\paragraph{Dicke model}
The third column of Fig.~\ref{Fig:4_DynamicalAverages} shows the time averages of (c) scaled $\langle n\rangle/n_{j}$ values and (f) simulation fidelities for the system sizes labelled in Fig.~\ref{Fig:4_DynamicalAverages}~(c), and for $t = 200~(2\pi g^{-1})$.
Here, we observe a threshold in time-averaged dynamics even at the smallest size $j = 0.5$, the Rabi model.
In the Appendix~\ref{App:RabiThreshold}, we include the dynamical evolutions of various quantities for the Rabi model, which also already show dynamical signatures suggestive of quantum chaos, such as the destruction of quasiperiodicity.

The Dicke model provides a very interesting case study into what kind of system-size effects influence whether a genuine transition to quantum chaotic dynamics can be observed.
Clearly, a Trotterisation threshold (and quantum chaos beyond it) can only exist when there are non-commuting terms in the Trotter decomposition.
Otherwise, the decomposition would be exact and show regular dynamics for any $\tau$ (since we have chosen integrable target model parameters).
In the Trotterised Dicke model, the non-commutation between the TC and anti-TC component unitaries in each Trotter step is introduced primarily by single-qubit gates, and not the cavity operators.
This suggests that the quantum chaotic dynamics in this model is mainly driven by interactions of the spin-$j$ component Hilbert space.
We further support this intuition with the numerical observations that, beyond the threshold: (a) the entropy in the spin-$j$ component converges to the maximum value as $j$ increases (Appendix~\ref{App:otherSignatures}), and (b) the maximum photon numbers and participation ratio values (PR; defined in Appendix~\ref{App:otherSignatures}) are, respectively, linear and quadratic in the dimension of the spin-$j$ component $\dim_j = 2j + 1$ (Appendix~\ref{App:DickeNormCoefs}).
This is consistent with a notion that, despite strictly being infinite dimensional, the cavity mode Hilbert space explored by the Dicke model dynamics is effectively limited in size by the dimension of the spin.
In our results, we therefore normalise the photon numbers by $n_{j} = 7\times\dim_j$ (Fig.~\ref{Fig:4_DynamicalAverages}~(c)), and the PR by $(2\times\dim_j)^2$) (Appendix~\ref{App:otherSignatures}), the specific coefficients ($7$ and $2$) being identified empirically (see Appendix~\ref{App:DickeNormCoefs}).
The question of how to normalise the Hilbert-space dimensionality also arises in the context of the eigenvector statistics analysis (Sec.~\ref{Sec:QChaosOnset}), with results consistent with the somewhat empirical approach taken here.

\subsection{Eigenvector Statistics}\label{Sec:QChaosOnset}

In this section, we analyse the static properties of a Trotterised evolution, namely the eigenvector statistics of the Trotter step unitary.
So far, we have employed time evolutions and averages of dynamical quantities to show a range of Trotterisation behaviours, including the threshold, low system-size dependence, and dynamical signatures of quantum chaos beyond the threshold.
These dynamical analyses, however, are generally neither conclusive nor global.
That is, they are not able to provide unambiguous evidence of quantum chaotic Trotterised evolution without relying on additional random matrix theory (RMT) analysis, nor do they readily enable conclusions about Trotterised evolution across the full system Hilbert space.
Here, using eigenvector statistics, we demonstrate conclusive agreement with the predictions of quantum chaotic dynamics beyond the threshold, and since this approach probes properties of the full Trotterised unitary, it provides information about the global dynamics unrestricted by initial-state choice.

\subsubsection{Goodness-of-Fit Test for Eigenvector Statistics}

Dynamical signatures of quantum chaos are fundamentally consequences of the random matrix theory (RMT) properties of the eigenvectors and eigenvalues of a quantum chaotic system~\cite{HaakeF2018QSC}, known respectively as eigenvector~\cite{IZRAILEVFM1987CSEF, KusM1988UEV} and level-spacing statistics~\cite{BohigasO1984QCS, BerryMV1977LCRS} (see Appendix~\ref{App:RMTandQuantumChaos} for further details).
Reference~\cite{SiebererLM2019DQC} explored the level-spacing (eigenvalue) statistics through the mean eigenphase-spacing ratio, and a participation ratio derived from eigenvector statistics.
Unfortunately, these averaged and derived quantities do not unambiguously identify the presence (or absence) of quantum chaos for a particular Trotterised step unitary, without appealing to other model-specific information, which may or may not be available.
For example, for their particular Trotterised A2A-Ising model, Ref.~\cite{SiebererLM2019DQC} was able to exploit its well-understood classical limit as a kicked top, an approach which is obviously not available in general.

In this work, we instead carry out the comparison against RMT statistics at the full-distribution level.
Furthermore, we focus on eigenvector statistics, because while it may be more straightforward to measure level spacings directly in natural quantum systems, this is generally not also true for quantum simulators, and unlike with eigenvectors, there is no generic closed-form solution for calculating the expected statistics for RMT level spacings (see Appendix~\ref{App:RMTandQuantumChaos}).
In experimental quantum simulators, spectral information gained via Fourier transform and Ramsey-like measurements can provide access to eigenvalue statistics~\cite{RoushanP2017ssl}, and recent theory work has begun exploring ways to probe both eigenvalue and eigenvector statistics via the spectral form factor~\cite{JoshiLK2022pmq, VasilyevDV2020mqs}.

While this is not the first time the RMT properties of physical models have been studied at the full-distribution level, we demonstrate that the usual approach of making a visual comparison of the eigenvector statistics against the above RMT distribution can be misleading.
We overcome this problem by combining the eigenvector statistics with a reduced chi-squared goodness-of-fit test, and use a reduced chi-squared goodness-of-fit test statistic, $\XsqRMT$, to clearly and objectively demonstrate the onset of quantum chaos in the Trotterised dynamics as a function of step size.
As a measure of full distribution-level agreement, $\XsqRMT$ provides objective, statistically rigorous RMT comparison for individual unitary matrices, avoiding the ambiguities that can arise when relying on individual distribution moments.

Here, we calculate the eigenvector statistics of the Trotter step unitary (TSU), $U_{\tau}$, by expanding its eigenvectors $\ket{\phi_{i}}_{\tau} = \sum_{k=1}^{\mathcal{D}}c_{ik}\ket{k}$ in a non-quantum chaotic basis $\{\ket{k}\}$ (a good generic first choice being the eigenbasis of the model's free Hamiltonian---i.e., without the coupling terms---see App.~\ref{App:RMTandQuantumChaos} for more details).
The squared moduli of complex coefficients $\eta = \abs{c_{ik}}^2$ then show certain distributions for quantum chaotic systems~\cite{IZRAILEVFM1987CSEF,KusM1988UEV,HaakeF1990RMEF}.
Depending on the time-reversal symmetry of the system, there arise three (main) physically relevant random matrix ensembles~\cite{DysonFJ1962RMT1}, known for random unitaries as the circular-orthogonal ensemble (COE), circular-unitary ensemble (CUE), and circular-symplectic ensemble (CSE).
Their corresponding reduced probability densities are:
\begin{eqnarray}
	\widetilde{P}_{\text{COE}} (\eta) &=& \frac{\Gamma \left(\frac{\mathcal{D}}{2}\right)}{\Gamma 
	\left(\frac{\mathcal{D}-1}{2}\right)}\frac{(1-\eta)^{(\mathcal{D}-3)/2}}{\sqrt{\pi \eta}}, \\
	\widetilde{P}_{\text{CUE}}(\eta) &=& (\mathcal{D} - 1)(1 - \eta)^{\mathcal{D}-2}\text{, and}\\
	\widetilde{P}_{\text{CSE}}(\eta) &=& (\mathcal{D} - 1)(\mathcal{D} - 2)\eta(1 - \eta)^{\mathcal{D} - 3},
\end{eqnarray}
where $\mathcal{D}$ is the dimension of the Hilbert space, and $\Gamma(x)$ is the Gamma function, or generalized factorial.
We denote these distributions in our figures by the three letter acronyms for the corresponding classes.
Relevant details about these distributions, choosing an appropriate basis $\{\ket{k}\}$ that reveals the system's time-reversal symmetry class, the relation between RMT and quantum chaos, and other technical factors are discussed in the Appendix~\ref{App:RMTandQuantumChaos}.

For DQS Hamiltonians possessing COE time-reversal symmetry (like the A2A-Ising and Dicke models studied here), generic first-order Trotterisations of the form described in Eq.~(\ref{Eq:TrotterisationDef}) will automatically break or obscure that symmetry, even when represented in the target Hamiltonian's symmetric basis (see Appendix Sec.~\ref{App:refBasis}).
For two-part Trotterisations, however, provided the parts are each also time-reversal symmetric with the same symmetric basis, the first-order TSU is guaranteed to possess a time-reversal symmetry with a new symmetric basis, and should thus still be compared against COE statistics~\cite{HaakeF2018QSC}.
The unitary's COE statistics may not always be apparent in the computational basis, but assuming the TSU is written in the symmetric basis of the individual Hamiltonian parts, a symmetric basis can always be found via a simple basis transformation which corresponds to converting the TSU from a first-order to a symmetric second-order Trotterisation.
This is how we analyse the A2A-Ising and Dicke models, whose individual Hamiltonian parts $H_x$ and $H_z$, defined in Eq.~(\ref{Eq:IsingHamiltonian}), and $H_{TC}$ and $H_{ATC}$, defined in Eqs.~(\ref{Eq:TCHamiltonian}) and~(\ref{Eq:ATCHamiltonian}), respectively, are already real and thus invariant under complex conjugation in the computational basis.


\begin{figure*} 
	\centering
	\includegraphics[scale=0.94]{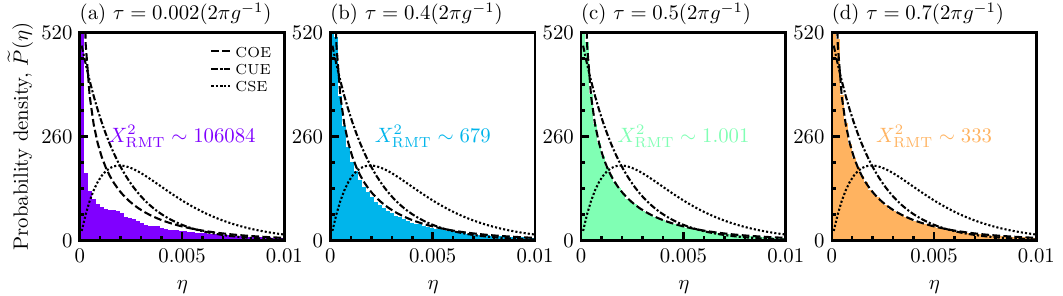}
	\caption{Histograms showing eigenvector statistics for the $j=256$ A2A-Ising Trotter step unitary (TSU) for representative step sizes in different regimes: (a) $\tau = 0.002~(2\pi g^{-1})$ (before the threshold), (b) $\tau = 0.4~(2\pi g^{-1})$ (\emph{on} the threshold), (c) $\tau = 0.5~(2\pi g^{-1})$ (quantum chaotic regime beyond the threshold), and (d) $\tau = 0.7~(2\pi g^{-1})$ (stable region in the quantum chaotic regime).
	Agreement with random matrix theory (RMT) distributions, shown for different time-reversal symmetries by black dashed and dotted lines in each plot, provides strong evidence for quantum chaotic dynamics.
	However, while the visual agreement and disagreement in the quantum-chaotic and regular regimes, respectively, in (a) and (c) seem clear, apparently similar \emph{visual agreement} in (d) would then be misleading, as it corresponds to a stable region where the behaviours of dynamical signatures clearly deviate from quantum chaotic, and even (b) shows reasonable visual agreement at a step size in the transition where the dynamics still exhibits clearly quasiperiodic behaviour.
	By contrast, the reduced chi-squared goodness-of-fit test statistic $\XsqRMT$, compared here against the COE distribution and labelled in coloured text on each figure, provides quantitative and objective evidence for agreement or disagreement with RMT and successfully predicts the observation of quantum chaotic dynamics in each case: for example, $\XsqRMT$ shows clear disagreement with quantum chaotic dynamics in both transition regions (b) and even stable regions (d).}
	\label{Fig:5_A2AIsingEigenVecHistograms} 
\end{figure*}


\subsubsection{Failure of visual comparisons versus $\XsqRMT$}

Figure~\ref{Fig:5_A2AIsingEigenVecHistograms} shows the eigenvector statistics of the A2A-Ising TSU calculated in the $J_{z}$ basis for $j = 256$ and, respectively, for the step sizes: (a) $\tau = 0.002~(2\pi g^{-1})$ before the threshold, (b) $\tau = 0.4~(2\pi g^{-1})$ on the threshold, (c) $\tau = 0.5~(2\pi g^{-1})$ in the quantum chaotic regime beyond the threshold, and (d) $\tau = 0.7~(2\pi g^{-1})$ on a stable region in the quantum chaotic regime.
Recall that collective spin behaviour in the A2A-Ising DQS restricts the system evolution into a total-spin conserved subspace $\mathcal{D}_{j} = 2j + 1$ of the full Hilbert space $\mathcal{D}_{N} = 2^{N}$ of $N$ spin-1/2 systems.
It therefore cannot display many-body quantum chaos of the full $N$-spin system, but can still exhibit the quantum chaos of a single spin $j$ within each symmetry subspace.
The standard RMT distributions in Fig.~\ref{Fig:5_A2AIsingEigenVecHistograms} are therefore plotted for $\mathcal{D}_{j} = 2j + 1$.
Because the A2A-Ising decomposition in Eq.~\eqref{Eq:A2AIsingDecomp} has non-symplectic time-reversal symmetry, it should show COE statistics~\cite{SiebererLM2019DQC}.
While Figure~\ref{Fig:5_A2AIsingEigenVecHistograms} panels (a) and (c) show clear visual disagreement and agreement, respectively, with the expected COE distribution, the apparent visual \emph{agreement} in Fig.~\ref{Fig:5_A2AIsingEigenVecHistograms}~(d) is highly misleading, as this corresponds to a stable region, where the quantum chaotic dynamics are clearly interrupted.
To characterise this agreement in a more quantitative manner, we introduce a reduced-chi-squared goodness-of-fit test statistic $\XsqRMT$ to provide a generic quantitative signature for agreement with the expected RMT eigenvector statistics, with agreement requiring $\XsqRMT \sim 1$ (see Appendix~\ref{App:EigenVecTheory} for the definition and more details).
The level of agreement is quantified by comparing $\XsqRMT$ with confidence regions for the reduced chi-squared distribution:
these scale in size roughly inversely with the square root of the chi-squared degrees of freedom (here, $\text{DOFs} \sim \text{histogram bins}$, which we set proportional to $\mathcal{D}^2$).
The values of $\XsqRMT$ comparing the histogram plots with the relevant COE distribution in Figs~\ref{Fig:5_A2AIsingEigenVecHistograms}~(a--d) are, respectively, (a) $\sim 106084$, (b) $\sim 679$, (c) $\sim 1.001$, and (d) $\sim 333$ (compared with a $95\%$ confidence region---approximately $\pm 2\sigma$---of 0.983--1.017 for $j=256$), successfully identifying not only agreement and disagreement in the quantum-chaotic and regular regimes, but also clear disagreement with quantum chaotic eigenvector statistics within the stable region.
Even Figure~\ref{Fig:5_A2AIsingEigenVecHistograms}~(b) arguably also agrees reasonably well visually, but $\XsqRMT$ clearly demonstrates there is still strong disagreement with the target COE distribution at this point in the middle of the transition.

In general, RMT can be used to make specific quantitative predictions about the expected distributions of eigenvector coefficients for random unitary matrices, but does not provide any guidance as to the eigenvector distributions expected for non-random matrices.
Similarly, the $\XsqRMT$ goodness-of-fit test enables quantitatively rigorous analysis of whether the observed distribution for a particular unitary is consistent or inconsistent with the distribution predicted by RMT.
However, while $\XsqRMT$ sufficiently greater than 1 can definitively rule out agreement with a particular model, the goodness-of-fit test makes no generic prediction about what $\XsqRMT$ should be in the case of disagreement.
In quantum chaotic systems, eigenvectors are generally quite evenly distributed across all basis states, typically giving $\eta \sim 1/\mathcal{D} \ll 1$).
Except for very small systems, RMT probability densities rapidly approach exponentially small values for increasing $\eta$, with negligible support across most of $\eta$'s allowed range ($1/\mathcal{D} \ll \eta \leq 1$).
Eigenvectors for regular system dynamics, however, commonly contain larger components ($\eta \gg 1/\mathcal{D}$) which can greatly distort $X^2$ goodness-of-fit comparisons against RMT distributions unless significant care is taken in the way the calculated values of $\eta$ are binned to form a histogram.
At the same time, these are exactly the components we want to retain sensitivity to in order to detect deviations from quantum chaotic behaviour.
In Appendix~\ref{App:EigenVecTheory}, we provide a detailed initial recipe for how to implement the $\XsqRMT$ goodness-of-fit test in a robust and rigorous way, which involves dividing up the main part of the comparison RMT probability distribution into equal probability bins ($\sim 10$ counts per bin) and capturing the exponentially vanishing tail region in a single, low-probability tail bin ($\ll 1$ count) that enhances sensitivity to small regions of stable dynamics in Hilbert space.
This provides an important first step in the development of objective statistical techniques for identifying RMT, which we will study further in future work.

The $\XsqRMT$ goodness-of-fit test statistic provides a single and (compared to visual inspection) meaningful value for a histogram plot, and therefore also allows us to analyse how agreement with RMT eigenvector statistics varies as a function of different parameters.
In the following, we analyse the eigenvector statistics for the Trotter step unitaries of each DQS as a function of step and system sizes.
Full eigenvector statistics histograms for some relevant step and system sizes are shown in Appendix~\ref{App:IsingEigenVecs} for A2A-Ising, in Appendix~\ref{App:HeisebergEigVec} for Heisenberg, and in Appendix~\ref{App:DickeEigTruncation} for Dicke Trotterisations.
Note that $\XsqRMT$ values for the Dicke DQS are obtained by setting the default cavity truncation to $\dim_{\rm c} = \dim_j+1$ (except for $j=1/2$), and we discuss this further after presenting the results.


\begin{figure*}[ht!]
	\centering
	\includegraphics[scale=0.94]{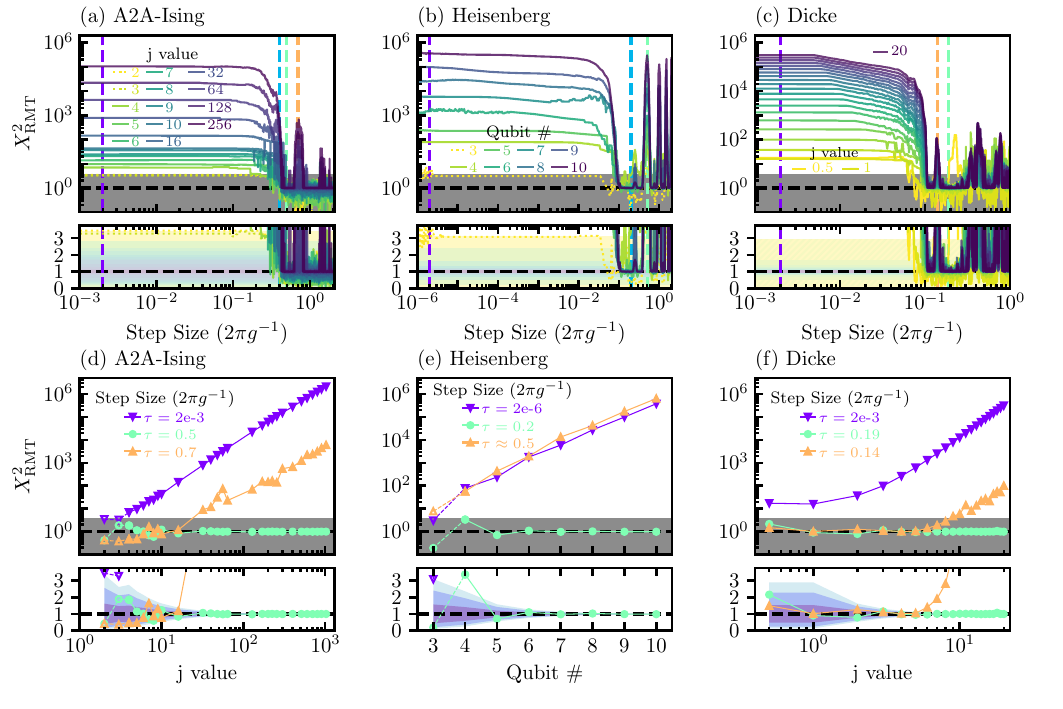}
	\caption{The reduced chi-squared goodness-of-fit test statistic $\XsqRMT$, calculated from the Trotter step unitary (TSU) $U_{\tau}$, quantifying each model's agreement with RMT eigenvector statistics (here, against COE).
	(a--c) $\XsqRMT$ versus step size (x-axes) for the system sizes given in the legends for A2A-Ising TSU, Heisenberg TSU, and Dicke TSU, respectively.
	(Sizes in panel (c) include $j=1/2$ and all integer $j$ between 1 and 20.)
	The lower panels zoom in on the grey shaded regions from the main panels, plotted on a linear $y$ axis, with the shaded regions, coloured by system size with colours matching (but paler than) the corresponding line colour, indicating the $99\%$ confidence regions for the chi-squared distribution, to illustrate where statistically reliable conclusions can be drawn. (Note: The $j=1$ region in (c) is shown in "hatched" form to ensure the $j=1/2$ region is also visible.)
	(d--f) $\XsqRMT$ versus system size (x-axes) for representative step sizes chosen in regular, quantum chaotic, and stable regions, with extra panels below zooming in on the range around $\XsqRMT \sim 1$, with the $70\%$, $95\%$ and $99\%$ confidence regions (purple, blue and light blue) of the chi-squared distributions defining regions of rigorous statistical agreement with RMT.
	The step sizes, recorded in the legends, are marked as vertical dashed lines in (a--c).
	The colours also match those of the histograms plotted in Figs~\ref{Fig:5_A2AIsingEigenVecHistograms}~(a--d) (for the A2A-Ising TSU).
	From (a--c), the goodness-of-fit test statistic not only shows the clear transition threshold observed in dynamical quantities, but also identifies stable regions.
	Figures (d--f) show how these features depend on system sizes:
	For example, in the A2A-Ising and Dicke models, quasiperiodic dynamics for the stable regions are not observed to emerge from surrounding quantum chaotic regions until larger system sizes, and we see here that $\XsqRMT$ is able to track this behaviour successfully.
	In the Heisenberg model, the quasiperiodicity of the stable regions is so marked that $\XsqRMT$ is comparable in the stable regions to the pre-threshold non-quantum-chaotic regime.}
	\label{Fig:6_chiSquareForAll} 
\end{figure*}


\subsubsection{$\XsqRMT$ Demonstration of Quantum Chaos}

Figure~\ref{Fig:6_chiSquareForAll}~(a--c) shows the $\XsqRMT$ test statistic as a function of step size, respectively, for (a) the A2A-Ising TSU (against COE), (b) the Heisenberg TSU (against COE), and (c) the Dicke TSU (against COE).
In each model, $\XsqRMT$ shows clearly different qualitative behaviour in the step-size regions before and after the threshold observed in Trotterised dynamics.
For large enough system sizes, we consistently observe a sudden drop in $\XsqRMT$ at the threshold, and $\XsqRMT \sim 1$ for extended step-size regions after the threshold (aside from stable regions), showing quantitatively rigorous agreement with RMT (defined by the shaded regions representing the $99\%$ confidence regions for each system size) that provides strong, objective evidence of the onset of quantum chaotic dynamics.
Under the same conditions, we also consistently observe $\XsqRMT \gg 1$ at small step sizes, exhibiting clear disagreement with RMT expectations for step sizes before the threshold, where the DQS produces quasiperiodic (non-quantum-chaotic) dynamics.
These results illustrate that each of these models undergoes a dramatic transition from regular to quantum chaotic dynamics at a position that matches the threshold identified from dynamical signatures, and which already sets in at quite modest system sizes.
Across all models and multiple orders of magnitude in Hilbert-space dimension, once the system reaches sufficient size, we do not observe any significant variation with system size of the position of onset for $\XsqRMT \sim 1$.

%

We also use $\XsqRMT$ to carry out a thorough study of how the onset of quantum chaos varies with system size, and particularly at smaller system sizes more easily realisable with current experimental processors.
Characterising quantum chaos at small system sizes is a significant challenge.
Due to the small number of discrete eigenfrequencies in a unitary matrix, even a Haar-random (CUE) unitary will give rise to significant quasiperiodicity in system dynamics.
And in the case of a $\XsqRMT$ goodness-of-fit test, the limited number of eigenvector components available to form the ``observed'' histogram counts for the unitary matrix under study strongly restricts the statistical confidence that the $\XsqRMT$ test can achieve.
It becomes harder to retain either sufficient counts in each histogram bin to form a statistically rigorous $X^2$ test statistic (convention requiring counts per bin $\gtrsim 10$), or enough bins (the degrees of freedom in the $X^2$ test) to be able to meaningfully distinguish between the shapes of different probability distributions.
In general, the generic approach we have defined in Appendix~\ref{App:EigenVecTheory}---dividing up the target RMT distribution (discounting the exponentially vanishing tail region) into \emph{at minimum} 5 equal probability bins with $\gtrsim 10$ expected counts per bin ($\mathcal{D}^2 \geq 50$ components, after accounting for any relevant symmetries)---ensures a good chance of successfully characterising low-dimensional systems without compromising statistical validity.
For the current analysis, however, we ideally also want the $\XsqRMT$ value for a target DQS model ($\tau \rightarrow 0$) to lie well outside the standard confidence regions of the chi-squared distribution (shown for $99\%$ confidence), to be able to clearly distinguish quantum chaotic Trotterised dynamics from regular target dynamics.
This requires at least $j \geq 4$ for A2A-Ising and $N\geq 4$ for Heisenberg.
For the Dicke model, however, it is possible to obtain reliable results even for $j=1/2$, due to the extra flexibility offered by the cavity mode degree of freedom.

Note: In Figure~\ref{Fig:6_chiSquareForAll}, we do include results for a few limited cases---for qualitative comparison purposes---which have fewer components than is ideal (flagged in Figs~\ref{Fig:6_chiSquareForAll}a and~\ref{Fig:6_chiSquareForAll}b using dotted lines and in Figs~\ref{Fig:6_chiSquareForAll}d and~\ref{Fig:6_chiSquareForAll}e with ``empty'' markers and dashed lines), calculating $\XsqRMT$ using $\sim 5$ counts per bin across at least 5 bins.
For such small system sizes, however, statistically rigorous conclusions will likely require more sophisticated statistical analysis.
For example, for the method used here, in the $j=2$ A2A-Ising model ($\mathcal{D}=5$, with five 5-count main bins plus a tail bin, see App.~\ref{App:chiSquaredTestDetail}), even the ideal target model is barely distinguishable from RMT distributions, with $\XsqRMT \sim 3.45$ lying inside the conventional $99\%$ region for a $\chi^2$ distribution with 5 degrees of freedom.

To explore more quantitatively how the quantum chaos $\XsqRMT$ test statistic varies with system size, in Figures~\ref{Fig:6_chiSquareForAll}~(d--f), we plot $\XsqRMT$ as a function of system size for three indicative step sizes in the regular (pre-threshold), quantum-chaotic (beyond-threshold), and stable-revival regions (given in the legend for each model).
At small (non-quantum chaotic) step sizes, we see that $\XsqRMT$ scales rapidly with system size, at the same time as the standard chi-squared confidence regions shrink closer to one (1), illustrating how regular dynamics deviates more strongly from quantum chaotic dynamics with increasing system sizes.
While the specific scaling is necessarily model dependent, it also depends on how the number of histogram bins is chosen.
With our method, which scales the number of bins---and hence the number of chi-squared degrees of freedom---linearly with the number of independent RMT components, each model converges to a power-law scaling, with $\log\left[\XsqRMT\right]$ proportional to \emph{dimension}: $(2j{+}1)$ for A2A-Ising, $2^N$ for Heisenberg, and $(2j{+}1)(2j{+}2)$---see below---for Dicke.
For the stable-island step size, $\XsqRMT$ also rapidly increases to $\XsqRMT\gg1$ for each model, and converges to a similar scaling as observed in the regular region.
But while the stable islands are visible at any system size in the Heisenberg DQS, they do not tend to show up until larger sizes in the A2A-Ising and Rabi-Dicke models.
In clear contrast, however, for the post-threshold step size, we find $\XsqRMT\sim1$ at any system size large enough for $\XsqRMT$ to meaningfully discriminate regular from quantum chaotic dynamics.
The bottom panels show a zoomed-in view around $\XsqRMT\sim1$, along with the $70\%$, $95\%$ and $99\%$ confidence regions for the chi-squared distribution.
These results show that the eigenvector components for these three models exhibit quantitative and rigorous statistical agreement with RMT-predicted distributions, providing a rigorous demonstration of quantum chaotic dynamics in the post-threshold regime.

\subsubsection{Finite Truncation of the Cavity Dimension in the Rabi-Dicke DQS}

While all three models involve explicitly finite spin-chain systems, the Dicke model also includes the formally infinite-dimensional cavity degree of freedom.
This introduces additional complications when analysing the eigenvector statistics of the Dicke TSU arising from the numerical need to truncate the infinite-dimensional cavity at finite dimension $\dim_{\rm c}$.
It is non-trivial and is not in the scope of this paper to provide a rigorous, theoretical explanation of the effects of cavity Hilbert-space truncation on eigenvector statistics, but here we study it in some detail numerically (see Appendix~\ref{App:DickeEigTruncation} for the full discussion).
At a high level, we find results consistent with an argument that, due to the single-qubit gates in the Dicke Trotterisation, the Dicke model's effective total spin component $j$ is the main driver of quantum chaotic dynamics.
Specifically, we observe that the beyond-threshold Dicke TSUs show agreement with RMT distributions only for $\dim_{\rm c} \sim \dim_j$, for both eigenvector statistics, as well as another strong signature of quantum chaos from RMT, the level-spacing statistics~\cite{BerryMV1977LCRS, BohigasO1984QCS, LucaD2016ETH, HaakeF2018QSC}.
Any agreement is lost in both of these statistics when $\dim_{\rm c} \gg \dim_j$.
These results are consistent with our observations from dynamics that the maximum photon-number build-up is linear in $\dim_j$, and that PR is quadratic in $\dim_j$, along with the fact that the reduced entropy of the spin $j$ subsystem is nearly maximised post-threshold, a sign of near-maximal spin-photon entanglement.
Thus, we consider for now that quantum chaotic dynamics and the breakdown of Trotterisation in the Dicke model are primarily driven and limited by effects arising within the Hilbert space of the spin component.
Based on this, we truncate $\dim_{\rm c}$ at ${\dim_j}+1$ when analysing statistical RMT properties of the Dicke TSU.
In doing so, we are effectively analysing the statistics of the Dicke model in a restricted subspace that is within reach of the model system's interactions, in a similar way to what is established practice (if more rigorously defined) for systems with physical symmetries such as parity.
(Note: For $j=1/2$, we instead use $\dim_{\rm c}=\dim_j+4=6$, large enough to allow reliable $\XsqRMT$ analysis, but still within the range where larger spins show good RMT agreement.)

 
\begin{figure*} 
	\centering
	\includegraphics[scale=0.94]{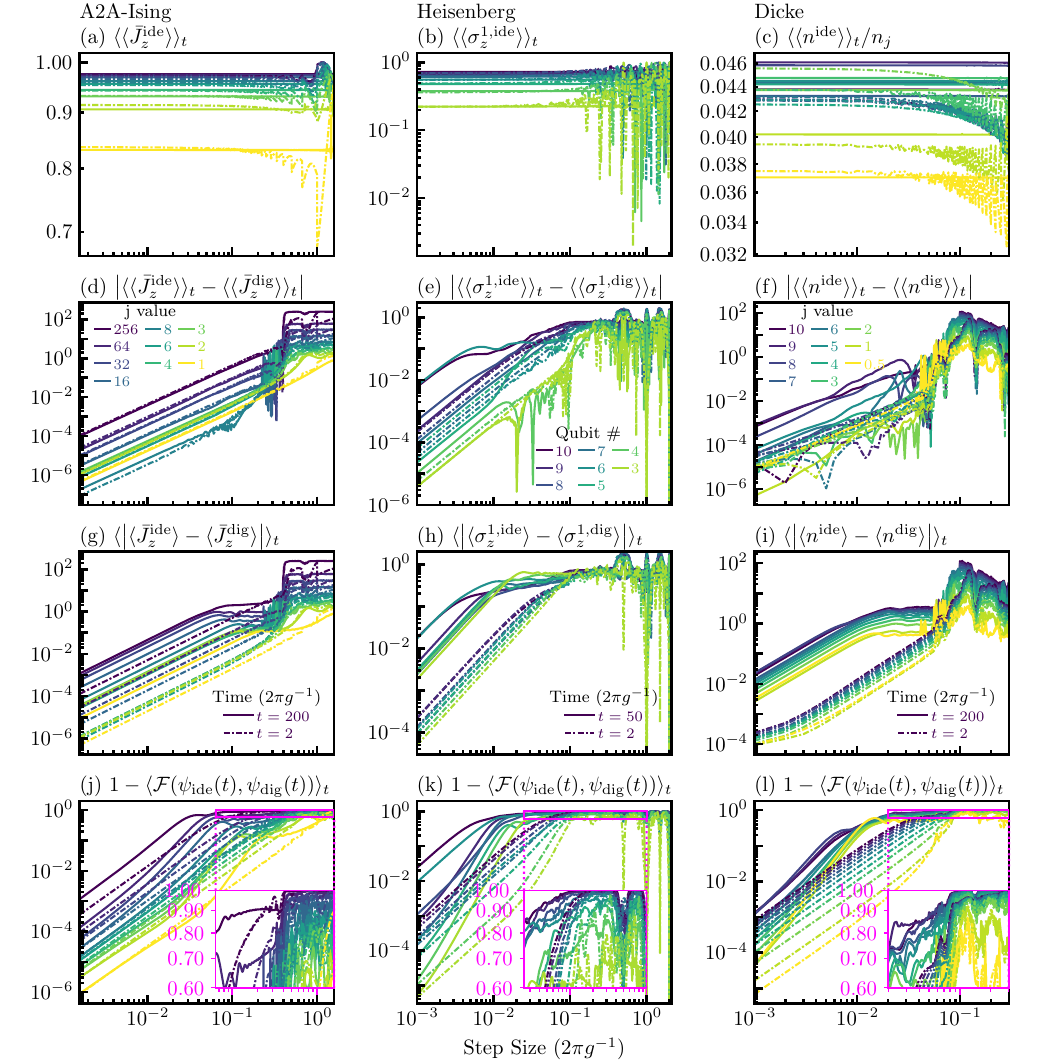}
	\caption{
	Trotterisation error analysis for (a--i) local-observable expectation values and (j--l) simulation fidelity for different system sizes and different time scales.
	Left to right, the columns show results for A2A-Ising, Heisenberg, and Rabi-Dicke models, respectively, with line colours corresponding to system sizes shown in the legends of (d--f), line styles indicating the averaging time (shown in the legend of (g--i)), and Trotter step sizes on the x-axes common to each column.
	The specific quantities plotted on the y-axes are given in the sub-figure titles, with $\langle .\rangle_{t}$ representing time averaging.
	The sub/super-scripts ide and dig represent quantities obtained, respectively, from the ideal and digital simulations.
	Local-observable errors are calculated for expectation value of magnetisation $\bar{J}_{z}$ (A2A-Ising), the first qubit polarization $\sigma_{z}^{1}$ (Heisenberg), and the photon number $n$ (Rabi-Dicke).
	(a--c) Raw temporal averages for expectation values $\langle\langle O^{\text{ide}}\rangle\rangle_{t}$ for coarsely sampled ideal dynamics plotted against step-size sampling intervals, $\tau$.
	Sampling errors for the ideal dynamics depend strongly on averaging time, but can be disentangled from Trotterisation errors by comparing the digital quantities $\langle O^{\text{dig}}(t)\rangle$ with coarsely sampled ideal $\langle O^{\text{ide}}(t)\rangle$.
	(d--i) The absolute difference between time-averaged expectation values for digital and ideal dynamics, $\abs{\langle\braket{O^{\text{ide}}}\rangle_{t}-\langle\braket{O^{\text{dig}}}\rangle_{t}}$.
	(g--i) Time average of the absolute, point-to-point difference between digital and ideal dynamics at each time step, $\langle \abs{\braket{O^{\text{ide}}}-\braket{O^{\text{dig}}}}\rangle_{t}$.
	(j--l) The simulation infidelity $1-\langle\mathcal{F}(\psi_{\text{dig}}(t), \psi_{\text{ide}}(t))\rangle_{t}$, is another point-to-point error metric.
	The quasiperiodic pre-threshold oscillations observed in the full simulation fidelity dynamics cause the threshold to be seen as a secondary jump in the time-averaged simulation infidelity, highlighted by the insets in (j--l).}
	\label{Fig:7_TrotterErrors}
\end{figure*}


\subsection{Trotter Error Analyses}\label{Sec:ErrorAnalyses}

In each model, by analysing the statistical properties of the Trotter step unitary and temporal evolutions (and averages) of various dynamical quantities, we have identified a range of apparently universal performance behaviours and a threshold in Trotter step size beyond which the Trotterised dynamics become quantum chaotic.
Here, we analyse the Trotterisation errors in state fidelities and expectation values, and we identify the same threshold in each error measure and also examine the behaviour of the errors before and beyond the threshold.
We look at the Trotter errors for various system sizes and different time scales:
The shorter time scales considered can be achieved within accessible coherence times for current experimental platforms, demonstrating that studies of Trotter-error scaling with step size are within reach of current experiments.

Figure~\ref{Fig:7_TrotterErrors} shows our error analysis for the A2A-Ising, Heisenberg, and Dicke models.
We first analyse errors in the expectation values of local observables, using the same observables considered in previous sections.
The results in Figs~\ref{Fig:7_TrotterErrors}~(a--c) highlight an important subtlety:
namely, even for ideal dynamics, sampling the time evolution at fixed times $\tau$ can lead to errors in time averages at large enough $\tau$.
These time averages also depend strongly on the averaging time, with more pronounced errors arising at smaller step sizes for short averaging times.
Such undersampling errors obviously become more significant with a lower sampling rate.
Here, we introduce appropriate error measures to distinguish trivial sampling errors from the approximation errors caused by Trotterisation.

We start by considering the absolute difference between the temporal averages (as previously studied in Ref.~\cite{SiebererLM2019DQC})
\begin{equation}\label{Eq:absErr}
	\Delta O := \abs{\langle\langle O^{\text{dig}} \rangle\rangle_{t} - \langle\langle O^{\text{ide}} \rangle\rangle_{t}},
\end{equation}
where $\braket{O}$ is the expectation value of some operator $O$, with the superscripts (dig/ide) indicating expectation values obtained, respectively, from digital and ideal dynamics (sampled at every $\tau$), and the second layer of brackets $\braket{.}_{t}$ represent time averaging for total simulation time $t$.
Figures~\ref{Fig:7_TrotterErrors}~(d--f) show $\Delta O$ for each model.
In each model, $\Delta O$ shows an approximately increasing trend with step size leading up to the previously identified threshold positions, before flattening out after the thresholds.
(Note that the error values for the $\braket{J_{z}}$ expectations in Ising and $\braket{n}$ photon numbers in Dicke are not normalised with spin sizes to avoid removal of system size dependencies in the errors.)
In the A2A-Ising model (Fig.~\ref{Fig:7_TrotterErrors}~(d)), the errors reveal a reasonably clear threshold at the right position, although this is somewhat complicated by the appearance of additional noise prior to the threshold.
In the Heisenberg and Rabi-Dicke models (Fig.~\ref{Fig:7_TrotterErrors}~(e--f)), the situation becomes much less clear, with significant extra noise before and after the threshold, and clear irregular behaviour even at small $\tau$.
Moreover, while $\Delta O$ still shows a generally increasing trend with step size, it does not show an obvious scaling dependence with system size or total simulation time across any of the models.
So even though this measure does to some degree attempt to isolate the effect of Trotter errors from sampling errors, it nevertheless does not provide a good error measure for identifying differences between Trotterised and ideal time evolutions, especially at the finite times relevant to practical DQS.
For example, a small difference between the averages does not guarantee that the instantaneous time evolutions are actually close.

We therefore introduce and analyse an alternative time-averaged error, namely the average, point-to-point absolute difference between expectation values
\begin{equation}
	\delta O := \braket{\abs{\braket{O^{\text{dig}}} - \braket{O^{\text{ide}}}}}_{t},
\end{equation}
where the notation is the same as in Eq.~\eqref{Eq:absErr}.
Now, $\delta O$ captures the absolute deviations of Trotterised evolution from the ideal dynamics at every sample, and is only zero for identical trajectories.
It therefore both separates the Trotter errors from sampling errors, and reliably measures how close the full dynamical evolutions are (not just their averages).
From Figs~\ref{Fig:7_TrotterErrors}~(g--i), unlike for $\Delta O$, we observe that $\delta O$ is a simple analytic function of step size over a large range of step sizes (prior to a plateau region that appears at long time averages), suggesting that it can be perturbatively expanded (and potentially corrected) in step size, even at point-by-point detail for finite-time evolution.
Similarly, it also shows a clear increasing trend not only with step size, but also as a function of both total simulation time and, with some subtle variation, system size.
From Figs~\ref{Fig:7_TrotterErrors}~(g--i), we see that the average, point-to-point error also shows the thresholds much more clearly, especially in the A2A-Ising and Rabi-Dicke models, with the errors before the threshold being consistently smaller than the errors in the quantum chaotic regime.
Even in the Heisenberg model, where the near-threshold errors almost reach the levels observed in the quantum chaotic regime, the errors still show a clear change in step-size scaling beyond the threshold.

Another averaged, point-to-point error measure is the averaged simulation (state) infidelity (or simulation error) $1 {-} \langle\mathcal{F}(\psi_{\text{dig}}(t), \psi_{\text{ide}}(t))\rangle_{t}$, shown in Figs~\ref{Fig:7_TrotterErrors}~(j--l), and it shows the same behaviours observed in $\delta O$.
The lack of a sharp threshold in average simulation error has already been explained in the context of average fidelity as the result of the extra oscillatory behaviour of simulation fidelity before the threshold.
Consistently, it affects the long time averages much more than the short, and we identify the threshold as the residual, secondary jump in simulation errors highlighted by the insets of Figs~\ref{Fig:7_TrotterErrors}~(j--l).


\section{Discussion and Outlook}\label{Sec:Conclusion}


The power of a quantum simulator, as represented by the achievable complexity of tasks that it can solve, is typically limited by physical resources (e.g., fabrication costs or runnings costs per qubit, cooling budget per qubit), hardware performance (e.g., qubit noise and decoherence, gate errors, etc), and in the case of DQS, the intrinsic accuracy of the digitisation algorithm (e.g., Trotter errors).
In large-scale quantum simulators, these resource costs will be to some degree fungible, and balancing them will become an engineering optimisation to maximise simulation power.
For example, in fault-tolerant DQS, error correction will ultimately allow hardware performance limits to be surpassed by transferring the cost into increased consumption of physical resources.
But because these costs snowball as the noise to be corrected approaches the fault-tolerant threshold, directly optimising hardware performance will still remain crucial.
Similarly, methods for improving digitisation accuracy typically result in increased exposure to noise and decoherence.
For example, in architectures where gate times operate reasonably close to the speed of the control electronics, the total run time (and hence decoherence effects) can be increased because the time spent on buffers between gates can be significant.
Also, constructing complex interactions from simple primitive operations often requires additional transformation gates (e.g., as in the Heisenberg and Dicke models defined in Eqs~\ref{Eq:HeisenbergDecomposition} and \ref{eq:DickeDecomposition}, respectively), adding both run time and gate errors.
Similarly, increasing the \emph{order} in Trotterisation algorithms, which typically requires the addition of negative-time unitary evolutions (at least beyond second order), also increases the laboratory run time required to achieve a given simulation time.
As a result, any improvements that can be achieved in terms of relaxing the digitisation requirements for a given simulation accuracy, will also directly decrease the decoherence and gate errors experienced by the processor.

\subsection{On the universality of Trotterisation performance}

In this work, we have focussed exclusively on digitisation errors, in particular Trotter errors, to study the ultimate performance limits that Trotterisation algorithms will face.
We numerically investigate the performance of three experimentally realisable Trotterisation models: the A2A-Ising, Heisenberg, and Dicke models.
Using a range of dynamical and static signatures (derived from state evolutions and unitary operators, respectively), we have demonstrated a marked degree of universality across DQS models for a range of performance behaviours, including the emergence of a performance threshold at modest system sizes, which is observable across all signatures we studied.
Firstly, the full dynamics (Figs~\ref{Fig:3_FullDynamicsAndFT}~(a--c)) and time averages (\ref{Fig:4_DynamicalAverages}~(a--c)) of local observables, delocalisation and other complementary derived quantities (Fig.~\ref{AppFig:10_OtherDynamicalSignatures} in Appendix~\ref{App:otherSignatures}), and new static signatures of quantum chaos based on RMT (Fig.~\ref{Fig:6_chiSquareForAll}), show that, with some exceptions at very small system sizes, all models studied exhibit regular, quasiperiodic behaviour prior to, and quantum chaotic dynamics after the threshold.
However, in addition to just the threshold, by studying in detail the full dynamics of the simulation fidelity (Figs~\ref{Fig:3_FullDynamicsAndFT}~(d--f)), we demonstrate that the widely different models considered also exhibit a broader range of shared behaviours:
a high-fidelity region at smaller steps and shorter times, with fast, but relatively stable oscillatory errors;
a second region between that and the threshold exhibiting additional, slower large-amplitude oscillations; and, within the post-threshold quantum chaotic regime, intermittently emerging stable regions where the quantum chaotic dynamics is interrupted and some degree of quasiperiodicity reemerges.
These new shared behaviours, as well as the threshold itself, become arguably even more apparent from studying the corresponding Fourier spectra (Figs~\ref{Fig:3_FullDynamicsAndFT}~(g--l)).

The models studied here encompass widely distinct system types and symmetries (kicked tops, nearest-neighbour spin chains with longitudinal and transverse external fields, and coupled spin-cavity systems with infinite dimension), and we also saw qualitatively similar behaviour across different parameter choices (not shown).
The marked universality of all behaviours suggests that there may exist underlying generic principles governing the behaviour of Trotterised DQS, which may help to predict features of large-scale simulations which are difficult to study classically.
More work is required to identify and understand these underlying principles.
Interesting open questions in this direction include, for example, studying whether these generic behaviours persist for different digitisation techniques, even more varied simulation models (e.g., in quantum chemistry), and non-integrable target systems.

\subsection{On interpretation of the Trotterisation threshold}

Focussing now directly on the threshold transition to quantum chaos, combining what we learn from across all our results provides valuable insight into the underlying nature of the breakdown in simulation performance.
From the full dynamics for each model (Fig.~\ref{Fig:3_FullDynamicsAndFT}), we observe clear quasiperiodic dynamics, consistent with the Trotterised system being non-quantum-chaotic, until immediately prior to the threshold (given in units of $2\pi/g$: $\tau_{\rm A2A{\mhyphen}Ising} \sim 0.4$, $\tau_{\rm Heisenberg} \sim 0.09$ and $\tau_{\rm Rabi{\mhyphen}Dicke} \sim 0.08$).
The pre-threshold oscillatory behaviour is particular clear and sudden for the local observables (Fig.~\ref{Fig:3_FullDynamicsAndFT}~(a--c)) (and the participation ratio (Fig.~\ref{AppFig:10_OtherDynamicalSignatures}~(a--c)) in the Appendix~\ref{App:otherSignatures}).
For the simulation fidelity in Fig.~\ref{Fig:3_FullDynamicsAndFT}~(d--f) (and the perturbation fidelity (Fig.~\ref{AppFig:10_OtherDynamicalSignatures}~(d)) and entropy of entanglement (Fig.~\ref{AppFig:10_OtherDynamicalSignatures}~(e--f)) in the Appendix~\ref{App:otherSignatures}), the periodicity of the oscillations starts changing prior to the threshold (as seen in the Fourier transform shown in Fig.~\ref{Fig:3_FullDynamicsAndFT}~(g--l)), but clear oscillations continue until the threshold.
This understanding is also consistent with the Trotter error analysis shown in Fig.~\ref{Fig:7_TrotterErrors}, and particularly the full quantum state error calculated from the simulation fidelities Fig.~\ref{Fig:7_TrotterErrors}~(j--l).
Focussing on the plots with shorter time averaging, where the simulation-fidelity time averages are less complicated by the behaviour in the near-threshold, quasiperiodic region, we observe that the full quantum state errors follow an analytic, increasing trend up until just prior to the threshold.
In other words, the Trotter error grows steadily with step size, which matches the intuition suggested by considering the Trotterised (Floquet) Hamiltonian as a convergent Floquet-Magnus expansion, with the zeroth-order solution reproducing the target dynamics and perturbatively increasing correction terms~\cite{HeylM2019QLTE, SiebererLM2019DQC}.
And while the errors (state infidelity) may still become significant in this region, they are in some sense still well behaved:
from a given step size, you can expect to improve performance by reducing the step size, and worsen performance by increasing it.
This provides crucial support to the notion that, prior to the threshold, the Trotterised model implements a meaningful and controllable approximation of the underlying target dynamics.

Beyond the threshold, our new model-agnostic, objective, static $\XsqRMT$ signatures based on eigenvector statistics (Fig.~\ref{Fig:6_chiSquareForAll}) demonstrate that, once the system is large enough, the Trotter step unitary shows quantitatively rigorous agreement with RMT distributions.
This provides robust, statistically rigorous evidence that the Trotterised dynamics become quantum chaotic beyond the threshold.
And since we chose target systems and parameters corresponding to integrable dynamics, we know that it is not the target dynamics ($H_M$ in Eq.~\eqref{Eq:HamiltonianErrors}), but the increasingly dominant Trotter errors that give rise to the observed quantum chaos.
But if the Trotter step unitary becomes a \emph{completely random matrix}, the intuition of Trotterised DQS as approximate simulation of one particular target model breaks down entirely, since its behaviour (across some finite region in step size) is no more consistent with one model than any other (subject to underlying symmetries).
This combination of results shows how we can make a direct connection between the onset of digitisation-induced quantum chaotic behaviour in DQS and the empirical breakdown in meaningful simulation performance.
And while we do not make any rigorous connection between this threshold and the formal radius of convergence in the corresponding Floquet-Magnus expansion, we nevertheless identify the quantum chaotic threshold as an \emph{emergent} radius of convergence for simulation performance.

Here, it is worth noting again the importance of introducing the $\XsqRMT$ goodness-of-fit test statistic as an objective, quantitative measure of likelihood that an individual Trotter step unitary belongs to a given class of random unitary matrices.
As discussed in Sec.~\ref{Sec:QChaosOnset} and illustrated in Fig.~\ref{Fig:5_A2AIsingEigenVecHistograms}, the visual comparisons usually applied to identify distribution-level agreement with RMT predictions are completely inadequate to reliably and accurately identify whether genuine agreement is present.
This work establishes the use of statistically rigorous, full-distribution comparison techniques as a valuable new approach for developing better ways to study random matrices and quantum chaotic dynamics through eigenvector and eigenvalue statistics.
This has important applications across various areas of quantum physics, including quantum computing.
Yet there are many important open questions.
For example, while exploring the ideal (non-Trotterised) Dicke model to find operating parameters giving regular dynamics, we found regimes where good agreement could be identified with eigenvalue but not eigenvector statistics, and also vice versa.
In such cases, which signature, if either, can be considered to be more definitive?
And while using $\XsqRMT$ has already provided significant value and insight for this work, what is the optimal statistical approach to take?

\subsection{On size dependence of the threshold}

In this work, we have explored two different questions about size-dependence of the quantum chaotic threshold: 
How does the position of the threshold depend on system size?
And what size of system is required to see quantum chaotic dynamics?

One of the most consistent outcomes from the comprehensive exploration carried out in this work, across multiple models and numerous different signatures, in addition to the fact that the existence of the threshold is observed consistently across \emph{all} models and signatures, is that its location, again across all signatures, appears to depend very little across orders of magnitude increases in system size, rapidly converging to a fixed, rather sharply defined location.
We see this effect very strongly in time-averages of wide array of physical quantities, including local observables and simulation fidelities and related digitisation errors (Figs~\ref{Fig:4_DynamicalAverages} and~\ref{Fig:7_TrotterErrors}), and various dynamical signatures of quantum chaos and delocalisation like participation ratio, perturbation infidelity and subsystem entropies (Appendix\ Fig.~\ref{AppFig:10_OtherDynamicalSignatures}), as well as statistical RMT properties of global TSUs (Fig.~\ref{Fig:6_chiSquareForAll} and Appendix\ Fig.~\ref{AppFig:20_DickeTruncationDep}).
Yet despite the distinctness and broad applicability of these observations, which are also supported by initial indications from the results of previous work in Refs~\cite{HeylM2019QLTE, SiebererLM2019DQC} which first identified the existence of a possible quantum chaotic threshold, they are perhaps surprising when considered in the context of results from other areas such as Floquet physics.

In the context of DQS, Trotterisation can be interpreted as implementing a periodically driven, time-dependent Hamiltonian, and the digitised dynamics understood in terms of an effective Floquet Hamiltonian~\cite{HeylM2019QLTE, SiebererLM2019DQC}.
As discussed in Sec.~\ref{Sec:Background}, the Floquet-Magnus approximation in the general case breaks down beyond a formal radius of convergence $\tau^\ast$, and the inverse scaling of a rigorous sufficient condition for convergence with system size~\cite{Casas2007MEC, Moan2008CMS, BLANES2009MEA} (if tight) would imply that realising accurate simulations for very large systems could require vanishingly small Trotter step sizes.
This could have significant, fundamental implications for the practicality of future large-scale DQS, but results from recent works~\cite{HeylM2019QLTE, SiebererLM2019DQC, TakashiI2018hit} do not appear to conform to these expectations.
We have extended the work on this issue, in terms of the range of both models and characterisation tools, and have seen minimal if any movement of the threshold, certainly much less than the almost 3 orders of magnitude variation in Hilbert space size studied.
Indeed, the threshold arguably becomes more fixed in position as size increases, with possible movement observed more at low system sizes where the thresholds are less sudden and harder to locate precisely.
These observations seem to contradict the expectation that $\tau^\ast \propto N^{-1}$ for the radius of convergence of the Floquet-Magnus expansion.

Recent asymptotic results from Floquet physics~\cite{Abanin2015, Mori2016, Kuwuhara2016, Abanin2017} suggest that any sufficiently generic, fixed, interacting quantum many-body system, when subjected to periodic driving, will heat up indefinitely, which is a signature of quantum chaotic dynamics.
Thus, in the thermodynamic size limit and asymptotically long times, generic Trotterised DQS may also exhibit quantum-chaos-driven heating, which suggests that a quantum-chaotic breakdown in simulated dynamics may already be observed for step sizes smaller than thresholds identified at short times.
It is not yet understood what system sizes will bring these heating effects into play, but as already noted in Refs~\cite{HeylM2019QLTE, SiebererLM2019DQC}, this heating takes place exponentially slowly in the fast driving regime which usually characterises a DQS operating in the pre-threshold regime.
It therefore seems likely that the universal Trotterisation behaviours observed here will continue to be relevant
for sizes and times applicable to NISQ-era and early fault-tolerant processors, where error mitigation and resource optimisation will be so critical.
And even if systems and run times do eventually reach large enough scales, they can still be exponentially suppressed by increased digitisation.
That is, for specific, long, yet finite simulation times, the Trotterisation threshold can persist even in the thermodynamic limit, because heating will only become effective for large enough $\tau$.

Our work also raises a number of interesting open questions in this area of asymptotic heating and Trotterisation thresholds.
In the context of quantum many-body physics, sufficiently generic systems are, with vanishingly infrequent exceptions, non-integrable (or quantum chaotic).
For integrable systems, however, which occur with measure zero and arise from fine-tuning of system parameters and symmetries, the asymptotic heating results~\cite{Abanin2015, Mori2016, Kuwuhara2016, Abanin2017} do not necessarily hold~\cite{TakashiI2018hit}.
In this work, to focus on digitisation-induced quantum chaos, we specifically considered systems and parameters which remain integrable even for very large system sizes.
Such systems may support a full performance threshold for Trotter errors even in the thermodynamic limit.
This obviously needs to be studied in more detail, but such a result, even if it does not hold completely universally, would provide valuable evidence that DQS can be carried out reliably, even at very large scales, which will be critical, if future large-scale quantum computers are to realise the promising applications of most interest.
Another very important point is that these asymptotic heating results have only been studied in the context of systems with bound energy spectra, like networks of spins or fermions.
It is a completely open question what might happen for systems involving bosonic degrees of freedom, which give rise to very different phenomena under periodic driving, such as parametric resonance.
Systems like the Rabi-Dicke model considered here could provide interesting testbeds for exploring these questions, with potential implications for both fundamental physics and quantum processors utilising bosonic degrees of freedom.

At a more practical level, while the threshold behaviours studied in this work are visually quite clear, an important open challenge for understanding the size dependence of the threshold is to develop a way to quantitatively characterise where the threshold sits.
Our $\XsqRMT$ test statistic provides a promising way to attack this problem, since it provides a direct quantitative measure of statistical agreement with RMT in a single number, but exploring the nuances in how to identify the exact threshold location is left as an important open direction for future work.
For example, while the thresholds observed in $\XsqRMT$ in Fig.~\ref{Fig:6_chiSquareForAll} are relatively sudden, and get sharper with increasing system size, they still appear to span a small finite region in step size, broader than is observed in the full dynamics.
Further study of these effects are required, but we speculate this may arise because $\XsqRMT$ is defined from the global unitary.
That is, it would be consistent with the notion of mixed-phase regions in phase space, well known in the context of the kicked top \cite{SiebererLM2019DQC, NeillC2016EDQS}, for the position of the chaotic threshold to show some dependence on system initial state.
Indeed, it would not be surprising for such an effect to be quite significant, although our initial explorations of this possibility for the Heisenberg model in Appendix\ Fig.~\ref{AppFig:16_HeisenbergOtherQubs} observed a rather stable threshold.
Also, while a goodness-of-fit test statistic value of $\XsqRMT \sim 1$ is known to indicate good agreement with the target probability distribution, values greater than 1 do not uniquely specify the manner of disagreement.
Further study is therefore required to understand whether there is a way to identify the smallest step sizes where the threshold starts.
Finally, we observe that $\XsqRMT$ appears to exhibit some strong sensitivity to step size near the Trotterisation threshold, which may arise from the observed sensitivity of the $\XsqRMT$ test statistic to small changes in the larger eigenvector component counts that are signatures of non-quantum chaotic dynamics, or may be consistent with the known sensitivity of quantum chaotic dynamics to Hamiltonian perturbations.
Often observable in the narrow threshold transition, this effect creates further ambiguity in identifying a single, precise threshold.
Developing a better understanding of the detailed nature of the DQS quantum chaos transition and an objective way to precisely describe the threshold position are important open topics for future DQS research.

The challenge of unambiguously identifying the onset of quantum chaotic dynamics, and whether that is even possible, becomes even more pronounced at small system sizes.
This is a question that demands consideration of quantum chaos as a distinct phenomenon, when it is often interpreted only as a small-system manifestation of asymptotic classical phenomena.
Yet this perspective is crucial for studying DQS, where use cases of interest almost by definition involve systems which cannot be described accurately by generalising to a well-defined classical limit (which does not usually exist).
In this work, interpreting the onset of quantum chaotic dynamics in a model of interest as entering a parameter regime where that model produces unitaries that effectively sample from a universal RMT ensemble, our quantitative $\XsqRMT$ method for characterising how well a unitary's eigenvector statistics agrees with RMT predictions provides a statistically rigorous way to separate identification of quantum chaos from any connection with a classical limit.
More specifically, our results permit the following observations:

Across all models we considered, distinct post-threshold regions where $\XsqRMT\sim1$ were found for all system sizes large enough for $\XsqRMT$ to allow statistically meaningful conclusions to be obtained, and in the same regions, the destruction of apparent visual quasiperiodicity in the dynamics of different physical quantities was also observed.
These behaviours were found already for very modest system sizes across all models.
For the Rabi-Dicke model, our results identified finite step-size regions beyond the threshold for all system sizes where $\XsqRMT\sim1$ (within rigorous expected statistical limits defined by the chi-squared distribution), even down to $j=1/2$ for the Rabi model (e.g., as illustrated in Figure~\ref{Fig:6_chiSquareForAll}).
This agrees with observations across a range of time-averaged dynamical signatures (Figs~\ref{Fig:4_DynamicalAverages} and~\ref{Fig:7_TrotterErrors}), as well as from clear threshold behaviours in full dynamics (shown for the Rabi model in Appendix\ Fig.~\ref{AppFig:17_RabiDynamics}).
For both Ising and Heisenberg models, the same behaviours were found for $j\ge4$ and $N\ge4$, respectively.
At smaller sizes, these models still exhibited recognisably different behaviours in the pre- and post-threshold regions across all signatures, including both dynamical signatures and $\XsqRMT$.
Time averages of common dynamical signatures of quantum chaos such as participation ratio, perturbation infidelity and subsystem entropies do show an observable, though smooth response (see Appendix\ Fig.~\ref{AppFig:10_OtherDynamicalSignatures}), but the lack of a sharp threshold makes interpretation of the slow variations in time-averaged local observables and simulation fidelities inconclusive, and destruction of quasiperiodicity was not visually apparent until the larger system sizes.
For these models and sizes, $\XsqRMT$ was unable to differentiate even the target ($\tau\rightarrow0$) dynamics with confidence from RMT predictions.

These results illustrate numerous challenges involved with identifying quantum chaotic dynamics at small system sizes.
As discussed in Sec.~\ref{Sec:QChaosOnset}, even completely random unitaries will exhibit significant apparent visual quasiperiodicity at the smallest system sizes, due to their small number of discrete eigenfrequencies, making interpretation based on dynamics difficult.
At the same time, the limited number of eigenvector components available for constructing an $\XsqRMT$ goodness-of-fit test statistic prevents even analytically regular target dynamics being distinguished from RMT predictions with any statistical confidence, and it is even more difficult to study level-spacing statistics.
In addition, stable island features in the post-threshold region, if visible, tend to be wider, and further confine the regions where quantum chaotic dynamics could potentially be observed.
In future work, we will explore more sophisticated, ensemble-based statistical analysis methods to try and address some of these challenges.

\subsection{On scaling of Trotter errors with step size}

Ultimately, resource optimisation is the goal that motivates most theoretical DQS research.
A key driver for our work is that a better understanding of the detailed behaviour of Trotter errors in the pre-threshold regime most relevant for useful DQS may allow advanced quantum processors to maximise computational capacity by better balancing Trotter errors against the effects of decoherence and gate noise.
References~\cite{HeylM2019QLTE, SiebererLM2019DQC} show, at least for specific cases in the Ising model, that certain error signatures for observables agree pre-threshold with the values predicted by a perturbative Floquet-Magnus expansion of the system dynamics.
Rather surprisingly, however, the specific signatures studied in these works did not generally show any clear scaling of errors with other system parameters such as size or simulation time, suggesting a rather striking difference in behaviour between errors in local observables and state fidelity signatures calculated from the full quantum state.
In Fig.~\ref{Fig:7_TrotterErrors}, we broaden the observation that observables errors are perturbative functions of step size to a range of different models.
We also show, however, that well-defined, point-to-point errors (Figs~\ref{Fig:7_TrotterErrors}~(g--l)) do indeed scale sensibly with both system size and simulation time, showing they remain controllable across all models up to just prior the quantum chaotic threshold, and are also quite consistent with observed state fidelity scalings.
Indeed, by identifying two distinct behavioural regimes in the pre-threshold dynamics (Fig.~\ref{Fig:3_FullDynamicsAndFT}), we identify a DQS performance region in which the simulation error (global state infidelity) can be expected to also show well-behaved perturbative scaling with Trotter step size, and our Fourier analysis of the full fidelity dynamics highlights an interest avenue for pursuing a better understanding of the underlying cause of these pre-threshold behaviours.
As suggested first in \cite{HeylM2019QLTE}, this means that measuring results for a range of larger step sizes can potentially allow more accurate values to be predicted by extrapolating the observations back to smaller step sizes, which then don't need to be measured explicitly.
Furthermore, provided these errors can be explained perturbatively, calculating the correction terms in a perturbative expansion like the Floquet-Magnus expansion may provide insight into how to \emph{correct} these errors at an algorithmic level, to further compensate and reduce the observed Trotter errors without moving to smaller Trotter step sizes.
Critical to these new observations was the introduction of more nuanced time-averaged errors that distinguished reliably between sampling errors and Trotter errors, and identify all real dynamical errors that do not always show up in infinite-time or different time-averaged error signatures.
An interesting open question that requires further investigation is why, in some models,
the longer-time Trotter errors appear to saturate to a finite error considerably below the value saturated by quantum chaotic dynamics beyond the threshold.

\subsection{Towards experiments and potential applications}

In future work, and especially in the design of future experiments, it will also be important to study the interplay between Trotter errors and noise.
At the simplest level, to be able to experimentally observe the onset of quantum chaotic dynamics at the Trotterisation threshold, the decay and dephasing times will need to be significantly longer than the quantum chaotic collapse time.
For the strongly quantum chaotic systems demonstrated here, this occurs already after very few Trotter steps (Fig.~\ref{Fig:3_FullDynamicsAndFT}), but to distinguish this reliably from quasiperiodic dynamics in the regular (pre-threshold) regime, it will obviously be necessary to observe coherent dynamics over a longer duration.
In this context, decoherence can be modelled straightforwardly using a traditional master equation in the ``lab frame'' of the simulator system, and simple initialisation and measurement errors can be modelled heuristically via appropriate choices of starting states and measurements POVMs.
By contrast, simple gate errors such as calibration errors and parameter drift errors can be modelled using random noise parameters and Monte-Carlo techniques.
The effects of gate errors, however, can be expected to raise more subtle questions for future study, since hardware errors such as gate errors and residual couplings can already lead to quantum chaotic dynamics~\cite{GeorgeotB2000QCQC, GeorgeotB2000QCQC2, SilvestrovECQC, Shepelyansky2001QCQC, BraunD2002QCQA, Benenti2003SEP}, even in analogue quantum systems.
Note that, based on our work in this paper, some experimental results previously observed in Ref.~\cite{LangfordNK2017DQRS} for the quantum Rabi model, can be interpreted as providing preliminary experimental verifications of our new results, indicating that the performance required to investigate these effects is already within reach of current experiments.

\subsection{The broader role of Random Matrix Theory}

In this work, use of random matrix theory has been crucial in establishing quantum chaos as the underlying universal (across models) and global (throughout Hilbert space) cause of the Trotterisation performance breakdown threshold first noticed in the context of Ising DQS in Refs~\cite{HeylM2019QLTE, SiebererLM2019DQC}.
Here, it is worth putting this in the context of
how RMT is also arising in other areas of quantum information.
In recent years, the classical computational complexity of the random matrices underpinning quantum chaos~\cite{LeoneL2021QCQ} has motivated a range of pioneering experiments designed to implement random unitary circuits, like quantum supremacy and information scrambling~\cite{BoixoS2018QSNT, Arute2019QS, WuY2021QASC, ZhuQ2021QCA60, LandsmanKA2019QIS, MiX2021QIS}.
For example, in contrast to our main results, where target models are explicitly integrable and quantum chaotic dynamics arise from simulation errors, quantum supremacy experiments target the accurate simulation of quantum chaotic target dynamics specifically because it pushes the classical complexity.
This provides further motivation for many open questions around how to optimise DQS of a target model which is itself quantum chaotic, a situation which is likely to arise often in real DQS applications.
In Appendix~\ref{App:DickeChaos2Chaos}, we note already that Trotterisation thresholds may still be observable for quantum chaotic target models, but quantum supremacy experiments illustrate that this context merits much more detailed investigation.
For example, does the universality of Trotterisation performance across system types observed here extend to Trotterisations of quantum chaotic target models?
Do similar thresholds appear in other contexts where hardware-error-induced quantum chaos appears~\cite{GeorgeotB2000QCQC, GeorgeotB2000QCQC2, SilvestrovECQC, Shepelyansky2001QCQC, BraunD2002QCQA, Benenti2003SEP}, and (how) are these situations connected?
We anticipate that the new analytical tools we have introduced, like the goodness-of-fit test statistic $\XsqRMT$, will provide invaluable insight into such questions.

In a different direction, while we can connect pre-threshold Trotter errors to perturbative corrections in the Floquet-Magnus expansion~\cite{HeylM2019QLTE, SiebererLM2019DQC} and identify the threshold as an emergent radius of convergence, it remains an open question to find some mechanistic origin for the threshold and the onset of quantum chaotic dynamics beyond it.
Given the key role of RMT in establishing the classical computational complexity of quantum supremacy experiments~\cite{BoixoS2018QSNT, LeoneL2021QCQ}, we suggest that RMT could perhaps be used to build an information theoretic understanding of the breakdown of DQS performance in terms of computational complexity.

Finally, in this work, we show that digitisation-induced quantum chaotic dynamics can be realised in quite modestly sized systems.
While this heralds a breakdown in performance in DQS, it is intriguing to speculate whether this can be used as a resource in other contexts, such as open-system simulations (using the deep connection between quantum chaos and thermalisation in closed quantum systems~\cite{DeutschJM1990QSM, SrednickiM1994CQT, LucaD2016ETH}), or quantum metrology~\cite{FidererLJ2018QMQC}.



\section{Acknowledgments}


Authors thanks G. J. Milburn, R. L. Mann, M. J. Bremner, T. Srivipat, K. Modi, D. Braak, J. F. Ralph, and D.~W.~Berry for useful discussions.
Numerical simulations are realised using a Python library, called QuanGuru~\cite{QuanGuru}, developed in our group at UTS.
The data and the Python code that support the findings of this study are available from the corresponding authors upon reasonable request.

This work has been funded by the Australian Research Council Future Fellowship of N.K.L.\ (FT170100399) and Discovery Project (DP210101367).
C.K., A.M.\ and F.H.\ acknowledge the supports from UTS President's scholarship, international research scholarship, and Sydney Quantum Academy.
J.P.D.\ acknowledges support from the University of Technology Sydney, Chancellor's Postdoctoral Research Fellowship (UTS, CPRDF).
L.M.S.\ acknowledges support from the Austrian Science Fund (FWF): P 33741-N.
P.H.\ acknowledges support from Provincia Autonoma di Trento, the ERC Starting Grant StrEnQTh (project ID 804305), the Google Research Scholar Award ProGauge. This project has benefited from Q@TN, the joint lab between University of Trento, FBK-Fondazione Bruno Kessler, INFN-National Institute for Nuclear Physics and CNR- National Research Council.
M.H.\ and P.Z.\ both acknowledge funding from the European Research Council (ERC) under the European Union's Horizon 2020 research and innovation programme, under grant agreements No.~853443 and No.~817482 (Pasquans), respectively. 
M.H.\ also acknowledges support by the Deutsche Forschungsgemeinschaft (DFG) via the Gottfried Wilhelm Leibniz Prize program.
P.Z.\ also acknowledges support by the Simons Collaboration on Ultra-Quantum Matter, which is a grant from the Simons Foundation (651440, P.Z.).

\clearpage


\appendix



\section{Random Matrix Theory and Quantum Chaos}\label{App:RMTandQuantumChaos}


Dynamical signatures of quantum chaos are manifestations of certain random matrix properties in the eigenvectors of quantum chaotic systems.
Here, we provide an overview of quantum chaos and its relation with random matrix theory (RMT).
We also discuss the technical subtleties and nuances involved in the calculation of the quantum chaos signatures used in this paper, and we use quantum kicked-top models of different symmetries to explain these for both statistical and dynamical signatures.

\subsection{Random Matrix}

A random matrix has all or most of its entries as independent random numbers satisfying certain probability laws, and random matrix theory is concerned with the statistical properties of (all or a few of) the eigenvalues and eigenvectors of such matrices.
Wigner used RMT analyses to explain certain spectral properties of heavy nuclei~\cite{WignerEP1955CVID}, motivating much subsequent theoretical work, and Dyson showed that physically relevant random matrices fall (mainly) into one of three universality classes according to their symmetries~\cite{DysonFJ1962RMT1}. 
These classes are categorised into circular or Gaussian ensembles for unitary and Hermitian random matrices, respectively. 
Ensembles are categorised according to the Dyson parameter $\beta$ into orthogonal (COE/GOE, $\beta = 1$), unitary (CUE/GUE, $\beta = 2$) and symplectic (CSE/GSE, $\beta = 4$), related to degrees of freedom in random entries and degeneracies of the matrices~\cite{KusM1988UEV}.

\subsection{Quantum Spectra of Classically Chaotic Systems}\label{App:eigenValStat}

Over the years, many studies connected RMT with quantum chaos~\cite{BerryMV1977LCRS, BohigasO1984QCS, IZRAILEVFM1987CSEF, KusM1988UEV, LucaD2016ETH, HaakeF2018QSC}, by showing that the distribution of energy levels in the Hamiltonians of classically chaotic systems, commonly referred to as the eigenvalue or level (spacing) statistics, adhere to Wigner-Dyson statistics~\cite{BohigasO1984QCS, LucaD2016ETH, HaakeF2018QSC}. 
For $\mathcal{D} = 2$, these can be calculated for different random matrix ensembles to be:
\begin{equation}\label{AppEq:WignerDyson}
	P(s) = A_{\beta} s^{\beta} e^{-B_{\beta} s^{2}}
\end{equation}
where s is the spacing between neighbouring levels, and the coefficients
\begin{equation}
	A_{\beta} = \begin{cases}
		\pi/2 & \beta = 1 \\
		32/\pi^{2} & \beta = 2 \\
		2^{18}/3^{6}\pi^{3} & \beta = 4
	  \end{cases}
\end{equation}
and
\begin{equation}
	B_{\beta} = \begin{cases}
		\pi/4 & \beta = 1 \\
		4/\pi & \beta = 2 \\
		64/9\pi & \beta = 4
	  \end{cases}
\end{equation}
are obtained by normalising $P(s)$ while keeping the mean level spacing fixed to 1~\cite{HaakeF2018QSC}.
For other Hilbert space dimensions $\mathcal{D}\neq2$, the Wigner-Dyson distributions do not have closed forms, but are generally understood to agree reasonably closely with the $\mathcal{D} = 2$ case at the level of \emph{local} level-spacing fluctuations (see Ref.~\cite{LucaD2016ETH, sandro2016nonlinear} for further details).
In Appendix~\ref{App:kickedTopStatic}, we discuss the nuances involved with calculating the local level spacing statistics from the full spectrum, and illustrate them with the quantum kicked top.

It is conjectured by Bohigas-Giannoni-Schmit (BGS conjecture)~\cite{BohigasO1984QCS} that, even for single particle quantum systems that are chaotic in the classical limit, level spacings exhibit Wigner-Dyson statistics.
For quantum systems with an integrable classical limit, Berry-Tabor conjectured that the level spacings exhibit Poisson statistics~\cite{BerryMV1977LCRS}.
There are a few counterexamples to both of these conjectures, including: (i) a classically chaotic system without Wigner-Dyson statistics in its quantised spectrum~\cite{BogomolnyEB1992CBAG}, (ii) a 1D integrable system with Wigner-Dyson statistics~\cite{WuHua1990GOES}, and (iii) an integrable system without Poisson statistics~\cite{Casati1985EVI}.
Additionally, these statistics can be tricky to interpret in a context where the natural classical limit of a quantum system is not clear.
Therefore, the existence of Wigner-Dyson statistics in level spacings is neither a necessary nor a sufficient condition of quantum chaotic dynamics, but is still a strong signature of quantum chaos.

\subsection{Eigenvector Statistics}\label{App:EigenVecTheory}

Despite the popularity of using level (eigenvalue) statistics as a signature for quantum chaos, the statistics of the dynamics' eigenvectors offer important advantages for identifying the RMT behaviour of quantum chaotic dynamics.
For example, contrary to the Wigner-Dyson distributions (e.g., Eq.~\eqref{AppEq:WignerDyson}), the statistical distributions for the eigenvector components of a random matrix (see Eqs~(\ref{AppEq:RMTdist:COE}--\ref{AppEq:RMTdist:CSE}) below) have a rigorously defined, closed-form solution for \emph{arbitrary} dimension $\mathcal{D}$~\cite{IZRAILEVFM1987CSEF, KusM1988UEV, HaakeF1990RMEF}.

Eigenvectors $\{\ket{\psi_i}_{U}\}$ of a random unitary matrix $U$ are also random (column) matrices, and once they are expanded in a generic-basis $\{\ket{k}\}$, as $\ket{\psi_i}_{U} = \sum_{k=1}^{\mathcal{D}}c_{ik}\ket{k}$ with the constraint $\sum_{k=1}^{\mathcal{D}}|c_{ik}|^2 = 1$, the squared moduli (component populations/probabilities) $\eta = |c_{ik}|^2$ obey a certain distribution depending on the universality class~\cite{IZRAILEVFM1987CSEF,KusM1988UEV,HaakeF1990RMEF}. As stated in the main text, the reduced probability densities corresponding to these classes are given as
\begin{eqnarray}
	\label{AppEq:RMTdist:COE}
	\widetilde{P}_{\text{COE/GOE}} (\eta) &=& \frac{\Gamma \left(\frac{\mathcal{D}}{2}\right)}{\Gamma 
	\left(\frac{\mathcal{D}-1}{2}\right)}\frac{(1-\eta)^{(\mathcal{D}-3)/2}}{\sqrt{\pi \eta}} \\
	\label{AppEq:RMTdist:CUE}
	\widetilde{P}_{\text{CUE/GUE}}(\eta) &=& (\mathcal{D} - 1)(1 - \eta)^{\mathcal{D}-2}\\
	\label{AppEq:RMTdist:CSE}
	\widetilde{P}_{\text{CSE/GSE}}(\eta) &=& (\mathcal{D} - 1)(\mathcal{D} - 2)\eta(1 - \eta)^{\mathcal{D} - 3},
\end{eqnarray}
where $\mathcal{D}$ is the dimension of the Hilbert space, and $\Gamma(x)$ (the Gamma function) is the generalized factorial.
In the figures of this paper, we denote these (and Wigner-Dyson) distributions by the appropriate three letter acronym, depending on whether the unitary or the Hamiltonian is used to calculate the eigenvectors (and eigen-phases/values).

The quantitative eigenvector statistics analysis in this work compares the eigenvector statistics of each Trotter step unitary against the universal RMT distributions from Eqs~(\ref{AppEq:RMTdist:COE}--\ref{AppEq:RMTdist:CSE}) using the following recipe:
\begin{enumerate}[parsep=0mm,itemsep=1mm,leftmargin=5mm]
\item Choose an appropriate reference basis $\{\ket{k}\}$ in which to write the unitary $U$ (discussed below).
\item Calculate the eigenvectors $\ket{\psi_i}_{U} = \sum_{k=1}^{\mathcal{D}}c_{ik}\ket{k}$ of $U$ and, from these, follow established procedures to calculate a set of component populations $\eta = |c_{ik}|^2$ \cite{HaakeF2018QSC,IZRAILEVFM1987CSEF,KusM1988UEV}.
Depending on the unitary's time-reversal symmetry or other physical symmetries, additional pre-processing is required to define the appropriate set of component populations $\{\eta\}$ from the set of raw eigenvector components $\{c_{ik}\}$ (discussed further in Appendix Section~\ref{App:kickedTopStatic}).
\item Using the target distributions from Eqs~(\ref{AppEq:RMTdist:COE}--\ref{AppEq:RMTdist:CSE}), calculate a set of bin edges with which to create a histogram of component populations.
The procedure in Appendix Section \ref{App:chiSquaredTestDetail} is used here to ensure that quantitative histogram comparisons are sufficiently well behaved (e.g., by ensuring that most bins, on average, contain equal and sufficient counts).
\item Using these bin edges, create a histogram for the unitary's eigenvector component populations $\{\eta\}$.
\item Compare, using a chi-squared goodness-of-fit test, with the corresponding target histogram (calculated using the same bin edges).
\end{enumerate}

\subsubsection{Choosing the reference basis}\label{App:refBasis}

The question of how to choose the reference basis is important and a little subtle.
It turns out that using an arbitrary basis will almost always lead to eigenvector statistics satisfying the probability distribution of one of the randomness universality classes:
Even a non-random basis will give random eigenvector statistics when expanded in a random basis, and an arbitrary basis will almost always be random.
So the reference basis needs to be chosen in such a way that the eigenvector statistics satisfy one of the universality classes \emph{if and only if} the underlying unitary matrix derives from non-integrable/quantum chaotic dynamics (i.e., is a random matrix), and this means that the reference basis must itself not be random.
The easiest way to ensure this is to use the eigenbasis of an integrable system unitary as the reference basis:
the eigenvector statistics of an integrable system will not satisfy RMT distributions when expanded in the eigenbasis of another integrable system, whereas the eigenvector statistics of a non-integrable system, using the same reference basis, will.
Specifically, the free, uncoupled evolution of a quantum system (without any non-interacting terms) is integrable by construction.
The eigenvectors of the uncoupled system can therefore reliably be used to define a non-random reference basis, and this is the approach taken in this work. 

It must be acknowledged that there is some risk of falling into a circular argument here, but with care, it can be avoided.
One way to ensure that the reference basis derives from an integrable system is to test the robustness/sensitivity of eigenvectors to small perturbations~\cite{LucaD2016ETH, FeingoldM1986MEC,HaakeF1990RMEF}.
Under small perturbations, the eigenvectors of a regular quantum system only negligibly differ from the unperturbed basis, while the eigenvectors of perturbed quantum chaotic systems become random matrices in the unperturbed basis~\cite{LucaD2016ETH, FeingoldM1986MEC}.
Note that the dynamical distinctions between regular and quantum chaotic systems also derive from this robustness (sensitivity) of the eigenvectors of integrable (quantum chaotic) systems to small perturbations~\cite{HaakeF2018QSC}.
Therefore, eigenvector statistics are used in this paper as the defining property of quantum chaotic dynamics.

Particular care may need to be taken with the choice of reference basis if the eigenvector statistics are observed to follow CUE statistics.
Due to the time-reversal symmetry of COE unitaries, the derivation of the COE eigenvector statistics in Eq.~\ref{AppEq:RMTdist:COE} starts by assuming a symmetric unitary with real eigenvectors~\cite{HaakeF2018QSC}.
Such unitaries can still exhibit CUE eigenvector statistics, however, when calculated in a basis where the eigenvectors are complex.
When the eigenvector statistics of a unitary are observed to be CUE, it is therefore then important to calculate its \emph{eigenvalue} statistics to verify that they also follow the corresponding Wigner-Dyson statistics for CUE matrices.
If the eigenvalue statistics are instead found to follow COE Wigner-Dyson statistics, the correct COE \emph{eigenvector} statistics can be recovered by identifying an additional, non-random basis transformation that leaves the unitary symmetric, giving real eigenvectors, as assumed in the derivation of Eq.~\ref{AppEq:RMTdist:COE}.
Such a symmetric basis and basis transformation are easily constructed if the unitary's (or Hamiltonian's) time-reversal symmetry operator can be identified~\cite{HaakeF2018QSC}.
The Trotterised Heisenberg and Dicke models discussed in the next sections provide examples to illustrate this process.
Note that, if it is only necessary to identify the presence or onset of quantum chaos in general, without needing to know what precise class of RMT dynamics is observed, these additional steps are generally not required.

\subsubsection{Chi-squared goodness-of-fit test for agreement with RMT}\label{App:chiSquaredTestDetail}

As illustrated in Fig.~\ref{Fig:5_A2AIsingEigenVecHistograms}, comparing eigenvector histograms with the expected statistical distributions visually can be very misleading.
To make this comparison both quantitative and objective, we therefore characterise the agreement of a given histogram with RMT predictions using a reduced chi-squared $\chi_{\nu}^{2}$ goodness-of-fit test.
Specifically, we define a reduced RMT goodness-of-fit test statistic, following the usual definition~\cite{Mood}, according to:
\begin{equation}
	\XsqRMT = \frac{1}{M - 1}\sum_{i=1}^{M}\frac{(n_i^{\rm o} - n_{a,i}^{\rm e})^2}{n_{a,i}^{\rm e}},
\end{equation}
where $a\in\{{\rm COE, CUE, CSE}\}$ is chosen according to the time-reversal symmetry satisfied by the system dynamics under study (in our case, the Trotter step unitary).
Here, $M$ is the number of bins used to create the eigenvector statistics histogram from the appropriate set of unitary eigenvector component populations $\{\eta\}$, $n_i^{\rm o}$ is the observed number of counts in the $i^{\rm th}$ bin, and $n_{a,i}^{\rm e}$ is the expected number of counts in the $i^{\rm th}$ bin calculated from the corresponding universal RMT distributions from Eqs~(\ref{AppEq:RMTdist:COE}--\ref{AppEq:RMTdist:CSE}).
If the eigenvector statistics agree with the appropriate universal RMT probability distribution, the residuals in the $\XsqRMT$ test statistic will reduce to independent random variables, and the value for $\XsqRMT$ should be distributed according to a chi-squared probability distribution.
If needed, this comparison can be formulated as a hypothesis test to determine whether to reject the \emph{null hypothesis} that the eigenvector statistics do agree with RMT predictions.

It may not always be known in advance how to choose the appropriate RMT ensemble to check for quantum chaos, since time-reversal symmetries are not always obvious for complex system dynamics.
Often, however, there is an obvious transition in the value of $\XsqRMT$ between regular and quantum chaotic regimes, regardless of which RMT ensemble is used to calculate it.
Once a quantum chaotic region is identified, an initial guess can be made by visually comparing the histogram with the different distributions from Eqs~(\ref{AppEq:RMTdist:COE}--\ref{AppEq:RMTdist:CSE}), or alternatively by calculating $\XsqRMT$ for each of the main ensembles and seeing which gives a value closer to 1 (noting that COE dynamics may appear to be CUE when analysed in a non-symmetric basis, as discussed above).
There may also be situations where it is useful to track the $X^2$ test statistic for more than one distribution, for example if the observed universality class transitions from one to another~\cite{RegnaultN2016fts}.
Generally, we will simply leave this choice implicit and write $\XsqRMT$, but occasionally, we may write explicitly, e.g., $X^2_{\rm COE}$.

Another detail which can prevent $\XsqRMT$ from reaching 1 in a quantum chaotic region is the presence of other symmetries in the quantum dynamics, such as parity or rotational symmetries.
If the computational basis states are symmetry eigenstates, then symmetry subspaces can be easily identified, because the unitary will already be block-diagonal in the computational basis.
In such cases, comparisons with RMT should only analyse the statistics of eigenvector components within these eigenspaces (that is, disregarding the blocks of zeros).
If the symmetry eigenbasis is rotated away from the computational basis, however, this symmetry may not be obvious from looking at the unitary, or even at the Hamiltonians generating the evolution.
When an unidentified symmetry exists, the additional structure will not be accounted for, and $\XsqRMT$ will not reach the value of 1 that corresponds to full agreement with RMT.
Analysing $\XsqRMT$, and finding it reaches a limiting value that is greater than 1, can therefore even assist in flagging the presence of an unidentified symmetry.

\paragraph{Choosing bins in $\eta$ to define the histogram} 
To calculate $\XsqRMT$ values reliably and accurately, bin edges should be chosen carefully, following good statistical practices.
When histograms are plotted for the purposes of visual comparison (e.g., see Fig.~\ref{AppFig:8_KickedTopEigAndChi}), it is common to choose bins that have equal widths in $\eta$.
However, because the RMT distributions from Eqs~(\ref{AppEq:RMTdist:COE}--\ref{AppEq:RMTdist:CSE}) decay rapidly to zero as $\eta \rightarrow 1$ (exponential in the Hilbert space dimension $\mathcal{D}$), creating the eigenvector component histogram using \emph{equal-width} bins gives rise to expected counts ($n_{a,i}^{\rm e}$) that also decay rapidly. 
Because the contribution to $\XsqRMT$ from each individual bin scales as $1/n_{a,i}^{\rm e}$, even individual instances of very low counts in bins where $\eta\gg1/\mathcal{D}$ (indicating the observation of eigenvectors which are substantially localised in the computational basis, as is expected for non-quantum chaotic unitaries) can give rise to terms, exponentially large in the system dimension, that contribute so dominantly to $\XsqRMT$ that $\XsqRMT$ becomes completely insensitive to the degree of agreement in the domain where there is actual support for the target RMT distribution.
An approach which provides a more robust characterisation of the overall distribution is to use equal-\emph{probability} bins with widths chosen to ensure expected counts meet good statistical heuristics (typically $\gtrsim 10$ per bin).
This is the approach we take, with one adaptation:

Our aim is to use $\XsqRMT$ to differentiate, at the full distribution level, between regular and quantum chaotic dynamics, but we also want to retain enough sensitivity to detect the onset of stable islands, the localised regions of regularity observed in dynamical performance signatures (e.g., see Fig.~\ref{Fig:3_FullDynamicsAndFT}).
To do this, our test statistic needs to be sensitive to low numbers of individually localised components in the long, vanishingly small tail of the RMT distributions where $\eta \rightarrow 1$, and we realise this by capturing this long tail in a last, low-probability bin.
The probability of detecting a count in this last ``tail'' bin given genuine RMT statistics can then be tuned to ensure that counts in this tail bin produce a $\XsqRMT$ contribution which is detectable, but not dominant over the contribution to $\XsqRMT$ made by the main ``body'' of the histogram (all bins except the last).
We describe below how we choose the tail bin probability $p_{a,{\rm tail}}^{\rm e}$ in this work, but note that most appropriate method may be different for different situations.
With that caveat, we developed the following constructive procedure to choose bin edges in a manner which only depends on the choice of RMT ensemble, the dimension of the unitary and the choice of tail bin probability:
\begin{itemize}
    \item Choose the number of equal-probability bins in the main body of the histogram, $M-1$, such that the expected number of counts per bin $\mathcal{D}^2/(M{-}1)\sim10$.  
    For small dimensions, we do not find that fewer than 5 bins provides a meaningful comparison.
    \item[] (For a few exceptional cases with fewer total counts, we show $\XsqRMT$ results with $\sim 5$ counts per bin (still $\gtrsim 5$ bins) for qualitative comparison purposes only, but we do not find those results allow very statistically significant conclusions. We did not consider systems with fewer than 25 counts.)
    \item For the chosen RMT ensemble, calculate $\eta_{\rm tail}$ to set the tail bin ($\eta_{\rm tail}{<}\eta{\le} 1$) expected counts to a value that ensures a) that the probability of quantum chaotic unitaries producing a count in the tail bin is very small ($n_{a,{\rm tail}}^{\rm e} \ll 1$), and b) that tail counts arising from non-quantum chaotic dynamics do not cause the tail contribution to $X_{RMT}^2$ to swamp the accompanying contribution from the main body of the histogram.  
    \item Calculate the remaining bin edges ($\eta_0, \ldots, \eta_{M-1}$) to divide the remaining histogram probability ($1-p_{a,{\rm tail}}^{\rm e}$) into $M-1$ equal-probability bins.
\end{itemize}

\paragraph{Identifying agreement with RMT when there is a tail bin}
We specifically choose the tail bin probability $p_{a,{\rm tail}}^{\rm e}$ such that the expected counts $n_{a,{\rm tail}}^{\rm e} = p_{a,{\rm tail}}^{\rm e} \mathcal{D}^2 \ll 1$, and the chance that a unitary sampled from the chosen target distribution would produce a tail bin count through statistical variation is vanishingly small.
For quantum chaotic systems, we therefore expect the observed number of tail bin counts $n_{\rm tail}^{\rm o}=0$, and the $\XsqRMT$ contribution from the tail bin to be $X^2_{\rm tail}=n_{a,{\rm tail}}^{\rm e}/(M{-}1) \approx 0$.
And since $n_{a,{\rm tail}}^{\rm e}\ll1$, the addition of a tail bin does not change the expected counts per bin predicted for the main body of the distribution ($n_{a,{\rm body}}^{\rm e}=\mathcal{D}^2/(M{-}1)$), and we can therefore still expect that unitaries sampled from the chosen target distribution will give $\XsqRMT \sim 1$.

\paragraph{Size dependence of $\XsqRMT$ for regular dynamics}
To understand how $\XsqRMT$ might be expected to depend on system size when the system is not completely quantum chaotic, consider a simple toy model unitary $U$ which exhibits mostly quantum chaotic dynamics with a small regular region of Hilbert space: that is, with an eigenvector matrix $V=I^{(k\mathcal{D})} \oplus V^{(1-k)\mathcal{D}}_{\rm RMT}$ (with $k\ll1$ and $k\mathcal{D}\in \mathbb{N}$), where $V^{(1-k)\mathcal{D}}_{\rm RMT}$ is the eigenvector matrix for a quantum chaotic unitary $U^{(1-k)\mathcal{D}}_{\rm RMT}$.
This gives rise to a small number $k\mathcal{D}$ of eigenvector components that are much larger than would be predicted by RMT statistics ($\eta\gg1/\mathcal{D}$) and a corresponding increase (by $k\mathcal{D}(\mathcal{D}{-}1)$ counts) in the number of very small eigenvector components (here $\eta=0$).
Breaking the body of the histogram into $M-1$ equal-probablity bins, we can calculate the expected $\XsqRMT$ contributions to be $X^2_{\rm body} \approx k^2\mathcal{D}^2(M{-}2)/(M{-}1)=k^2(M{-}2)n_{a,{\rm body}}^{\rm e}$, and $X^2_{\rm tail}=k^2 \mathcal{D}^2/(M{-}1)n_{a,{\rm tail}}^{\rm e}=k^2 n_{a,{\rm body}}^{\rm e}/n_{a,{\rm tail}}^{\rm e}$.
Thus, we would expect the tail contribution to $\XsqRMT$ to be roughly equal to the $\XsqRMT$ contribution from the body, if we choose $n_{a,{\rm tail}}^{\rm e}\sim1/(M{-}2)$.
We see a similar outcome for a completely trivial $U$ with $V=I^{(\mathcal{D})}$, where we have instead $X^2_{\rm body} = (M{-}2)n_{a,{\rm body}}^{\rm e}$ and $X^2_{\rm tail}=n_{a,{\rm body}}^{\rm e}/n_{a,{\rm tail}}^{\rm e}$.
For less trivial examples of non-quantum chaotic unitaries, we anticipate $\XsqRMT$ will scale broadly similarly with system size, but have observed that the prefactors are highly model dependent, and can lead to much larger tail contributions than what is predicted for simplistic models.
For this work, we therefore use a more empirical, explicitly model-dependent approach to choose the tail bin probability.

\paragraph{Choosing the tail bin probability}
Since the domains of $\eta$ where the RMT probability distributions have non-negligible support all vary dramatically with the effective system dimension $\mathcal{D}$ (after accounting for symmetries and choice of RMT ensemble), the value for $\eta_{\rm tail}$ must be chosen separately for each system size.
On the other hand, our choice of $\eta_{\rm tail}$ should \emph{not} depend on Trotter step size, nor, ideally, in any direct way on the unitaries being analysed.
Analysis of $\XsqRMT$ for our chosen models with pure equal-probability bins, shows they all give $\XsqRMT\gg1$ before the threshold, and $\XsqRMT\sim1$ after the threshold, but except for the Heisenberg model (with the highly symmetric parameters we chose), were almost completely insensitive to the stable regions which appear so clearly in the dynamical signatures we studied (e.g., Fig.~\ref{Fig:3_FullDynamicsAndFT} (a--c)).
As we discuss further in Appendix~\ref{App:IsingStableRegions}, this is because the regular dynamics in the less highly symmetric models emerges only in more localised regions of phase space, and the modest effect this produces on the counts in the last equal-probability bin makes only a tiny contribution to $\XsqRMT$ overall.
It is to resolve precisely this discrepancy, that we include the last, low-probability tail bin to make our $\XsqRMT$ directly sensitive to the relatively small numbers of the larger eigenvector components which are a defining feature of regions of regular dynamics.

The simple toy calculations above show that the tail contribution $X^2_{\rm tail}$ can be effectively scaled in size by orders of magnitude by tuning the value of $n_{a,{\rm tail}}^{\rm e}$, as determined by $\eta_{\rm tail}$. 
To choose $\eta_{\rm tail}$ in a systematic way which is model specific, allows fair and meaningful comparison of trends across system sizes, and independent of the Trotterisation data we are analysing, including the step size, we do so by analysing the eigenvectors of the underlying Trotter-error-free full Hamiltonian.
Specifically, we numerically search for the value of $\eta_{\rm tail}$ which sets $X^2_{\rm tail}=X^2_{\rm body}$ for the eigenvector statistics of the ideal full Hamiltonian for each model and system size.
This ensures that the large pre-threshold values of $\XsqRMT\gg1$ depend strongly on the overall RMT (dis)agreement across the full distribution, without being dominated by $X^2_{\rm tail}$, while also fixing across varying system size the relative sensitivity of the $\XsqRMT$ signature to regions of localised dynamics in proportion to the total degree of localisation in the ideal underlying model.
This strikes a balance that maintains sensitivity to regions of dynamical stability without obscuring the full-distribution RMT statistics.

\emph{Note on analysing the target Hamiltonian for the Heisenberg model:}
Because of the highly symmetric parameters chosen for the Heisenberg model we have studied in this work, the $\tau=0$ target model exhibits an additional spatial inversion symmetry which is not present in the Trotterised model.
This additional symmetry makes the resulting dimensionality of the RMT statistical analysis different from that of the Trotterised model, which changes the anticipated scalings of $X^2_{\rm tail}$ and $X^2_{\rm body}$.
To avoid this parameter-specific quirk, we instead set the relative scale between the tail and body contributions based on a Trotterised model with a very small step size.
To ensure we still choose a scale that is independent of the specific Trotterisation trends we are studying, we set $\tau=10^{-8}$, which is small enough to provide a good representation of the asymptotic small $\tau$ limit, but large enough that the results are not strongly impacted by numerical imprecision.

\subsection{Quantum Kicked Top}\label{App:KickedTop}


\begin{figure*}[ht]
	\centering
	\subfloat{\includegraphics[scale=0.925]{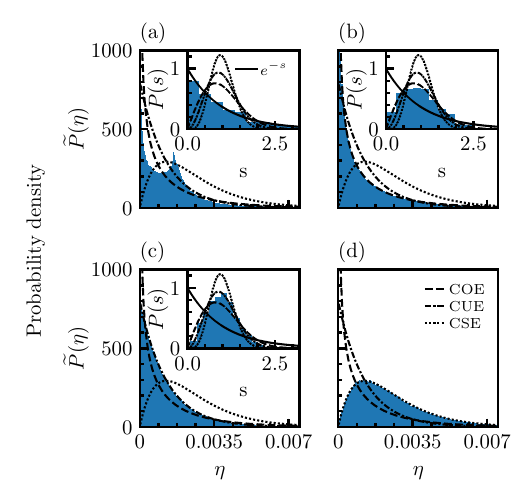}}
	\subfloat{\includegraphics[scale=0.925]{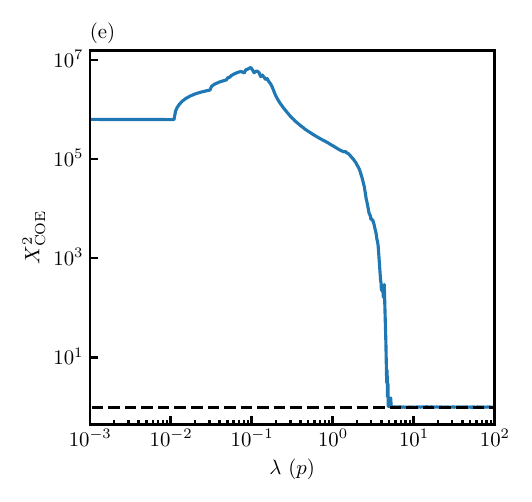}}
	\caption{Eigenvector and eigenvalue (insets) statistics for quantum kicked top with (a) regular parameters (i.e. weak kick) and COE, (b) chaotic (i.e. strong kick) and COE, (c) chaotic and CUE, and (d) chaotic and CSE, and (e) the $\XsqRMT$ test statistic for COE eigenvector statistics as a function of kick strength $\lambda$.
	The dashed and dotted lines in (a)--(d) show RMT statistics for COE, CUE and CSE universality classes, and the solid line shows Poissonian statistics.
	$\XsqRMT$ shows three main regions: (i) regular dynamics $\lambda \lesssim 1$, (ii) $1 \lesssim \lambda \lesssim 7$ transition region with mixed-phase space and stable regions, and (iii) $\lambda \gtrsim 7$ quantum chaotic system.
	$j = 399.5$ for COE, CUE, and CSE, and the other parameters are given in Appendix~\ref{App:KickedTop}.
	The dashed line in (e) indicates $\XsqRMT=1$.}
	\label{AppFig:8_KickedTopEigAndChi} 
\end{figure*}


The kicked top is one of the best-studied models, both in classical and quantum chaos, and is extensively used to demonstrate the connection between the two~\cite{HaakeF1987CQCKT}.
Here, we use quantum kicked tops of different symmetries to provide concrete examples of both the dynamical signatures of quantum chaos and the RMT properties discussed above (Fig.~\ref{AppFig:8_KickedTopEigAndChi}).
Quantum kicked-kicked tops of COE and CSE symmetries are given by the Floquet operator:
\begin{equation}\label{AppEq:kickedtopUnitary}
	F = e^{-iV}e^{-iH_{0}},
\end{equation}
with 
\begin{gather}
	H_{0} = pJ_{y} \\
	V = \lambda J_{z}^{2}/2j,
\end{gather}
for COE, and 
\begin{gather}
	H_{0} = pJ_{z}^{2}/j \\
	V = k\left[ J_{z}^{2} + k^{\prime}\left( J_{x}J_{z} + J_{z}J_{x} \right) + k^{\prime\prime}\left( J_{x}J_{y} + 
	J_{y}J_{x} \right) \right]/j,
\end{gather}
for CSE.
The kicked top with CUE symmetry is obtained by the following Floquet operator:
\begin{equation}
	F = \exp(-ik^{\prime}J_{x}^{2}/2j)\exp(-ikJ_{z}^{2}/2j)\exp(-ipJ_{y})
\end{equation}
For further discussion on standard quantum kicked-top models, see Refs~\cite{HaakeF1987CQCKT, KusM1988UEV}.
The parameters used for the numerical simulation below are (all with $j = 399.5$): $p = 1$, and $\lambda = 0.01$ (regular) \& $\lambda=10$ (chaotic) for COE; $p=1.7$, $k=6$ and $k^{\prime}=0.5$ for CUE; and $p=4.5$, $k=1.5$, $k^{\prime} = 2$, and $k^{\prime\prime} = 3$ for CSE.
In the dynamical simulations, we use the COE symmetry with the initial state $\ket{j, m = j-1}$, and the perturbation fidelity decay signature is obtained by comparing the quantum kicked tops with kick strengths of $\lambda$ and $1.001\lambda$.

\subsubsection{Static Signatures}\label{App:kickedTopStatic}

Here, we discuss the procedures and nuances involved in calculating the level-spacing and the eigenvector statistics, illustrating them with quantum kicked-top examples.

\paragraph{Level-spacing statistics}
In this paper, for the quantum kicked-top models here and the DQS models in the main text, we mostly calculate the level spacing statistics using the eigenphases $\{\phi_{U}\}$ of a unitary evolution operator $U(t)$, which are unique up to a $\{\phi_{U}\pm 2n\pi\}$ difference for any integer $n$.
This means that, for a given unitary operator, there are infinitely many corresponding Hermitian operators with the same eigenvectors but different $\{\phi_{U}\pm 2n\pi\}$ eigenvalue spectrums.
This is an important limitation for level-spacing statistics.
Moreover, regardless of the non-uniqueness, an unfolding procedure is required to remove system-dependent secular variations and make the level spacings of any eigen-phase/value spectrum comparable with Wigner-Dyson distributions. 
Full details on level spacing statistics and unfolding are already found in many references~\cite{HaakeF2018QSC, LucaD2016ETH, sirca2012computational, sandro2016nonlinear}, but the main calculation steps are:
\begin{itemize}
	\item Sort the eigen-values/phases and group them into symmetry-reduced sub-spaces.
	\item Unfolding: renormalise the local density of states by a smooth function to have both global and local unit-average level spacings.
	\item Calculate the level spacings for each symmetry-reduced sub-spaces and create histograms with these spacings.
	\item Normalise the area under the histogram to 1 and compare it with Wigner-Dyson distributions.
\end{itemize}
We can also superimpose the statistics of all these sub-spaces to have better resolution in the histogram plots~\cite{HaakeF1987CQCKT}.

Insets in Fig.~\ref{AppFig:8_KickedTopEigAndChi} show the level spacing statistics for (a) COE with regular, (b) COE with quantum chaotic, and (c) CUE with quantum chaotic parameters.
For regular parameters, it closely follows the Poisson statistics, and chaotic cases follow the corresponding Wigner-Dyson distribution of each class. 
They are super-imposed histograms of two sub-spaces.
CUE and COE unitaries given above are symmetric under a $\pi$ rotation around the y-axis, and two symmetry reduced sub-spaces are determined by even and odd eigenstates of $e^{-i\pi J_{y}}$~\cite{HaakeF1987CQCKT}.

Obviously, it is also relatively straightforward to define a goodness-of-fit $X^2$ test statistic to quantify agreement with Wigner-Dyson level-spacing statistics.
For this paper, however, we have not included any such calculations, because they tend to produce noisier statistical comparisons than the corresponding eigenvector-based test statistics due to there being:
quadratically fewer eigenvalues than eigenvector components, imperfections introduced by the unfolding procedure, and some additional disagreement expected from the small approximation involved with extrapolating Wigner-Dyson statistics for high dimensions from the analytic solutions calculated for two-level systems.
In Appendix Sec.~\ref{App:chiSquaredTestDetail} above, we note cases when it can be still important to analyse eigenvalue statistics.

\paragraph{Eigenvector statistics}\label{App:eigenVecKicked}
The eigenvector statistics are analysed using the prescription in Appendix~\ref{App:EigenVecTheory},
noting the subtleties that arise when the system has degeneracies or additional physical symmetries.
For example, when analysing eigenvector statistics for the CSE kicked top, which exhibits Kramers degeneracies, the additional eigenvector freedom introduced by these degeneracies can be taken into account by summing the eigenvector component populations for the degenerate eigenstates.
This procedure decreases the number of components $\eta$ by half,
but this is already taken into account in the derivation of the CSE distribution in Eq.~\eqref{AppEq:RMTdist:CSE}~\cite{KusM1988UEV}.
If the system has an additional symmetry, such as the $R_y(\pi)$ rotational symmetries present in the COE and CUE kicked top models described above, the eigenvector statistics are analysed using only the eigenvector components within the symmetry subspaces.
In such cases, the distributions in Eqs~(\ref{AppEq:RMTdist:COE}--\ref{AppEq:RMTdist:CUE}) must be calculated using the decreased degree of freedom, here $\mathcal{D}/2$.
The same approach is used to account for the parity symmetries present in the Heisenberg and Dicke model DQS systems studied in the rest of this paper.

Figures~\ref{AppFig:8_KickedTopEigAndChi}~(a--d) show the eigenvector statistics, respectively, for (a) COE with regular parameters and $X^2_{COE}\sim 6.229\times10^5$, and (b) COE, (c) CUE and (d) CSE with quantum chaotic parameters and their respective reduced chi-squared goodness-of-fittest statistic values $X^2_{COE}\sim 0.995$, $X^2_{CUE}\sim 0.999$, and $X^2_{CSE}\sim 1.013$.
By comparing the $\XsqRMT$ test statistic against standard confidence regions for the chi-squared distribution---e.g., $0.985<X^2<1.016$ for $95\%$ confidence---we observe strong, statistically quantified agreement (disagreement) between the eigenvector statistics of quantum chaotic (regular) systems and the corresponding RMT distributions.
We further quantify the agreement/disagreement by calculating the $\XsqRMT$ goodness-of-fit test statistic between the eigenvector statistics of quantum kicked top of COE symmetry and the corresponding RMT distribution as a function of kick strength.
Figure~\ref{AppFig:8_KickedTopEigAndChi}~(e) shows that a sudden drop in $\XsqRMT$ begins around the critical kicking strength $\lambda\approx2p$, before reaching statistically significant values of $\XsqRMT\sim1$ for $\lambda\gtrapprox5p$, reliably marking the onset of completely quantum chaotic dynamics.
Next, we demonstrate that this behaviour correlates well with what is observed for dynamical signatures of quantum chaos.

\begin{figure*}[ht]
	\centering
	\includegraphics[scale=0.94]{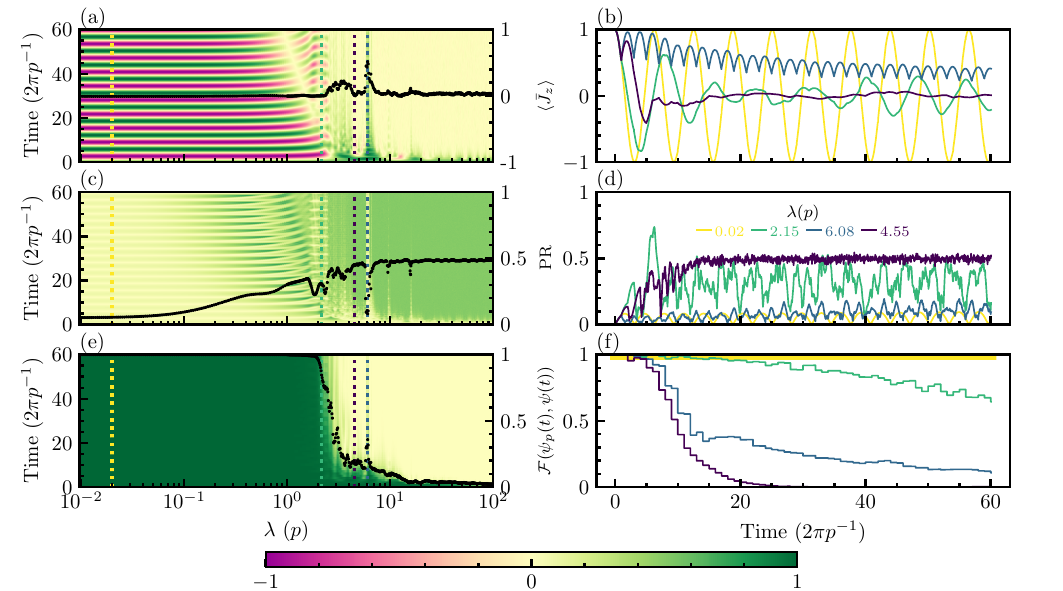} 
	\caption{Dynamical signatures of quantum chaos in quantum kicked top with COE and $j = 400$ (The other parameters are given in Appendix~\ref{App:KickedTop}).
	Dynamical evolutions of (a) (normalised) expectation value $\braket{\bar{J}_{z}} = \frac{1}{j}\braket{J_{z}}$, (c) participation ratio PR, and (e) fidelity under perturbation $\mathcal{F}(\psi_{p}(t), \psi(t))$ show quantum chaotic signatures for sufficiently large kicking strength $\lambda$.
	Black dots are temporal averages with y-axes on the right.
	(b), (d), and (f) show time traces, respectively, from (a), (c), and (d) for the kick-strengths values given in the legend of (d) and also marked by the line colours on colour plots.
	expectation values lose quasiperiodicity (a)\&(b), system gets delocalised (c)\&(d), and fidelity decays exponentially (e)\&(f) in the chaotic regime, while they all recover in the stable region ($\lambda = 6.08 (p^{-1})$).
	}\label{AppFig:9_KickedTopDynamical}
\end{figure*}

\subsubsection{Dynamical Signatures}

Here, we demonstrate onset of quantum chaos in the dynamics of a quantum kicked top with COE symmetry as a function of kick strength.
Figure~\ref{AppFig:9_KickedTopDynamical} shows the time evolutions of (a) (normalised) expectation value $\braket{\bar{J}_{z}} = \frac{1}{j}\braket{J_{z}}$, (b) participation ratio (PR), and (c) fidelity under perturbation.
Again, beyond the critical kicking strength $\lambda\sim 2p$, matching the starting point of the sudden drop in $\XsqRMT$ observed in Fig.~\ref{AppFig:8_KickedTopEigAndChi}~(e), the system starts to show dynamical signatures of quantum chaos: (i) destruction of quasiperiodicity in expectation values, (ii) high delocalisation, (iii) rapid perturbation fidelity decay.
For $\lambda\gtrapprox5p$, we see another clear transition, with---apart from a couple of stable islands---apparently complete destruction of quasiperiodicity, near-maximal delocalisation of the spin quantum state, and an end to the region of sharpest decline in perturbation fidelity decay.
While the perturbation fidelity remains very high all the way up to $\lambda\approx2p$, we also observe a clear quasiperiodic region prior to this kick strength for both local observables and participation ratio, starting around $\lambda\gtrapprox0.7p$--$0.9p$.
Finally, the dynamical signatures show clear stable island effects beyond $\lambda\approx5p$, e.g., for $\lambda\approx6p$ and $\lambda\approx 10p$.
In this case, we do not observe these same stable islands in the $\XsqRMT$ test statistic, and attribute this to the fact that these stable regions are highly localised in Hilbert space and highly sensitive to initial state for high-$j$ spin values (see complementary discussion in Appendix Sec.~\ref{App:IsingStableRegions}), their effects being washed out in $\XsqRMT$, which is insufficiently sensitive to the stable revivals due to measuring a collective response across all initial states.
These observations are further supported by looking at individual time traces for key relevant kick-strengths, as shown in Figures~\ref{AppFig:9_KickedTopDynamical}~(d--f), which clearly illustrate both quantum chaotic (and regular) dynamics as well as the revival of regular dynamics within the stable region marked (and plotted) by blue.

\section{Other Signatures of Quantum Chaos}\label{App:otherSignatures}

 
\begin{figure*} 
	\centering
	\includegraphics[scale=0.94]{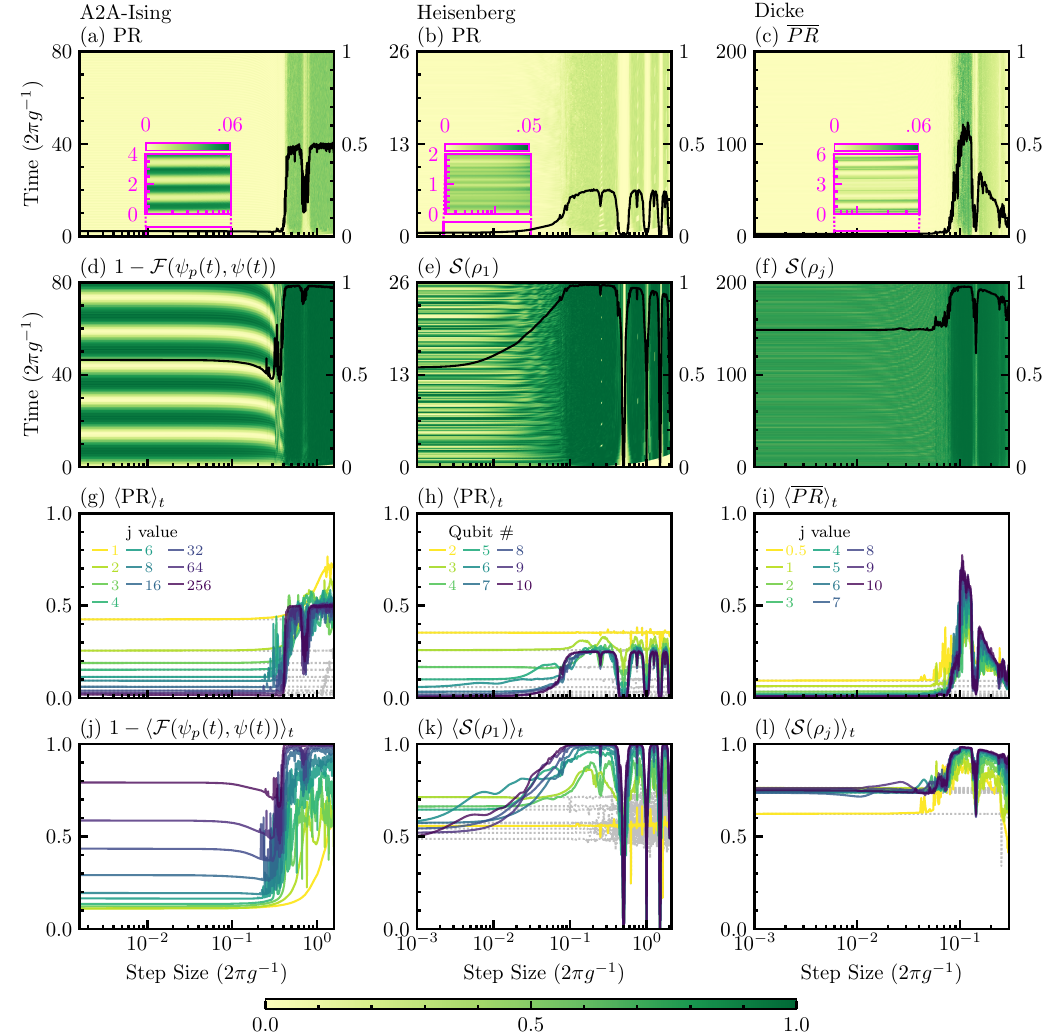}
	\caption{Other dynamical signatures of quantum chaos (and their time-averages for each model compared across system size), showing (a--c) (and (g--i)) delocalisation and (d--f) (and (j--l)) further complementary signatures.
	Left to right (and for color plots), columns show results for A2A-Ising (with $j = 64$), Heisenberg (chain of $N = 8$ qubits), and Dicke (with $j=6$) models, respectively, with x-axes common to each column showing step sizes.
	The y-axes of (a--f) are the simulation times, and the specific quantities plotted on the y-axes of (g--l) are given in the sub-figure titles, with $\langle .\rangle_{t}$ representing time averaging over periods $t = 200~(2\pi g^{-1})$ (A2A-Ising), $t = 50~(2\pi g^{-1})$ (Heisenberg), and $t = 200~(2\pi g^{-1})$ (Rabi-Dicke) and with line colours corresponding to system sizes given in legends in (g--i).
	The black lines (right y-axes) in (a--f) are the time averages of the colour plots.
	Dotted silver lines are time averages of sampled ideal dynamics, sampled at intervals of $\tau$, to illustrate the emergence of sampling effects at large step sizes, even in the absence of digitisation errors.
	In each model, dynamical quantities share a common Trotterisation threshold at a critical step size separating regular from quantum chaotic regimes.
	(a--c) Localised initial states rapidly delocalise beyond the threshold, quantified by participation ratios (PR, Eq.~\eqref{Eq:PR_def}).
	(d--f) further complementary dynamical signatures of quantum chaos exhibit the same threshold: e.g., infidelity under perturbation $1 - \mathcal{F}(\psi_{p}(t), \psi(t))$ (see Eq.~\eqref{Eq:FidelityDecay}) for Ising, and normalised sub-system entropies $\mathcal{S}(\rho_{i})$ (Eq.~\eqref{Eq:entropy_def}) for Heisenberg (first qubit) and Dicke (spin), are all rapidly maximised beyond the threshold.
	(g--i) Time-averaged participation ratios $\langle \text{PR}\rangle_{t}$ show clear delocalisation beyond the threshold.
	(j--l) Beyond the threshold, infidelity under perturbation $\langle 1 - \mathcal{F}(\psi_{p}(t), \psi(t))\rangle_{t}$ for Ising in (j), and normalised sub-system entropies $\langle\mathcal{S}(\rho_{i})\rangle_{t}$ for Heisenberg (first qubit) (k) and Dicke (spin) (l), rapidly approach maximum values for increasing system size.
	In each model, the dynamical signatures of quantum chaos (a--l) disappear in the stable regions, showing (at least partial) revival of regular dynamics.}
	\label{AppFig:10_OtherDynamicalSignatures}
\end{figure*}


In the main text, we used our observation of the clear qualitative destruction of quasiperiodicity to illustrate a transition to quantum chaotic dynamics as a function of step size.
In this section, we conduct a more quantitative study of this transition using a number of dynamical signatures underpinned by the connection of quantum chaos to ergodic dynamics and thermalisation.

For a closed system with a pure initial state, a concept of thermalisation can only be achieved in a context where the ergodic hypothesis holds, namely for dynamics where the time and state-space averages match for any macroscopic observable.
Quantum chaotic dynamics provide a natural mechanism to achieve this condition for quantum systems~\cite{LucaD2016ETH, DeutschJM1990QSM, SrednickiM1994CQT, NeillC2016EDQS}.
Specifically, quantum chaos drives a thermalisation-like process by dynamically delocalising the system state to span a large portion of the Hilbert space.
This process, which relates to the simple intuitive description of quantum chaotic dynamics as being exponentially sensitive to perturbations in system parameters, is reflected for composite quantum systems in an increase of sub-system entropy.
To study these concepts, in the following sections, we discuss example measures quantifying de-localisation, perturbation sensitivity, and entropy.

\paragraph{Delocalisation}
We first quantify the delocalisation of the time-evolved state $\ket{\psi(t)}$ in a given basis $\mathcal{B} = \{\ket{j}\}$ by means of the participation ratio (PR)~\cite{HeylM2019QLTE, SiebererLM2019DQC},
\begin{equation}\label{Eq:PR_def}
	\text{PR} := \frac{1}{\mathcal{D}}\left(\sum_{j=1}^{\mathcal{D}}\abs{\braket{\psi(t)|j}}^{4}\right)^{-1},
\end{equation}
where $\mathcal{D}$ is the dimension of the Hilbert space.
Quantum chaotic dynamics acts to create delocalisation, but to see this effect, it is important to choose an initial state and basis such that the state is initially localised.
In this definition, the normalisation factor ensures the PR lies between $\mathcal{D}^{-1}$ for a fully localised state and 1 for a fully delocalised state.
In the Dicke model, the cavity is infinite dimensional, and the choice of $\mathcal{D}$ is non-trivial. We here choose a finite $\mathcal{D} = (2\dim_j)^2$ (labelled $\overline{PR}$) and justify this choice in Sec.~\ref{Sec:SystemSize}.

Figures~\ref{AppFig:10_OtherDynamicalSignatures}~(a--c)~\&~(g--i) show that the temporal evolutions of PRs for A2A-Ising, Heisenberg, and Dicke models, and their time averages, are all clearly separated into localised and delocalised regions at the same step sizes as other quantities in Fig.~\ref{Fig:3_FullDynamicsAndFT}.
In A2A-Ising and Heisenberg Trotterisations, the PR beyond the threshold saturates to maximum values that can be well understood in terms of RMT symmetries and the choice of initial state.
The maximum PR value is related to the RMT class and the choice of basis $\mathcal{B} = \{\ket{j}\}$, and is limited to 0.5, as observed in Fig.~\ref{AppFig:10_OtherDynamicalSignatures}~(a), for A2A-Ising DQS (for further details see the discussions in Ref.~\cite{SiebererLM2019DQC}).
While the PR for the Heisenberg DQS saturates to $\sim 0.25$ (Fig.~\ref{AppFig:10_OtherDynamicalSignatures}~(b)) for the chosen initial state which evolves in one parity subspace only, the results in Appendix~\ref{App:HeisenbergAppendix_initialState} studying initial-state dependence for the Heisenberg model show that the PR often saturates to 0.5 for arbitrary states.

The maximum PR values in the Dicke model are more difficult to interpret due to our observation that the finite spin-ensemble dimension appears to limit how far the state can delocalise in the infinite dimensional cavity space (discussed further in later sections).
Nevertheless, it is clear that the system shows larger delocalisation beyond the threshold.
We also observe that the $\overline{PR}$ follows the same decaying trend beyond the stable region seen in the photon number.
Our numerical analyses do not provide an explanation for this behaviour, which we leave as an open question.

\paragraph{Fidelity decay}
Sensitivity to small perturbations in system parameters is one of the most widely appreciated characteristics of quantum chaotic dynamics.
This concept can be quantified via a variety of related signatures such as fidelity decay, Loschmidt amplitude and echo, and out-of-time order correlators (OTOCs)~\cite{PeresA1984SQM, EmersonJ2002FDEI, GORINT2006LEFD, GoussevA2012LE, Heyl2019QPT}, which are also useful in other contexts like quantum phase transitions and systems with broken time-reversal symmetries~\cite{GORINT2006LEFD, GoussevA2012LE, Heyl2019QPT}.
However, a drawback is that such signatures are reliable only if the perturbation itself does not show RMT distributions in its eigenvector statistics~\cite{EmersonJ2002FDEI}.

Here, we look at the perturbation fidelity, defined as the overlap
\begin{equation}\label{Eq:FidelityDecay}
	\mathcal{F}(\psi_{p}(t), \psi(t)) := \abs{\braket{\psi(0)|U_{p}^{\dagger}(t)U(t)|\psi(0)}}^{2}
\end{equation}
between two states evolved from the same initial state $\ket{\psi(0)}$ under a unitary $U(t)$ and a perturbed unitary $U_{p}(t)$~\cite{PeresA1984SQM}.
In the case of the quantum kicked top pertaining to the A2A-Ising model, we identify a reliable perturbation by applying $\exp(-iH_{z}0.05\tau)$ to the Trotter step unitary $U_{\tau}$ (see also Appendix~\ref{App:KickedTop}).
Figures~\ref{AppFig:10_OtherDynamicalSignatures}~(d)~\&~(j) shows that the fidelity decay, or rather, here, the complementary perturbation error (infidelity under perturbation) $1-\mathcal{F}(\psi_{p}(t), \psi(t))$, displays a regular oscillating signature before the threshold, and a characteristic exponential decay after, providing further strong evidence of the digitisation-induced transition to quantum chaotic dynamics.
This trend is observed also for very small perturbations, with the perturbation strength only affecting the oscillation and decay rates in the two regimes.

\paragraph{Sub-system Entropy}
For the Heisenberg and Dicke models, we illustrate an alternative signature of quantum chaotic dynamics that does not require perturbative comparisons, namely the increase of entanglement between subsystems, as characterised by the sub-system entropy~\cite{Furuya1998QDM, Bandyopadhyay2002SBE, MonasterioC2005EATC, NeillC2016EDQS}.

Here, we consider a normalised von Neumann entropy,
\begin{equation}\label{Eq:entropy_def}
	\mathcal{S}(\rho_{i}) = - \frac{1}{\ln(\mathcal{D}_{i})}\text{Tr} \left(\rho_{i}\ln(\rho_{i})\right),
\end{equation}
where $\rho_{i}:=\Tr_{\text{not} [i]}(\rho_{\mathrm{system}})$ is the reduced state of sub-system $i$ with dimension $\mathcal{D}_{i}$. 
The normalisation factor $\tfrac{1}{\ln(\mathcal{D}_{i})}$ rescales the usual entropy to be in $\left[0, 1\right]$ for any sub-system dimension, simplifying comparisons between different systems.
This signature can be measured experimentally via tomography of the sub-system state, but importantly does not require access to the full system state~\cite{NeillC2016EDQS}.
Figures~\ref{AppFig:10_OtherDynamicalSignatures}~(e--f)~\&~(k--l) show that sub-system entropies jump to much larger values beyond the digitisation threshold for both the Heisenberg model (qubit 1) and the Dicke model (the spin-$j$ sub-system).
In Sec.~\ref{Sec:SystemSize}, we show that this entropy approaches the maximum value asymptotically with system size.

The observations of delocalisation, fidelity decay, and increased sub-system entropies beyond the threshold provide strong evidence for the quantum chaotic dynamics, complementing the more quantitatively rigorous evidence using random matrix theory statistics provided in Sec.~\ref{Sec:QChaosOnset}.
Combined with the observation that time averages of expectation values saturate to the relevant Hilbert-space means for each model and observable, for $\braket{\bar{J}_{z}}$ in A2A-Ising (Fig.~\ref{Fig:3_FullDynamicsAndFT}~(a)), $\braket{\sigma_{z}^{1}}$ for Heisenberg (Fig.~\ref{Fig:3_FullDynamicsAndFT}~(b)), and cavity and qubit parity in the Dicke model (not shown), they also provide evidence for the idea that the digitisation-induced quantum chaos is driving ergodicity beyond the threshold.


\section{Ising Model}\label{App:IsingAppendix}


\begin{figure*}[ht]
	\centering
	\includegraphics[scale=0.94]{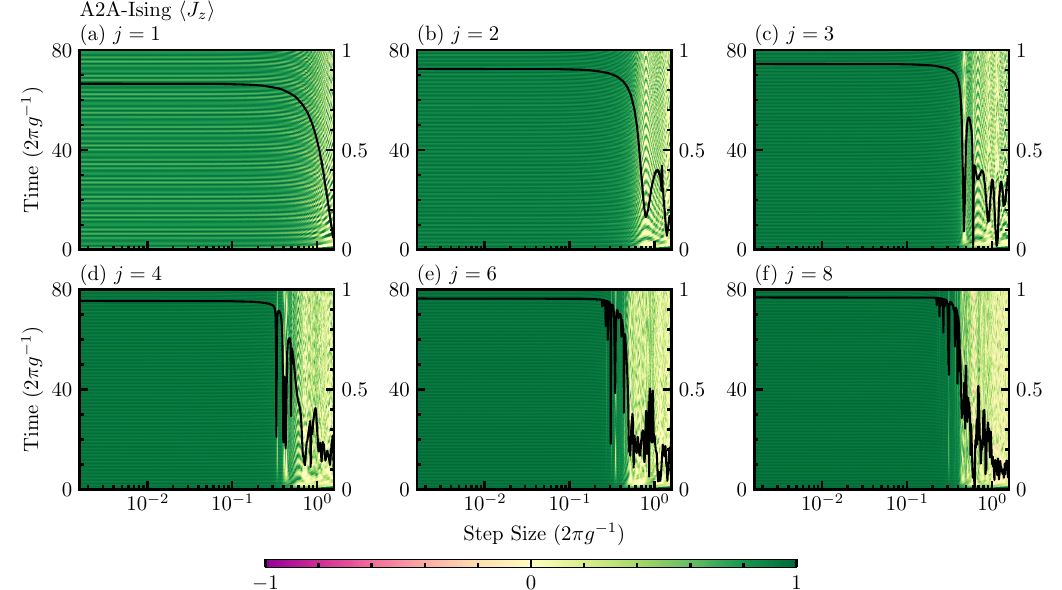}
	\caption{The emergence of a sharp Trotterisation threshold as the system size increases in A2A-Ising.
	The \emph{lack} of sudden thresholds in small sizes parallels the dynamical evolutions.
	(a), (b) and (c) show that the dynamics beyond the threshold still appears quasiperiodic for the small spin sizes $j= 1$, $j=2$ and $j=3$, which do not enable reliable conclusions based on RMT statistics analyses.
	The smallest system size for which $\XsqRMT \sim 1$ offers good statistical confidence is $j = 4$ and (d-f) shows that the dynamics beyond the threshold start losing visibly apparent quasiperiodicity for $j \ge 4$.
	The system parameters and the initial state are the same as in the main text of the paper.
	}\label{AppFig:11_A2AIsingDynamical}
\end{figure*}

\subsection{Dynamics in Small Systems}\label{App:IsingSmallSystems}

Here, we show dynamics for the magnetisation $\braket{\bar{J}_{z}} = \braket{J_{z}}/j$ for small $j$ values in A2A-Ising Trotterisation.
Figure~\ref{AppFig:11_A2AIsingDynamical} shows that the threshold gets sharper as the dynamics start to show the characteristic quantum chaos signature of destruction of quasiperiodicity beyond the threshold.
Even though a distinct change in behaviour at the approximate threshold position is still clear for any system size in Fig.~\ref{AppFig:11_A2AIsingDynamical}, the transition remains smooth at system sizes where quasiperiodicity still seems apparent beyond the threshold, and it gets sharper only for sufficiently large $j$ values, when visually identifiable quasiperiodicity after the threshold no longer appears to be observable.
In Fig.~\ref{AppFig:11_A2AIsingDynamical}, we observe that the sudden threshold and collapse of obvious visual quasiperiodicity start to appear for $j \geq 4$, at the same size a $\XsqRMT$ analysis showing $\XsqRMT \sim 1$ (see main text) starts to allow reliable conclusions to be drawn about agreement with RMT predictions.

\begin{figure*}[ht]
	\centering
	\includegraphics[scale=0.94]{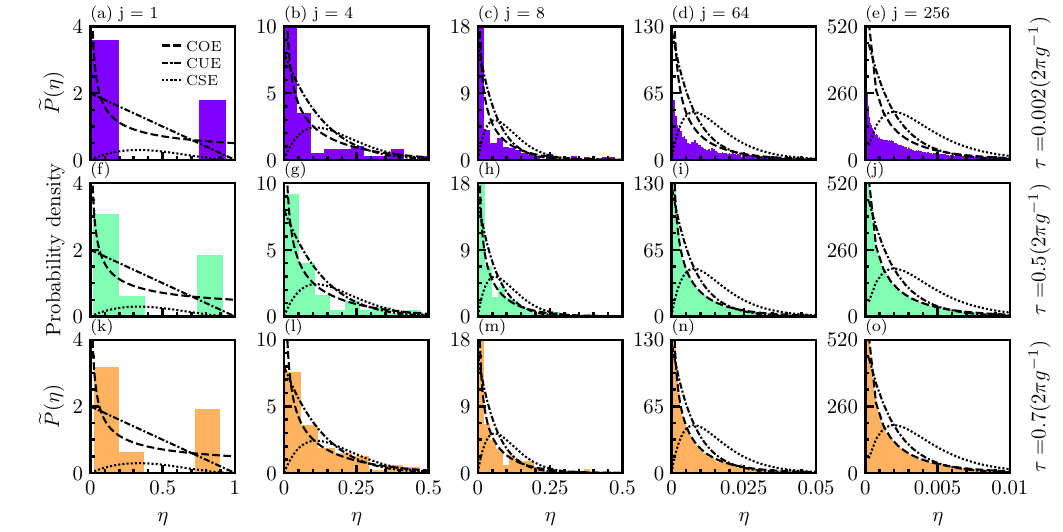}
	\caption{
	Eigenvector statistics for Trotterised A2A-Ising unitary.
	Columns show increasing $j$ values from left-to-right, and the rows, from top-to-bottom, are for step sizes before the threshold ($\tau = 0.002~(2\pi g^{-1})$), chaotic ($\tau = 0.5~(2\pi g^{-1})$), and stable revival (seen only in large system sizes) ($\tau = 0.7~(2\pi g^{-1})$) regions.
	The coloured lines are RMT distributions as shown by the legend.
	}\label{AppFig:12_A2AIsingEigHist}
\end{figure*}

\subsection{Eigenvector Statistics}\label{App:IsingEigenVecs}

In the main text of the paper, we show eigenvector statistics histograms for $j=64$ and calculate $\XsqRMT$ for the other system sizes.
Here, to illustrate how system size affects the histograms themselves, we provide example histograms for several $j$ values and three step sizes, which are chosen from regular (before the threshold), quantum chaotic (after the threshold), and stable revival regions.
Figure~\ref{AppFig:12_A2AIsingEigHist} shows the eigenvector statistics for different $j$ values on each column and different step sizes on each row: (top to bottom) before the threshold ($\tau = 0.002~(2\pi g^{-1})$), quantum chaotic ($\tau = 0.5~(2\pi g^{-1})$), and stable revival (seen only in large system sizes) ($\tau = 0.7~(2\pi g^{-1})$) regions, respectively.
The smallest system size, in Fig.~\ref{AppFig:12_A2AIsingEigHist}~(a), does not have enough components to produce meaningful statistics, and the visual comparisons in Figs~\ref{AppFig:12_A2AIsingEigHist}~(b--c) do not provide very clear conclusions.
The $\XsqRMT$ goodness-of-fit test statistic, on the other hand, provides consistent conclusions in terms of successfully identifying the disagreement/agreement of eigenvector statistics with RMT distributions (shown in Fig.~\ref{Fig:6_chiSquareForAll} (a,d)).
For the large system sizes, the regular and chaotic regions are clearly distinguishable in the eigenvector statistics, shown respectively for (d) $j = 64$ and (e) $j = 256$ in Figs~\ref{AppFig:12_A2AIsingEigHist}~(d--e).
However, the stable regions ($\tau = 4.5~(2\pi g^{-1})$) are visually indistinguishable from the quantum chaotic $\tau = 3.5~(2\pi g^{-1})$ case, and, once again, $\XsqRMT$ proves useful and identifies these weak islands (see main text).

\subsection{Stable Regions}
\label{App:IsingStableRegions}

\begin{figure*}[ht]
  \centering
  \includegraphics[width=\linewidth]{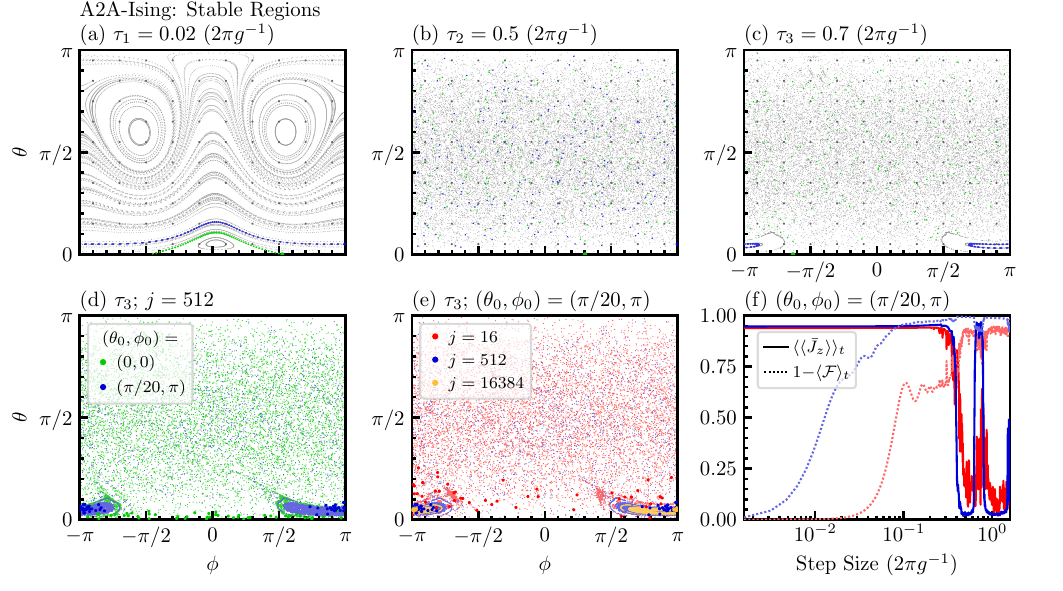}
  \caption{Connection between the appearance of stable regions in DQS of the A2A-Ising model and the re-emergence of stable islands in the classical phase space of the collective spin. Grey dots in panels (a--c) result from 200 iterations of the stroboscopic map Eq.~\eqref{eq:semiclassical_evolution} for (a) $\tau = 0.02~(2 \pi g^{-1})$, (b) $\tau = 0.5~(2 \pi g^{-1})$, and (c) $\tau = 0.7~(2 \pi g^{-1})$, and for 100 uniformly distributed initial states. 
  The initial states for the stroboscopic trajectories shown in green and blue are given by $\left( \theta_0, \phi_0 \right) = \left( 0, 0 \right)$ and $\left( \theta_0, \phi_0 \right) = \left( \pi/20, \pi \right)$, respectively.
  In (d), for both of these choices of $\left( \theta_0, \phi_0 \right)$, 50 initial states are sampled from the distribution corresponding to the spin-coherent state $\ket{\theta_0, \phi_0}$ for $j = 512$. Panel (e) shows the evolution for $\left( \theta_0, \phi_0 \right) = \left( \pi/20, \pi \right)$, again for initial states sampled from coherent-state distributions for the spin sizes $j = 16$, $512$, and $16384$ in red, blue, and orange, respectively. 
  In both (d) and (e), the Trotter step size is $\tau = 0.7~(2 \pi g^{-1})$.  
  The re-emergence of a stable island in panels (c--e) is connected to the revival of regular dynamics in a stable region as shown in (f) for $j = 16$ (red) and $512$ (blue).}
  \label{AppFig:13_IsingPhaseSpacePortraits}
\end{figure*}

In the A2A-Ising model, we demonstrate that the revivals in stability observed beyond the threshold can be connected with the familiar notion of stable islands in an otherwise chaotic classical phase space.
Here, the semiclassical limit is identical to the thermodynamic limit $j \to \infty$~\cite{HaakeF2018QSC}.
In particular, the commutators of rescaled spin components $X = J_x/j$, etc., vanish in the limit $j \to \infty$, so that $X$, $Y$, and $Z$ become effectively classical variables.
The stroboscopic classical evolution on the unit sphere of normalized vectors $\left( X, Y, Z \right)$ results from taking the semiclassical limit of the Heisenberg-picture evolution equations $J_{\mu}(t + \tau) = U_{\tau}^{\dagger} J_{\mu}(t) U_{\tau}$.
Using the shorthand $X' = X(t + \tau)$ and defining $W = \left( Y + i Z \right) e^{i \left( \omega_x + g_x X \right) \tau}$, the evolution equations read~\cite{SiebererLM2019DQC}
\begin{equation}
  \label{eq:semiclassical_evolution}
  \begin{split}
    X' & = \Re \! \left( \left( X + i \Re(W) \right)
      e^{i \left( \omega_z + g_z \Im(W) \tau \right)} \right), \\
    Y' & = \Im \! \left( \left( X + i \Re(W) \right)
      e^{i \left( \omega_z + g_z \Im(W) \tau \right)} \right), \\
    Z' & = \Im(W).
  \end{split}
\end{equation}

The dynamics are best visualised by parametrising the classical phase space through spherical coordinates $\theta$ and $\phi$ via $\left( X, Y, Z \right) = \left( \cos(\phi) \sin(\theta), \sin(\phi)
\sin(\theta), \cos(\theta) \right)$. 
This leads to the phase-space portraits shown in Fig.~\ref{AppFig:13_IsingPhaseSpacePortraits} for (a) $\tau = 0.02~(2 \pi g^{-1})$, (b) $\tau = 0.5~(2 \pi g^{-1})$, and (c) $\tau = 0.7~(2 \pi g^{-1})$.  
Grey dots in the figure correspond to 200 iterations of the stroboscopic map Eq.~\eqref{eq:semiclassical_evolution} for 100 uniformly distributed initial states; 
The initial states for the trajectories shown in green and blue are given by, respectively, $\left( \theta_0, \phi_0 \right) = \left( 0, 0 \right)$ and $\left( \theta_0, \phi_0 \right) = \left( \pi/20, \pi \right)$.  
In a quantum system with finite $j = 256$ as shown in Fig.~\ref{Fig:1_ThresholdConcept}, the above values of $\tau$ correspond to regular dynamics below the threshold, chaotic dynamics beyond the threshold, and the revival of regular dynamics in a stable region.  
For the classical phase space at the $\tau$ corresponding to a stable region in the quantum domain (Fig.~\ref{AppFig:13_IsingPhaseSpacePortraits} (c)), we also observe the re-appearance of a stable island, despite the phase space having already crossed over to be completely chaotic beyond the threshold (Fig.~\ref{AppFig:13_IsingPhaseSpacePortraits} (b)).  
The initial state described by the green dots, $\left( \theta_0, \phi_0 \right) = \left( 0, 0 \right)$, is chosen in Figs~\ref{AppFig:13_IsingPhaseSpacePortraits} (a--c) to match the initial state chosen for the dynamical DQS simulations in Fig.~\ref{Fig:1_ThresholdConcept}.
By contrast, the blue dots in Figs~\ref{AppFig:13_IsingPhaseSpacePortraits} (a--c) show the trajectories for an initial state with $\left( \theta_0, \phi_0 \right) = \left( \pi/20, \pi \right)$, which lies on the stable island for Trotter steps within the stable region.  
Even at the classical level, the regular dynamics for the stable island in Fig.~\ref{AppFig:13_IsingPhaseSpacePortraits} (c) are clearly different from the dynamics in Fig.~\ref{AppFig:13_IsingPhaseSpacePortraits} (a), which are closer to the target dynamics;
This accounts for the relatively poor revival of simulation fidelity in the stable region, as observed in the main text (see also Fig.~\ref{AppFig:13_IsingPhaseSpacePortraits} (f)).
However, while the classical $J_{z}$ component is relatively insensitive to the differences between these trajectories, we still see initial-state effects. 
Figure~\ref{AppFig:13_IsingPhaseSpacePortraits} (d) shows the classical trajectories for a distribution of initial states selected according to spin coherent states $\ket{\theta_0, \phi_0}$ with $j=512$ for the above states, $\left( \theta_0, \phi_0 \right) = \left( 0, 0 \right)$ and $\left( \theta_0, \phi_0 \right) = \left( \pi/20, \pi \right)$.  
The finite size of the spin coherent state leads to a finite overlap with both stable island and chaotic regions in classical phase space, but this becomes less pronounced with increasing spin as shown in Fig.~\ref{AppFig:13_IsingPhaseSpacePortraits} (e), for phase space trajectories corresponding to $j=16$, $j=512$ and $j=16384$ with $\left( \theta_0, \phi_0 \right) = \left( \pi/20, \pi \right)$.  
Simulations of the Trotterised A2A-Ising dynamics for the first two of these spin sizes ($j=16$ and $j=512$) are shown in Fig.~\ref{AppFig:13_IsingPhaseSpacePortraits} (f). 
The revival becomes more marked with increasing spin size for the initial state $\ket{\theta_0,\phi_0}=\ket{\pi/20,\pi}$, even while the simulation fidelity remains relatively low.

The quantum-classical correspondence principle suggests the following
  explanation for the changes observed in the local observable in Fig.~\ref{AppFig:13_IsingPhaseSpacePortraits} (f) around $\tau = 0.7~(2 \pi g^{-1})$:
Trotter errors are inevitably large for trajectories corresponding to chaotic dynamics that is ergodic within the entire classical phase space (as for a completely mixed state).
Trajectories within the stable islands of classical phase space follow regular
dynamics that is generally restricted to a more localised region, and which therefore exhibits a different overlap with the target dynamics, from that corresponding to an underlying classically chaotic trajectory.
Hence the re-emergence of stable islands in the A2A-Ising's classical phase space accompanies the revival of stability in the stable regions of the Trotterised quantum model, for initial states which overlap with the stable islands.
But note again with caution that, as in the case of the Trotterised Heisenberg model, the observed revivals of stability do not typically correspond to any revival in the quality of simulation with an underlying target model.


\section{Heisenberg Model}\label{App:HeisenbergAppendix}


\begin{figure*}
	\centering
	\includegraphics[scale=0.94]{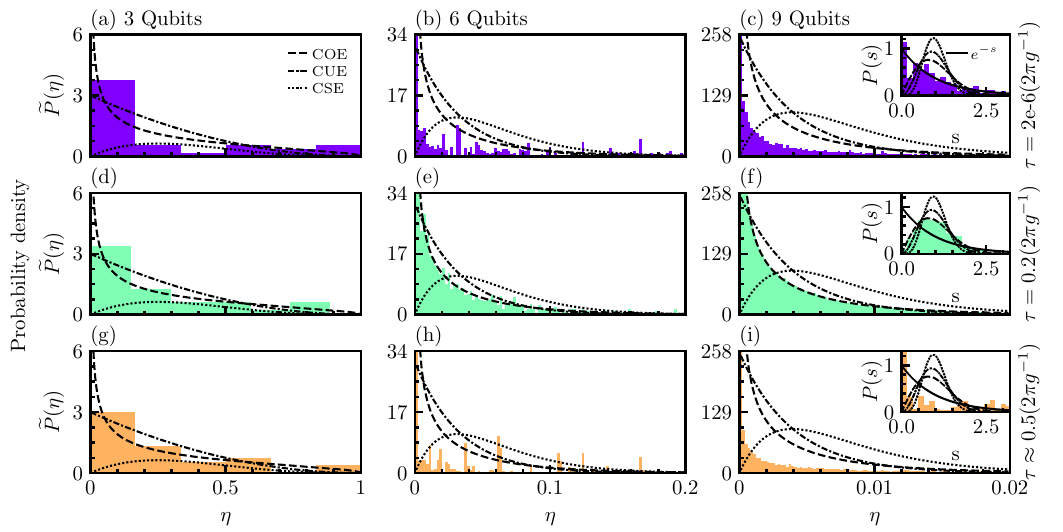}
	\caption{Eigenvector statistics for Trotterised Heisenberg unitary.
	Columns, left to right, show 3, 6, and 9 qubit chains, and rows, top-to-bottom, are for step sizes before the threshold ($\tau = 2e\text{-}6~(2\pi g^{-1})$), quantum chaotic ($\tau = 0.2~(2\pi g^{-1})$), and stable revival ($\tau \approx 0.5~(2\pi g^{-1})$) regions.
	6 qubits and 9 qubits show clear deviations from the RMT distributions (c) before the threshold and (i) in the stable revival region, and it closely follows the COE distribution in (f) quantum chaotic region.
	3 qubits cases in (a,d,g) do not have enough components to give good statistics. 
	The coloured lines are RMT distributions as shown by the legend.}
	\label{AppFig:14_HeisenbergEig}
\end{figure*}

\subsection{Eigenvector Statistics}\label{App:HeisebergEigVec}

For the Heisenberg model, direct analysis of the Trotter step unitary in the computational basis shows $\XsqRMT$ rapidly transitions to $\XsqRMT\sim1$ for CUE statistics at the Trotterisation threshold (not shown, data available).
As discussed in Appendix~\ref{App:refBasis}, however, while this clearly demonstrates the onset of quantum chaotic dynamics at the threshold, it is not sufficient to conclusively identify the correct symmetry class.
In this case, analysis of the basis-independent level-spacing statistics shows agreement with COE Wigner-Dyson statistics, showing the Heisenberg model actually \emph{does} exhibit time-reversal symmetry, despite $\XsqRMT$ agreeing with CUE in the computational basis.
To see the COE statistics revealed in the eigenvector statistics,
the Trotter step unitary has to be expressed in a basis where it is itself symmetric ($U_\tau=U_\tau^\top$).
Such a basis can be found by constructing basis vectors which are invariant under the unitary's time-reversal symmetry operator $T$~\cite{HaakeF2018QSC}.
Our Trotterised Heisenberg model has the following $T$:
\begin{equation}
  \label{eq:HeisenbergTRSOperator}
	T=\prod_{k=1}^{N-1}U_{yz}^{k,k+1}(\tau)\prod_{k=1}^{N-1}U_{xz}^{k,k+1}(\tau)\mathcal{R}_{z,\tfrac{\pi}{2}}^{\forall k}IK,
\end{equation}
where $\mathcal{R}_{z,\tfrac{\pi}{2}}^{\forall k}$ consists of locally applied single-qubit $\frac{\pi}{2}$ rotations around the $z$-axis, $K$ is the complex conjugation operator, and $I$ represents spatial inversion of the spin chain given by $I\sigma^{k}_{\mu}I^\dagger = \sigma^{N+1-k}_{\mu}$ where $\mu\in\{x,y,z\}$.
When expressed in the $T$-invariant basis found using this $T$ operator, the Trotter step unitary from Eq.~\ref{Eq:HeisenbergDecomposition} is symmetric.
(Note, however, that this $T$ matrix can only be used for $\tau\neq0$, because the $T$-invariant basis from this $T$ is not guaranteed to be full rank for $\tau=0$.)

So far, we have only presented the processed $\XsqRMT$ analysis for the Heisenberg Trotterisation eigenvector statistics. Here, Fig.~\ref{AppFig:14_HeisenbergEig} explicitly shows the full histogram data for 3, 6, and 9 qubits Heisenberg Trotterisations on each column and different step sizes on each row.
The step sizes shown in the columns of Fig.~\ref{AppFig:14_HeisenbergEig} are, respectively from top-to-bottom, from before the threshold ($\tau = 2e\text{-}6~(2\pi g^{-1})$), and from the quantum chaotic ($\tau = 0.2~(2\pi g^{-1})$) and stable revival ($\tau \approx 0.5~(2\pi g^{-1})$) regions.
(Note: Numerical diagonalisation becomes unstable at or near $\tau=0.5$ for our model parameters due to numerical imprecision, so we instead use $\tau=0.5000001$, which our analysis indicates is close enough to have ``converged'' to a limiting value for $\tau\rightarrow0.5$, but far enough from $\tau=0.5$ for numerical error to be neglected.)
Unlike in the A2A-Ising model, the stable revivals are so strong in the highly symmetric Heisenberg model considered here that the eigenvector statistics histogram plots for both 6- and 9-qubit models show clear agreement with RMT distributions in the quantum chaotic region (Figs~\ref{AppFig:14_HeisenbergEig}~(e,f)), and clear disagreement both before the threshold (Figs~\ref{AppFig:14_HeisenbergEig}~(b,c)) and on the stable island (Figs~\ref{AppFig:14_HeisenbergEig}~(h,i)).
By contrast, for the smallest system size in Fig.~\ref{AppFig:14_HeisenbergEig}, none of the histogram plots provides any clear conclusion, nor are there enough eigenvector components for $\XsqRMT$ to provide reliable statistical confidence.

As a final note, this Trotterisation has parity symmetry ($\Pi=\bigotimes_{j=1}^N \sigma_z^j$).
Therefore, the eigenvector statistics in Fig.~\ref{AppFig:14_HeisenbergEig} are calculated using the procedure discussed in Appendix~\ref{App:eigenVecKicked}.
Consequently, the number of independent components is reduced by half, so the theoretical lines in Fig.~\ref{AppFig:14_HeisenbergEig} are for $\mathcal{D} = 2^{N-1}$ for the case of $N$ qubits. 
This parity symmetry is removed when the longitudinal field $\sigma_{z}$ term is replaced with a transverse field $\sigma_{y/x}$ term. 
Then, the regular procedure can be followed, and we observed a good agreement with RMT distributions of the full dimension $\mathcal{D} = 2^{N}$.
In both cases, the results regarding the threshold and the onset of quantum chaos beyond it are qualitatively the same both in the eigenvector statistics and the dynamical signatures (not shown, data available).

\begin{figure*}
	\centering
	\includegraphics[scale=0.94]{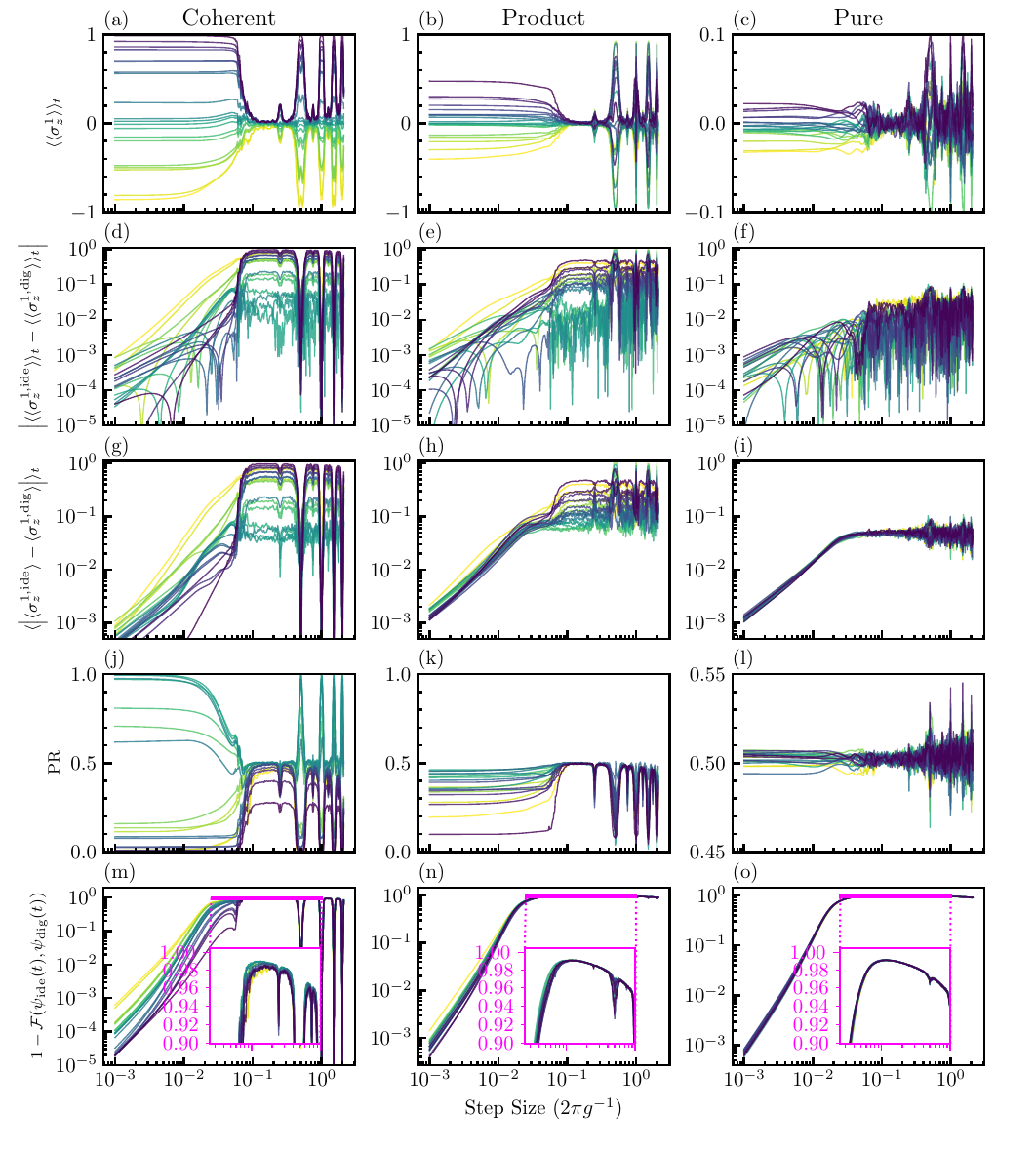}
	\caption{Threshold position for 9 qubits Heisenberg chain with various initial states.
	The columns, left-to-right, show the results for 20 spin-coherent, 20 product, and 20 pure initial states sampled uniformly at random.
	In each column, the same line colours correspond to the same initial state.
	Here, the second layer of brackets $\langle .\rangle_{t}$ represents time averaging.
	The rows, top-to-bottom, show time averaging of (a--c) the qubit polarization of the first qubit in the chain $\langle\langle\sigma_{z}^{1}\rangle\rangle_{t}$ for Trotterised evolution, (d--f) the absolute difference $\abs{\langle\langle\sigma_{z}^{1,\text{ide}}\rangle\rangle_{t}-\langle\langle\sigma_{z}^{1,\text{dig}}\rangle\rangle_{t}}$ between the time averaged expectation values obtained from the Trotterised ($\langle\sigma_{z}^{1,\text{dig}}\rangle$) and ideal dynamics ($\langle\sigma_{z}^{1,\text{ide}}\rangle$ sampled at every step size $\tau$), (g--i) time average of the point-to-point absolute difference $\langle\abs{\langle\sigma_{z}^{1,\text{ide}}\rangle-\langle\sigma_{z}^{1,\text{dig}}\rangle}\rangle_{t}$ at each step of the evolution, (j--l) participation ratio $\langle PR\rangle_{t}$ for the Trotterised dynamics, and (m--o) simulation infidelity $1-\langle\mathcal{F}(\psi_{\text{dig}}(t), \psi_{\text{ide}}(t))\rangle_{t}$.
	We find the threshold at the same position.
	The smooth deviations from the ideal value (i.e. the value at $\tau \rightarrow 0$) in the time averages of $\langle\langle\sigma_{z}^{1}\rangle\rangle_{t}$ start before the threshold for the states with larger Trotter errors, which is also explained as the result of the change in quasiperiodicity in the main text.
	}\label{AppFig:15_HeisenbergInitialState}
\end{figure*}

\begin{figure*}[ht]
	\centering
	\includegraphics[scale=0.94]{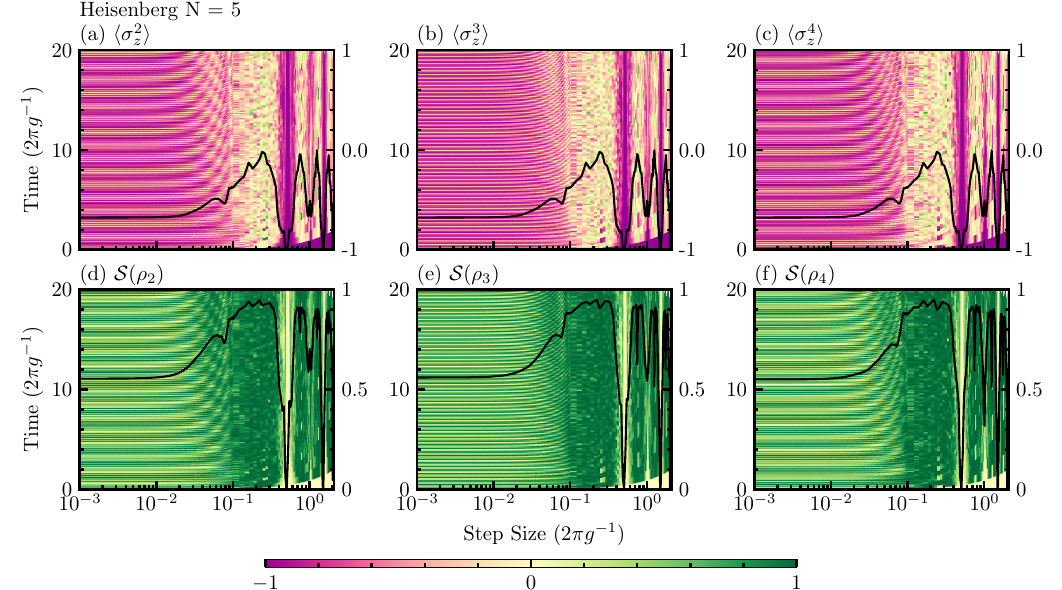}
	\caption{Dynamical evolutions of qubit polarization $\langle\sigma_{z}^{k}\rangle$ and sub-system entropy $\mathcal{S}(\rho_{k})$ for qubit $k=2,3,4$ in the $N=5$ Heisenberg chain.
	The black lines, with y-axes on the right, are the time averages of the colour plots.
	System parameters are the same as in the main text of the paper.}
	\label{AppFig:16_HeisenbergOtherQubs}
\end{figure*}

\subsection{Effect of Initial State on the Threshold}\label{App:HeisenbergAppendix_initialState}

The Trotterisation of the Heisenberg model shows certain initial-state-dependent effects in the main text of the paper.
Here, we analyse the dynamics of the 9-qubit Trotterised Heisenberg model for various initial states (of 3 types) and provide results for expectation values, PR values, and the Trotter errors.
The results show that the initial state does affect the size of the Trotterisation errors before the threshold and we note that the threshold might not seem as sudden for the states with larger errors.
Still, the breakdown of digitisation occurs at a distinct threshold largely independent of the initial state.

We sample 20 initial states uniformly at random from each spin-coherent, product, and pure state types, where the spin-coherent state
\begin{equation}
	\ket{\theta, \phi}_{N} = \otimes_{N}(\cos(\theta/2)\ket{1} + e^{-i\phi}\sin(\theta/2)\ket{0}),
\end{equation}
is just a subset of product states $\ket{N} = \otimes_{i=1}^{N}\ket{i}$, which is a subset of pure states.
Figure~\ref{AppFig:15_HeisenbergInitialState} shows various averaged quantities calculated from the digital simulations for $t = 20~(2\pi g^{-1})$. 
Even for initially highly delocalised states (and the pure state samples), there is a noticeable difference at the same threshold position, and we observe the same threshold in each quantity and initial state.

\subsection{Dynamics of the Other Qubits in the Chain}\label{App:HeisenbergAppendix_otherQubits}

In the main text of the paper, we use the first qubit of the chain to demonstrate the dynamics of the expectation values and subsystem entropies in the Trotterisation of the Heisenberg model, and we argue that the onset of quantum chaos and the threshold position is independent of this particular choice by using eigenvector statistics of the Trotter step unitary as well as sub-system independent dynamical quantities such as PR and simulation fidelity. 
Here, we provide the dynamics of the expectation values and the subsystem entropies for the second, third and fourth qubits of the chain of 5 qubits.
Figs~\ref{AppFig:16_HeisenbergOtherQubs}~(a--c) (and (d--f)) show the dynamics of expectation values (and subsystem entropies), respectively, for (a) (and (d)) second, (b) (and (e)) third, and (c) (and (f)) fourth qubit.
We clearly observe the same threshold and same behaviours observed for the first qubit.


\section{Dicke Model}\label{App:DickeAppendix}


\begin{figure*}
	\centering
	\includegraphics[scale=0.94]{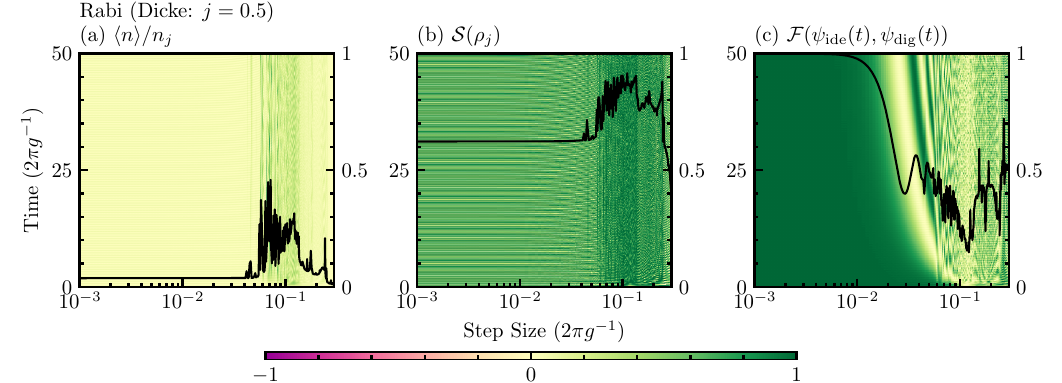}
	\caption{The sudden threshold in the Trotterised quantum Rabi evolution.
	Dynamical evolutions of the smallest case of the Dicke model ($j=1/2$ i.e. the quantum Rabi model) show the dynamical signatures of quantum chaos beyond the threshold.}
	\label{AppFig:17_RabiDynamics}
\end{figure*}

\subsection{Threshold in Rabi model}\label{App:RabiThreshold}

Here, we include the full dynamics of photon number, subsystem entropy, and simulation fidelity in the Rabi model (i.e., Dicke model with $j = 0.5$).
Figure~\ref{AppFig:17_RabiDynamics} shows that the Rabi model already shows a strong threshold in all these quantities, and we observe signatures of quantum chaotic dynamics beyond the threshold.
We see that the quasiperiodic dynamics of the (normalised) photon number is destroyed beyond the threshold (Figure~\ref{AppFig:17_RabiDynamics}~(a)), and that the subsystem entropy of the spin component is considerably larger than the pre-threshold value (Fig.~\ref{AppFig:17_RabiDynamics}~(b)), which is maximised for sufficiently large spin components in the Dicke model (Fig.~\ref{AppFig:10_OtherDynamicalSignatures}~(l)).
Finally, the threshold in simulation fidelity, shown in Fig.~\ref{AppFig:17_RabiDynamics}~(c), again appears as the secondary transition from (quasiperiodic) oscillatory-in-time to decaying behaviour.
Exploiting an extra flexibility afforded by the cavity degree of freedom (discussed in detail in Appendix\ Sec.~\ref{App:DickeEigTruncation}) to give a system size large enough to allow reliable statistics for $\XsqRMT$, we are also able to demonstrate statistically significant agreement with RMT, even in the eigenvector statistics of the Rabi model.
Due to its larger overall Hilbert space size at low spin values, unlike the other models, the Dicke model shows agreement with RMT predictions for all system sizes studied. 
It also shows certain additional cavity dimension truncation dependencies discussed in Appendix~\ref{App:DickeEigTruncation}.

\begin{figure}[ht]
	\centering
	\includegraphics[scale=0.94]{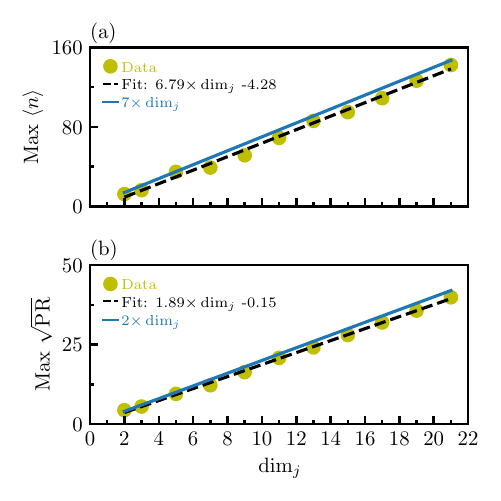}
	\caption{The maxima (over time) of (a) the photon numbers $\braket{n}$ and (b) the (unnormalised) PR values (shown as the square root) for dynamics in the quantum chaotic region, scale, respectively, linearly and quadratically (linearly for $\sqrt{\text{PR}}$) in $\dim_j$.
	In each sub-figure, the circles are the data points obtained numerically from the dynamical simulations, and the black lines are the numerical fits.
	The blue lines are the normalisation coefficients used in this paper.}
	\label{AppFig:18_DickeMaxVals}
\end{figure}

\subsection{Normalisation of the Photon Number and the Participation Ratio}\label{App:DickeNormCoefs}

In this section, we present numerical data supporting our choice of normalisation coefficients $n_{j}$ and $\mathcal{D}$ for, respectively, the photon numbers $\braket{n}$ and PRs in the Dicke model.
Figures~\ref{AppFig:18_DickeMaxVals}~(a--b) show that the maxima (over time) of (a) the photon numbers $\braket{n}$ and (b) the (unnormalised) $\sqrt{\text{PR}}$ values in the quantum chaotic regions are linear in $\dim_j$.
In this work, we use $n_{j} = 7\times\dim_j$ and $\mathcal{D} = (2\times\dim_j)^{2}$ to normalise the photon number and participation ratio, respectively, and the blue lines in Figs~\ref{AppFig:18_DickeMaxVals}~(a--b) show these values.
The exact choices used for the constant coefficients 7 and 2, which we selected based on the numerics, are not critical, as it is primarily their scalings that are important to enable a meaningful comparison between system sizes (e.g., see Fig.~\ref{Fig:4_DynamicalAverages} (c,f)).
The dynamical signatures are independent of the cavity truncation dimension $\dim_{\rm c}$, provided that $\dim_{\rm c}$ is large enough for the simulation to support all the components of the time evolving state.
Therefore, these observations (together with the maximised subsystem entropy of the spin component) suggest that the quantum chaos in this model is driven by the spin-$j$ component.
The eigenvector statistics analyses, in the next section, support this intuition as they show agreement with the RMT distributions only when the cavity truncation is $\dim_{\rm c} \sim \dim_j$.

\subsection{Eigenvector Statistics and the Effect of Truncated Cavity Dimension}\label{App:DickeEigTruncation}

In the numerical analyses of the Dicke DQS, the finite truncation of the cavity dimension poses a non-trivial problem.
Here, we provide a detailed numerical analysis of the truncation dependency in the eigenvector (and also eigenvalue) statistics of the Trotter step unitary of the Dicke model.
The results support and explain the cavity truncation choice used in the main text of the paper, giving some intuitive physical insight that suggests it is the spin component of the system that drives the quantum chaotic dynamics beyond the threshold.
They also provide a basis for future research into the subtle question of how to apply these statistical techniques with truncated dimensions in unbounded systems, which could provide a more rigorous understanding and explanation of these numerical results.

\begin{figure*}[ht]
	\centering
	\includegraphics[scale=0.90]{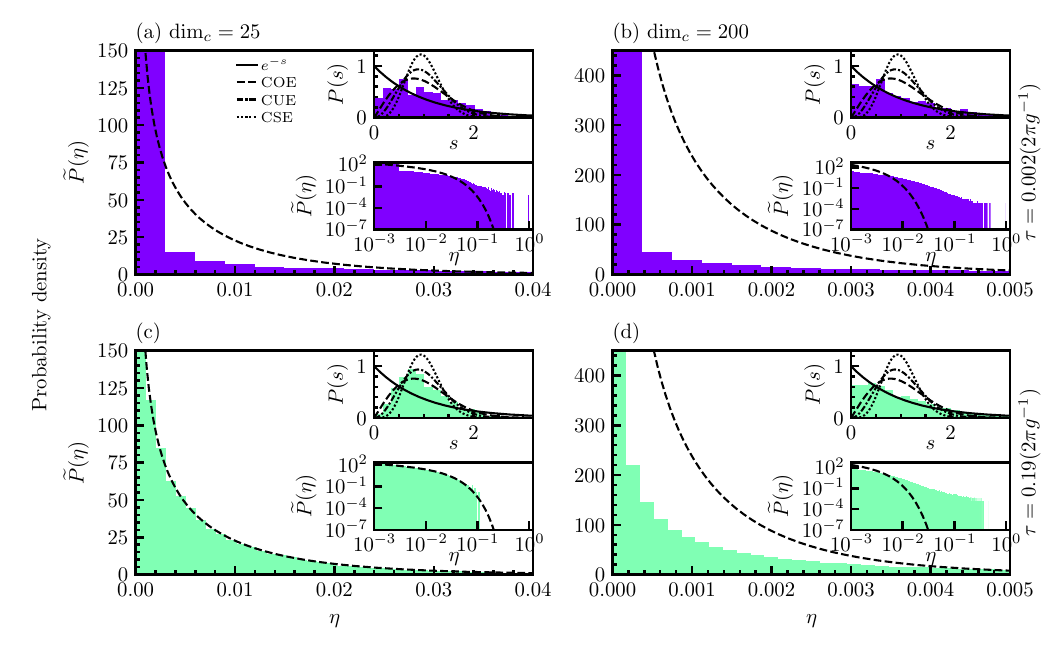}
	\caption{Eigenvector and eigenvalue statistics of Trotterised Dicke unitary with $j = 6$, and the system parameters are the same as the main text of the paper. In each panel (a--d), the main histogram plots show the eigenvector statistics, upper insets show the corresponding eigenvalue statistics, and the lower insets show the \emph{tail} of the main histogram plots.
	Before the threshold $\tau = 0.002~(2\pi g^{-1})$ neither the eigenvector nor the eigenvalue statistics show an agreement with RMT distributions for cavity dimension truncations (a) $\dim_{\rm c} = 25$ and (b) $\dim_{\rm c} = 200$.
	Beyond the threshold $\tau = 0.19~(2\pi g^{-1})$, both the eigenvalue and the eigenvector statistics are observed to be \emph{close} to RMT distributions for (c) $\dim_{\rm c} = 25$, and such agreements are lost both in the eigenvalue and the eigenvectors statists for (d) $\dim_{\rm c} = 200$.}
	\label{AppFig:19_DickeEigHists}
\end{figure*}

\subsubsection{Level-spacing and Eigenvector Statistics}

We first show the $j=6$ Dicke DQS eigenvector and eigenvalue (or level-spacing statistics) histogram plots, for different step sizes and cavity dimensions, to illustrate the effect of cavity dimension truncation.
The eigenvector statistics are analysed in a basis where the Trotter step unitary is symmetric (see Appendix~\ref{App:refBasis}).
As our Trotterisation is based on a two-part Hamiltonian, this is achieved by applying a unitary transformation $U=U_{TC}^\dagger (\tau/2)$, which transforms the Trotter step unitary to the following symmetric form:
\begin{equation}
	U_\tau \rightarrow U U_\tau U^\dagger = U_{TC}(\tau /2)U_{ATC}(\tau)U_{TC}(\tau /2).
\end{equation}
The level-spacing statistics (discussed in Appendix~\ref{App:eigenValStat}) complement these results as an additional static signature of quantum chaos.
For both eigenvector and eigenvalue analyses, the Dicke model parity symmetry must also be taken into account by analysing the statistics within each parity subspace ($\Pi = e^{i\pi(a^{\dagger}a + J_{z} + j)}$).
As discussed in Appendix~\ref{App:kickedTopStatic}, the relevant RMT distribution dimensions are therefore half the total size of the truncated space $\mathcal{D} = \frac{1}{2}\dim_j\times \dim_{\rm c}$.

Figure~\ref{AppFig:19_DickeEigHists} shows the level-spacing statistics in the upper-right insets, and the eigenvector statistics in the main figures with its \emph{tail} highlighted in log scale in the lower-right insets.
In Figs~\ref{AppFig:19_DickeEigHists}~(a--b), we use the step size $\tau = 0.002~(2\pi g^{-1})$ (i.e. before the threshold), and as expected, we see that neither the eigenvector nor the eigenphase statistics show agreement with RMT distributions.
The eigenphase statistics are actually Poisson-distributed for this step size, which is a signature of regular dynamics according to the Berry-Tabor conjecture~\cite{BerryMV1977LCRS}.
For the step size $\tau = 0.19~(2\pi g^{-1})$ (beyond the threshold) and a cavity dimension $\dim_{\rm c} = 25$, both the eigenvector and eigenphase statistics show agreement with RMT distributions, shown in Fig.~\ref{AppFig:19_DickeEigHists}~(c).
However, once $\dim_{\rm c}$ exceeds a certain value (that increases with $j$, as seen in the next section), agreement with RMT is lost in both of these cases; shown for $\dim_{\rm c} = 200$ in Fig.~\ref{AppFig:19_DickeEigHists}~(d) where the level-spacing statistics are actually closest to a Poisson distribution.
With the caveat of needing to choose an appropriate cavity truncation $\dim_{\rm c}$, eigenvector and level-spacing statistics both show corresponding agreement or disagreement with RMT beyond or before the threshold, and show the same behaviour in various different parameter regimes (not shown).

\begin{figure*}
	\centering
	\includegraphics[scale=0.94]{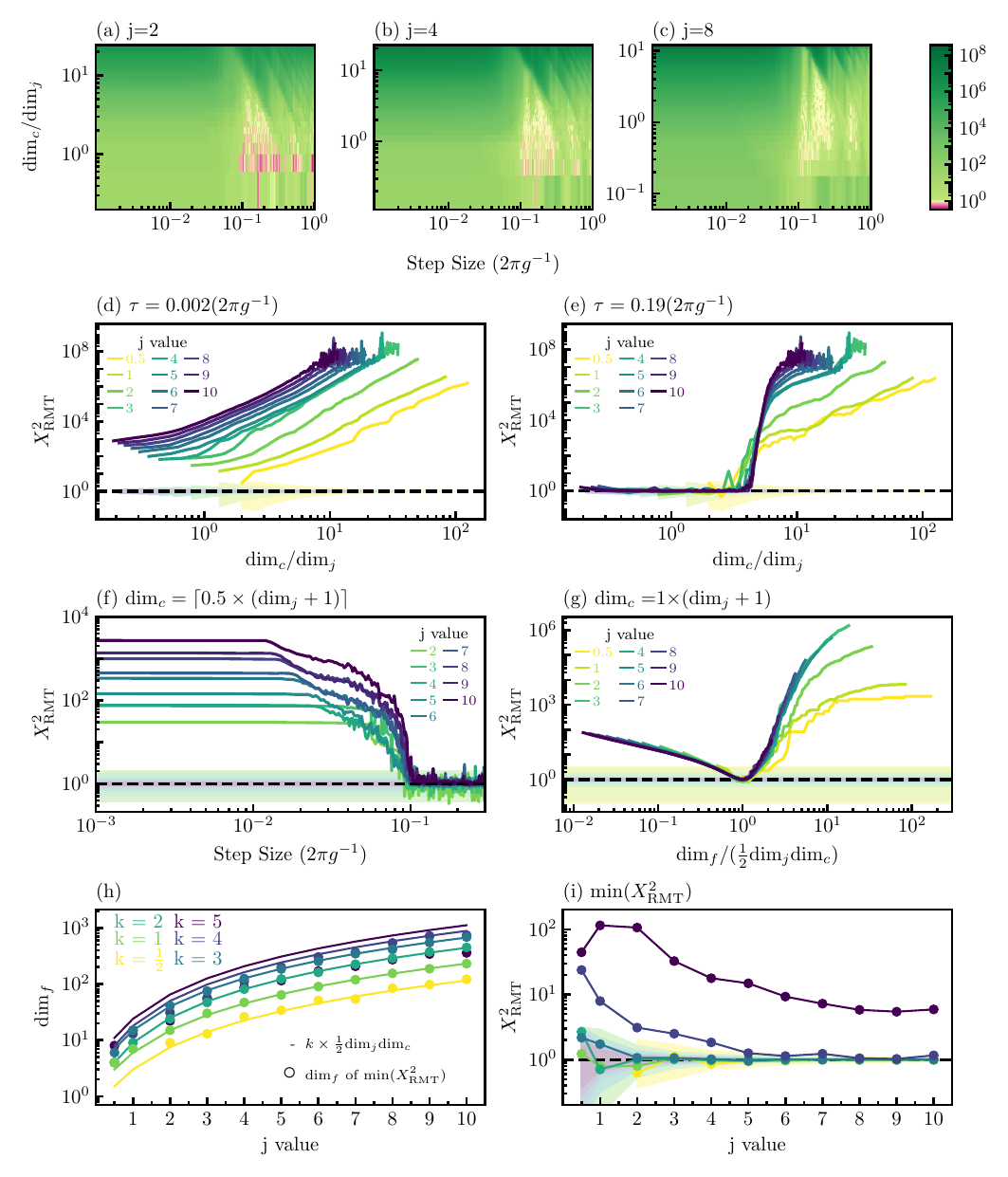}
	\vspace*{-0.35in}
	\caption{Dependency of eigenvector statistics on cavity truncation for Trotterised Dicke models with a range of $j$ values: (a--f) showing our results are reasonably robust, and (g--i) exploring the effective RMT system dimension.
	(System parameters are the same as in the main text.)
	(a--c) $\XsqRMT$ versus cavity truncation dimension $\dim_{\rm c}$ and Trotter step size $\tau$ for three different spin sizes, showing that the $\XsqRMT$ threshold is roughly independent of cavity truncation dimension, with $\XsqRMT$ not approaching near 1 before the threshold for any cavity truncation (shown in (d) for more spins $j$ for $\tau = 0.002~(2\pi g^{-1})$), and $\XsqRMT\sim 1$ in a sizeable region around $\dim_{\rm c}/\dim_j\sim1$ in the quantum chaotic region beyond the threshold (shown in (e) for a step size $\tau = 0.19~(2\pi g^{-1})$).
	(f) $\XsqRMT$ versus step size with the cavity truncation dimension chosen to be $\dim_{\rm c} = \lceil 0.5\times(\dim_j+1)\rceil$, compared with the main text of the paper, where $\XsqRMT$ is presented for $\dim_{\rm c}=\dim_j+1$ for the same spin values.
	(g) Fixing the cavity truncation dimension to $\dim_{\rm c}=\dim_j+1=2j+2$, and comparing the resulting eigenvector statistics against COE distributions of different RMT fit dimensions $\dim_{f}$, $\XsqRMT$ only reaches 1 for $\dim_{f} \sim \frac{1}{2}\dim_j\times \dim_{\rm c}$.
	(h) Comparison of the COE fit dimension $\dim_f$ that minimises $\XsqRMT$ against the expected value of $\dim_f=\frac{k}{2}\dim_j\times\dim_{\rm c}$, and (i) the minimum value of $\XsqRMT$ reached, when the cavity truncation dimension is fixed to $k\times\dim_j+\mod(k\dim_j^2,2)$, for various $k$.
	(d--g,i) The shaded regions, coloured by system size with colours paler than the corresponding line colour, indicate the $99\%$ confidence regions for the chi-squared distribution, to illustrate where statistically reliable conclusions can be drawn.
	}\label{AppFig:20_DickeTruncationDep}
\end{figure*}

\subsubsection{Goodness-of-fit Analyses}

In this section, we analyse the detailed dependency of eigenvector statistics on the truncation value of cavity dimension via the $\XsqRMT$ goodness-of-fit test statistic.
In the previous section, we compared the eigenvector and eigenphase statistics histograms of the Dicke Trotter step unitary for two different step sizes and cavity truncation values.
Here, providing further illustration of its value and efficiency in analysing statistical behaviours for RMT, we calculate $\XsqRMT$ as a function of various parameters: namely cavity truncation $\dim_{\rm c}$, Trotter step size $\tau$, spin size $j$, and an RMT fit dimension $\dim_{f}$.
We use this fit dimension to understand the effective system dimension:
That is, with the other parameters fixed (including the cavity dimension), we instead sweep the dimension variable of the RMT distributions, using $\XsqRMT$ to compare the numerical eigenvector histograms to RMT distributions with dimension $\dim_{f}$ instead of $\mathcal{D} = \frac{1}{2}\dim_j\times \dim_{\rm c}$.

The main observation from the results in this section is that, to observe $\XsqRMT \sim 1$, the cavity truncation dimension should be $\dim_{\rm c} \sim \dim_j$.
Together with the behaviours of the dynamical quantities (discussed in the main text and Appendix~\ref{App:DickeNormCoefs}), these $\XsqRMT$ results suggest that quantum chaotic dynamics in the Dicke DQS is mainly driven by the spin component.
This is consistent with a simple intuition that the non-commutations in the Dicke DQS result from single qubit rotations that affect only the collective spin component.
We nevertheless note the importance of developing a more theoretically rigorous understanding of these numerical results as an open question for further investigation.

\paragraph{Observation of the RMT threshold in a truncated Hilbert space}
Figures~\ref{AppFig:20_DickeTruncationDep}~(a--c) show colour plots for $\XsqRMT$ calculated as a function of both step size $\tau$ and cavity truncation dimension $\dim_{\rm c}$, for three example spins $j$.
These results firstly show that quantum chaotic dynamics in the Trotterised Dicke model, as illustrated by agreement with RMT statistics, emerge only after a threshold step size that appears to be largely independent of cavity truncation dimension as well as spin $j$.
As this threshold is observable for a wide range of $\dim_{\rm c}/\dim_j$ ratios, across 1--2 orders of magnitude around $\dim_{\rm c}/\dim_j\sim1$ for the spin sizes shown, this shows that the quantum chaotic dynamics are not an artefact that is highly sensitive to a particular choice of cavity dimension.
Yet, across this range, as the $\dim_{\rm c}$ used to analyse $\XsqRMT$ is increased, the extent of the quantum chaotic region beyond the threshold decreases.
This supports the intuition that Trotter errors related to the spin-cavity interaction in the Dicke model drive quantum chaotic dynamics within a finite subspace of the full spin-cavity Hilbert space, with the size of the participating portion of the infinite-dimensional cavity Hilbert space increasing in proportion to the spin dimension $\dim_j$.
Given this interpretation, our aim when using $\XsqRMT$ in the rest of this paper is to analyse the extent to which Trotter errors are creating quantum chaotic dynamics within that finite domain of interest.
To that end, our results in the rest of this section aim to identify an appropriate way to choose $\dim_{\rm c}$ directly from $\dim_j$ without looking at the $\XsqRMT$ values coming out of the analysis.

\paragraph{Sensitivity to cavity truncation}
Next, we analyse $\XsqRMT$ as a function of $\dim_{\rm c}$ for a range of $j$ values at one step size before ($\tau = 0.002~(2\pi g^{-1})$) and one after ($\tau = 0.19~(2\pi g^{-1})$) the threshold.
In this case, we determine the RMT fit dimension from the total dimension of the truncated system, i.e. $\dim_{f} = \frac{1}{2} \dim_j\times \dim_{\rm c}$ (accounting for parity symmetry).
Figure~\ref{AppFig:20_DickeTruncationDep}~(d) shows, for a larger selection of spin sizes than above, that the eigenvector statistics before the threshold do not agree with RMT for any value of $j$ or $\dim_{\rm c}$.
In the quantum chaotic region beyond the threshold, however, we observe that $\XsqRMT \sim 1$ (with agreement defined by the shaded regions representing the $99\%$ confidence regions for each system size) for cavity truncations in a region around $\dim_{\rm c}/\dim_j$ that becomes more sharply defined with increasing $j$ (for the step size shown in Fig.~\ref{AppFig:20_DickeTruncationDep}~(e), the $\XsqRMT \sim 1$ region appears to be asymptoting towards $0.3\lesssim\dim_{\rm c}/\dim_j\lesssim 5$), before rapidly transitioning to a region of large $\XsqRMT$ for larger truncations.
Though we leave it as an open question to find a rigorous explanation, our results show that there is a relatively wide range of cavity truncations that give good agreement with RMT.
In Fig.~\ref{AppFig:20_DickeTruncationDep}~(f), we plot $\XsqRMT$ versus step size for various system sizes, with $\dim_{\rm c} = \lceil\frac{1}{2}(\dim_j+1)\rceil = \lceil \frac{1}{2}(2j+2)\rceil$, showing results that are qualitatively the same as the main text, where we set $\dim_{\rm c} = \dim_j$ for the same spin values, with the same threshold position and overall behaviour, as further illustration that our main results regarding $\XsqRMT$ for the Dicke model do not depend sensitively on $\dim_{\rm c}$.

\paragraph{Effective RMT dimension}
Having observed agreement with RMT for a region around $\dim_{\rm c} \sim \dim_j$ in Figs~\ref{AppFig:20_DickeTruncationDep}~(a--c) and (e), we now fix the cavity truncation dimension to $\dim_{\rm c} \sim \dim_j$ and calculate $\XsqRMT$ for the same step size chosen in Fig.~\ref{AppFig:20_DickeTruncationDep}~(e), while varying the fit dimension $\dim_{f}$ for various $j$ values.
Note, we choose $\dim_{\rm c}=\dim_j+1=2j+2$, because that ensures that $\mathcal{D}=\dim_{\rm c}\times\dim_j$ is even, and this makes it easy to properly account for the Dicke model's two-fold parity symmetry.
(For the smallest case of $j=1/2$, we instead choose $\dim_{\rm c}=\dim_j+4=6$, because that gives a unitary that has enough components for reliable $\XsqRMT$ analysis, even once parity symmetry eliminates half of them, while still lying within the $\XsqRMT \sim 1$ region from Fig.~\ref{AppFig:20_DickeTruncationDep}~(e).)
The results in Figure~\ref{AppFig:20_DickeTruncationDep}~(g) confirm that $\XsqRMT \sim 1$ is observed only for $\dim_{f} \sim \frac{1}{2}\dim_j\times \dim_{\rm c}$.
Qualitatively similar shaped curves (not shown) are observed for other choices of cavity truncation dimension $\dim_{\rm c}=k\times\dim_j+\,{\rm mod}(k\times\dim_j^2,2)$ ($\dim_{\rm c}=\lceil k\times\dim_j\rceil+\,{\rm mod}(\dim_j \times \lceil k\times\dim_j\rceil,2)$ for $k=1/2$), but the curves and their minima shift upwards as the cavity truncation increases.

To explore the impact of the chosen $\dim_{\rm c}$ on the RMT fit dimension $\dim_f$, we plot the $\dim_f$ which minimises $\XsqRMT$ against spin $j$ for $k$ ranging between $k=0.5$ and $k=5$ (Figure~\ref{AppFig:20_DickeTruncationDep}~(h))---and the minimum $\XsqRMT$ value reached at the optimal $\dim_f$ (Figure~\ref{AppFig:20_DickeTruncationDep}~(i)).
In Fig.~\ref{AppFig:20_DickeTruncationDep}~(h), we see that the optimised $\dim_f$ closely follows the expected trend of $\dim_f=\frac{1}{2}\dim_j\times\dim_{\rm c}$, whenever the cavity truncation dimension is in the regime where the eigenvector statistics show agreement with RMT at the optiimised $\dim_f$.
When this is not the case---that is, when the observed statistics do \emph{not} agree with RMT statistics---it is probably not surprising that the $\dim_f$ which minimises $\XsqRMT$ doesn't match the dimension of the truncated Trotterised unitary.

\begin{figure}[t]
	\centering
	\includegraphics[scale=0.94]{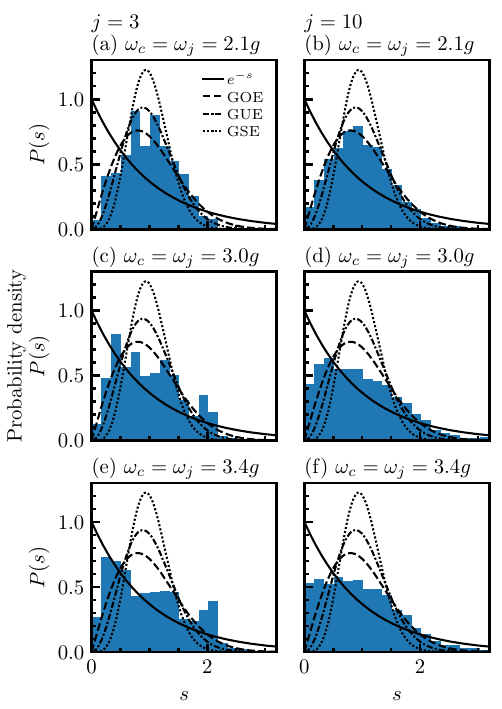}
	\caption{Finding \emph{non-chaotic} parameters of the Dicke Hamiltonian by calculating the level spacing statistics for $j = 3$ and $j = 10$ on each column.
	It is qualitatively insensitive to $\dim_{\rm c}$, which is set to 100 in these calculations.
	The dashed and dotted lines show RMT level spacing statistics for GOE, GUE and GSE universality classes, and the solid line shows Poissonian statistics.}
	\label{AppFig:21_DickeParameterChoice}
\end{figure}

\subsection{Quantum Chaos in Dicke model}\label{App:DickeChaos}

In order to focus on digitisation-induced quantum chaos, in the main text of the paper, we focus exclusively on \emph{regular} parameters of the Dicke Hamiltonian.
Here, we discuss how we choose these parameters, and also provide example results for the Trotterisation of the Dicke Hamiltonian with quantum chaotic parameters.

The Dicke model has previously been shown to be quantum chaotic for the coupling strengths $g$ around the critical coupling $g_{c} = \sqrt{\omega_{c}\omega_{j}}/2$ of the super-radiant phase transition (it becomes regular, if $g$ is either too much smaller or larger than $g_{c}$)~\cite{Clive2003CDM, CliveE2003QCD}.
For these regimes, these papers show that the level spacing statistics of the Dicke Hamiltonian follow Wigner-Dyson statistics.
They support these observations by obtaining a classical correspondence using Holstein-Primakoff transformations, together with certain approximations, and showing that the classical limit is chaotic for the parameters that give Wigner-Dyson statistics in the quantum case.

\subsubsection{Choice of the Parameters for the Main Text of the Paper}\label{App:DickeParamChoice}

We first show how we chose the Dicke model parameters used in the main text of the paper, by calculating the nearest-neighbour level-spacing statistics of the Dicke Hamiltonian for different $j$ values, and finding a parameter regime where it does not follow the Wigner-Dyson distribution.
To find a combination of spin and cavity frequencies far enough from the superradiant phase transition point of the Dicke Hamiltonian, we can move far into either the superradiant or normal phase regime.
Here we consider the latter regime, using resonant spin and cavity frequencies and gradually increasing the product of the frequencies $\omega_{c} \times \omega_{j}$ (and in turn the critical coupling $g_{c}$ of the superradiant phase transition) until Wigner-Dyson statistics are no longer observed.
Figure~\ref{AppFig:21_DickeParameterChoice} presents this analysis for $j = 3$ and $j = 10$, illustrating that the level statistics clearly do not agree with Wigner-Dyson statistics at both $j=3$ and $j=10$ by $\omega_{c} = \omega_{j} = 3.4g$.
Thus, any combination of spin and cavity frequencies that satisfy $\omega_{c} \times \omega_{j} \geq (3.4g)^{2}$ should be far enough from the $g\sim g_{c}$ quantum chaotic regime.

\begin{figure*}[ht]
	\centering
	\subfloat{\includegraphics[scale=0.94]{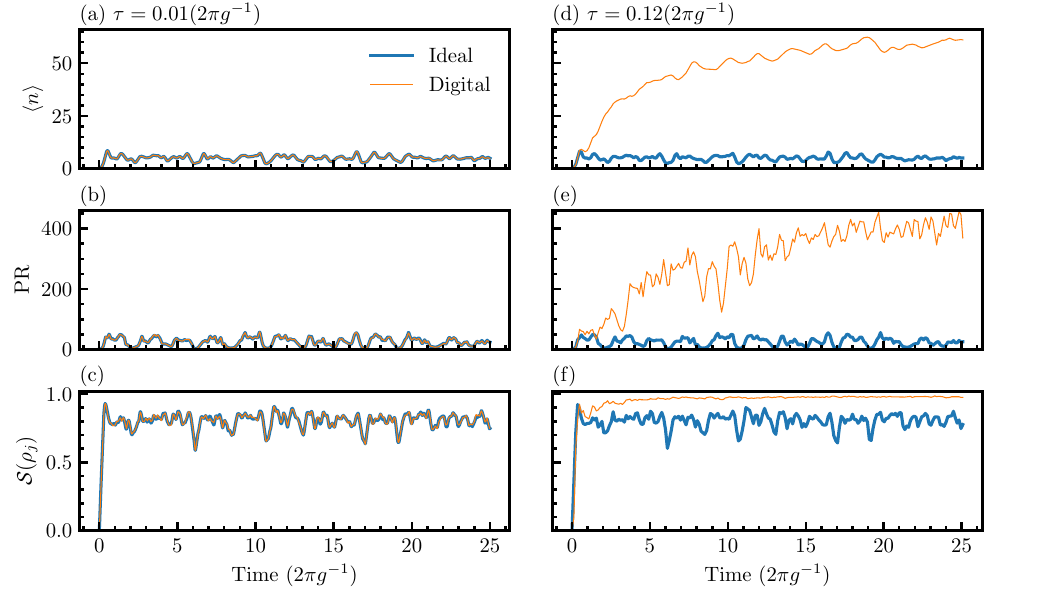}}
	\caption{Trotterisation of the Dicke model in its quantum chaotic regime ($0.5\omega_{c} = \omega_{j} = g = 1$ and $j = 6$).
	The dynamical evolutions of (a) and (d) $\braket{a^{\dagger}a}$ expectations, (b) and (e) PR, and (c) and (f) entropy of the reduced state of spin component $\mathcal{S}(\rho_{j})$ show more dominant quantum chaotic characteristic for digital simulation (orange) beyond the threshold (a--c).
	The digital simulation before the threshold (d--f) accurately simulate the quantum \emph{chaotic} Dicke.}\label{AppFig:22_IdealDickeChaos} 
\end{figure*}

\subsubsection{Trotterisation of the Dicke Model with Quantum Chaotic Parameters}\label{App:DickeChaos2Chaos}

While we study digitisation-induced quantum chaos---and hence a Dicke model with regular parameters---in the main text, we also observed qualitatively similar threshold behaviours for quantum chaotic parameters.
Detailed results for the quantum chaotic regime are not in the scope of this paper, but here we provide some example results for the Dicke model with the \emph{quantum chaotic} parameter choice of $0.5\omega_{c} = \omega_{j} = g$ ($j = 6$).
In Fig.~\ref{AppFig:22_IdealDickeChaos}, we compare the ideal and Trotterised dynamics for two step sizes, before and after the threshold, and observe that the Trotterised dynamics beyond the threshold show \emph{stronger} signatures of quantum chaos than the ideal, already quantum chaotic dynamics.
Figures~\ref{AppFig:22_IdealDickeChaos}~(a--c) show that ideal (blue) and Trotterised (orange) dynamics produce the same results before the threshold for (a) $a^{\dagger}a$ expectation, (b) PR, and (c) sub-system entropy of the spin component $S(\rho_{j})$.
We then make the same comparison beyond the threshold in Figs~\ref{AppFig:22_IdealDickeChaos}~(d--f), and observe that the Trotterised dynamics produce stronger dynamical signatures of chaos, namely:
(i) the destruction of quasiperiodicity in expectation values is much more prominent,
(ii) PR for DQS is an order of magnitude greater than ideal, and
(iii) entropy is maximised in DQS, larger than the ideal.

\bibliographystyle{quantum}

\providecommand{\noopsort}[1]{}\providecommand{\singleletter}[1]{#1}%

\end{document}